\documentclass[12pt]{article}
\usepackage{amssymb,amsmath,amsfonts}
\usepackage[dvips]{graphicx}
\graphicspath{{./ps/}}             
\usepackage{multirow}

\font\tenrm=cmr10 \font\sevenrm=cmr7 \font\fiverm=cmr5 
\font\teni=cmmi10 \font\seveni=cmmi7 \font\fivei=cmmi5
\font\tenex=cmex10
\font\tensy=cmsy10 \font\sevensy=cmsy7
\font\fivesy=cmsy5
  
\def\tenpoint{\def\rm{\fam0\tenrm}%
\textfont0=\tenrm \scriptfont0=\sevenrm \scriptscriptfont0=\fiverm
\textfont1=\teni \scriptfont1=\seveni \scriptscriptfont1=\fivei
\textfont2=\tensy \scriptfont2=\sevensy \scriptscriptfont2=\fivesy
\textfont3=\tenex \scriptfont3=\tenex \scriptscriptfont3=\tenex
   \normalbaselineskip=12pt%
   \let\sc=\eightrm	
\setbox\strutbox=\hbox{\vrule height8.5pt depth3.5pt width0pt}%
	\normalbaselines\rm}

\title{Universality and conformal invariance\\
for the Ising model in domains with boundary}
\author{
Robert P.\ Langlands\\
School of Mathematics, Institute for Advanced Study, \\
Princeton, NJ\ \ 08540, USA
\and
Marc-Andr\'e Lewis\\
Centre de recherches math\'ematiques, Montr\'eal, Canada, and\\
Laboratoire de Physique Th\'eorique et des Hautes Energies, \\
Universit\'es Pierre et Marie Curie (Paris VI) 
et Denis Diderot (Paris VII),\\
Paris, France\\
\and
Yvan Saint-Aubin\footnote{Supported in part by grants from the
Natural Sciences and Engineering Research Council of Canada
and the Fonds FCAR pour la formation de chercheurs
et l'aide \`a la recherche
(Qu\'ebec).}\\
Centre de recherches math\'ematiques and\\
D\'ept.\ de math\'ematiques et de statistique,\\
Universit\'e de Montr\'eal, C.P.\ 6128, 
succ.\ centre-ville,\\ Montr\'eal,
Qu\'ebec, Canada\ \ H3C 3J7
}
\date{\today}

\begin{document}

\maketitle

\begin{abstract}
The partition function with boundary conditions for 
various two-dimensional
Ising models is examined and previously unobserved
properties of conformal invariance and
universality are established numerically.
\end{abstract}

\bigskip

\begin{quote}
\end{quote}

\section{Introduction.}\label{intro}

\def\visible#1{#1}

 Although the experiments of this paper, 
statistical and numerical, were
undertaken in pursuit of a goal not widely shared, they may be of general interest
since they reveal a number of curious properties of the 
two-dimensional Ising model that had not been
previously observed. 

 The goal is not difficult to state. Although planar lattice models
of statistical mechanics are in many respects well understood
physically, their mathematical investigation lags far behind. Since these
models are purely mathematical, this is regrettable. It seems to us that 
the problem is 
not simply to introduce mathematical standards into 
arguments otherwise well understood; rather the statistical-mechanical consequences
of the notion of renormalization remain obscure.

 Our experiments were undertaken to support the view that the fixed
point (or points) of the renormalization procedure can be realized as
concrete mathematical objects and that a first step in any attempt to 
come to terms with renormalization is to understand what they are. We
have resorted to numerical studies because a frontal mathematical
attack without any clear notion of the possible conclusions
has little chance of success. We are dealing with a domain in which
the techniques remain to be developed.

 A fixed point is a point in a space of presumably an infinite number
of dimensions; so this point and
all other points of the space are defined by an infinite number of coordinates.
Some will presumably be superfluous, so that the total space is
realized as a submanifold of some larger coordinate space. The total space
will be the carrier, in some sense, of the renormalization transformation, 
but the transformation will not appear explicitly
in this paper. The point does! The implicit 
condition on each quantity serving as a possible coordinate of the fixed point
is that, at the fixed point itself, it remains invariant under renormalization
and that, at a critical point of any model
within the class considered, its value approaches a limit
under repeated renormalization because renormalization drives the critical
point to the fixed point.
Since repeated renormalization is in coarsest terms nothing more than passage 
to larger and larger blocks or to smaller and 
smaller mesh, the condition is that the quantity has a meaning as the mesh
length goes to zero, the dimensions otherwise remaining the same. 
For percolation this is a property of
crossing probabilities.
Our point of view is that any such quantity is a candidate as a
possible coordinate in the space of the fixed point. 
Rather than a single numerical quantity we can consider 
several at once, which amounts in the customary
mathematical way to considering objects lying in some given
space, finite-dimensional or infinite-dimensional, 
for example, a space of probability distributions, and if these
objects satisfy this criterion, thus if, for each model
at the critical point,
they tend to a limit as the mesh
goes to zero, 
then this limit or rather its coordinates
in the given space
can also serve as coordinates of the fixed point. 
Such objects are described in the paper.

 There are at least
two possibilities: one modeled on the considerations
for the free boson of \cite {L}; the other on the crossing probabilities
for percolation \cite{LPPS,LPS}. The second possibility was suggested to
us by Haru Pinson to whom we are grateful. Thus to each form 
$M$ of the Ising model (taken at the critical temperature) we will attach
two points ${\mathfrak p}_D(M)$ and ${\mathfrak p}_C(M)$,
each defined by an infinite
number of coordinates. Both are, in so far as this can be confirmed
by experiments, universal and conformally invariant in the sense of
\cite{LPS}. It is unlikely that these two points are independent.
One set of coordinates may well be deducible from another, but we
have not examined this possibility.

 Crossing probabilities may or may not be peculiar to a few models.
The evidence for their conformal invariance and universality is
not difficult to present and appears in Section \ref{trav}. There is, however,
one point to underline. The Ising model is
considered in regions that may be bounded or unbounded.
Crossing probabilities are defined for crossings within a region
that may or may not coincide with the region in which the model is
considered. It may be smaller. In contrast to crossing
probabilities for percolation, in which there is no
interaction, those for the Ising model
depend on both the region in which the crossings are allowed to
occur and the region of thermalization on which the Ising model
is considered. Conformal invariance refers to the simultaneous
action of a conformal mapping on the pair of regions.

The coordinates modeled on the free boson should,
on the other hand, be available for a large class
of models. Their definition is, in principle, quite general, but 
we have confined ourselves to the
Ising model.
The states $\sigma$ of the Ising model take values in the set $\{\pm 1\}$
which is contained in the set of all complex numbers $z$ with $|z|=1$.
This set in turn is covered by the line $z=\exp(2\pi ix)$. We simply develop
the circle on the line. We first assign to one site $p_0$
in the lattice the value $h(p_0)=0$ and then choose for all other $p$
the value $h(p)=m\pi$, $m\in{\mathbb Z}$, so that $\exp(ih(p)-ih(p_0))=\sigma(p)/\sigma(p_0)$.
Of course, there has to be more method than that. For example, for the square
lattice we introduce clusters: maximal collections of lattice sites of 
the same spin that are connected through bonds joining nearest-neighbor sites.
Each  cluster is surrounded by a curve constructed from edges of the dual
lattice and this curve separates it from all other clusters. To each of these curves
an orientation is assigned randomly and, for nearest neighbors $p$ and
$q$, we set $h(p)-h(q)=\pm\pi$,
the sign being determined by the relative orientation of
the bond from $p$ to $q$ and the curve it crosses. If it crosses no curve, then
$\sigma(p)=\sigma(q)$ and we take $h(p)=h(q)$. Thus to every state $\sigma$ are
attached several functions $h$, but $h$ determines 
$\sigma$ up to sign. For a finite lattice the measure on
$\mathfrak H$, the set of all possible $h$ is taken to be such that the measure, $m_I$, on $\mathfrak H$
assigns the same mass to all points lying above a given $\sigma$. Their sum
is one-half the mass of $\sigma$. 

 Fix now a bounded planar region $D$ and consider the Ising model in this region
with respect to a square lattice whose mesh $a$ approaches $0$. Since the model
is to be critical, the contribution to the Boltzmann weight of
a pair of neighboring spins is
$$
  e^{J\sigma_1\sigma_2},\quad \sinh(2J)=1,\quad J=.440687.
$$
Each $h$ is in effect a function
on the whole region if we take the value in the open square of side
$a$ about the site $p$ to be $h(p)$. (The ambiguity at the boundary is
disregarded here; it has to be confronted in various ways from experiment to
experiment.) If $C$ is a (smooth) curve in $R$, which can lie entirely in the interior
or run entirely or partially along the boundary, then we can restrict each function
in $\mathfrak H$ to $C$. This yields a set of points, each carrying a mass, in
the set ${\mathcal D}_C$ of Schwartz distributions on $C$, 
and thus a probability measure on  ${\mathcal D}_C$. Experiments to be described
in Sections \ref{dis1} and \ref{dis2}
suggest strongly that this measure has a limit as the mesh tends to $0$ and that
the limit is universal and conformally invariant. This is perhaps the most
important conclusion of the paper.

These measures have surprised the authors more than once.
When $C$ is the boundary,
the measure has a number of properties, to which we devote
considerable attention, that suggest it is gaussian. It
is not.

 There is no reason to restrict ourselves to planar
regions and we begin our study with the cylinder, because the ambiguities
at the boundary are then absent. A long cylinder
(effectively semi-infinite) is, provided we stay close 
to the end, to be regarded as equivalent
to a disk. The simplest conformally invariant distribution on the
set of distributions on the boundary of a disk is the gaussian
distribution with respect to the quadratic form defined by the
Dirichlet form. For a function $\varphi$ this form is obtained by extending 
$\varphi$ to a harmonic function $\tilde\varphi$ in the interior and then taking
\begin{equation}    
   \frac{g}{4\pi}\int\left(\left(\frac{\partial\tilde\varphi}{\partial x}\right)^2+
    \left(\frac{\partial\tilde\varphi}{\partial y}\right)^2\right)dxdy.
\label{eq:lagrangien}
\end{equation}
The first experiments described in Section \ref{dis1} strongly 
suggest that the measure given
by our construction is in some respects very similar to
this gaussian with a constant $g=g_B$ that appears to be
universal; the final experiments of that section
show, however, that it  differs
in important respects from a gaussian. If $C$ is interior, 
the measure on ${\mathcal D}_C$ is in no respect similar to a gaussian.

 Our construction is different from but not unrelated to familiar
constructions relating the Ising model to SOS models. For the 
Ising model on a triangular lattice
our construction is equivalent
in many respects to the usual one for the $O(1)$-model. In particular,
it is expected that in the plane
$$
   \langle (h(x)-h(0))^2\rangle \sim \frac2g\ln|x-0|.
$$
The constant $g=g_I$ is expected to be $4/3$, at least for the
triangular lattice, but the two constants $g_I$ and $g_B$
are not equal.
$$
        g_I=4/3,\qquad g_B=1.4710.
$$
Although $g_I$ is usually defined only for the triangular lattice,
it can be defined in general. We suppose that it is
universal, but we have not examined this
carefully. The pertinent experiments 
are briefly discussed in Section \ref{hh2}. The conclusion, which
will be reinforced more than once as we proceed, is 
that the comparison with the free boson
undertaken in this paper is quite different than
the usual one. 

 There is no reason that the two constants $g_I$ and $g_B$ should
be equal. One refers to interior behavior
in the bulk, the other to behavior
on the boundary. Moreover, as it turns out, they refer to different
aspects of a construction that leads to nongaussian measures with 
some gaussian behavior.
Although a departure from the conventional view,
it could be argued (we do not attempt to do so here)
that for questions of renormalization the constant $g_B$ may be
every bit as important as $g_I$, or, much better, that the distributions on
curves of Section
\ref{dis1} are at least as important for renormalization as asymptotic
behavior because renormalization, at least as it is often
presented, entails the fusion of bounded regions
along their boundaries (which may or may not partially coincide with that of
the region of thermalization). The measures on
${\mathcal D}_C$ were originally examined only for curves on or close
to the boundary. They appear, somewhat to
our surprise, to be of interest even in the
absence of a boundary. 
Indeed it may turn out, with hindsight, that the numerical arguments
towards the end of Paragraph 3.2 for the existence of nontrivial
and conformally invariant measures on interior curves are 
at least as important
as the other results, argued more elaborately and with more detail,
of Sections \ref{dis1} and \ref{dis2}.  

 Our point of view would not be at all persuasive if there were no sign
in our fixed point ${\mathfrak p}_D$, thus in the measures on ${\mathcal D}_C$,
of the critical indices $0$, $1/2$ and $1/16$. It is seen in Paragraphs
\ref{arbitre} and \ref{simconf} that these measures
do contain information about critical indices. Section
\ref{huit}, in which we
describe another manifestation of the
index $1/16$ as well as an interpolation of a formula of Cardy,
is also an essential part of the paper.
 
 The final section is less important. It contains a
few observations that provide some perspective on the definitions
of the paper.
First of all, the construction of $h$ is
by no means canonical. There are alternative constructions
decribed in Section \ref{alter}. We can 
allow jumps other than
$\pm\pi$, in particular several jumps 
$n\pi$, $n$ odd, with equal or different probabilities.
They lead to different values of $g_I$ and to a measure
on ${\mathcal D}_C$ with little resemblance to a gaussian.

 The possibility of not
using clusters in our sense but the clusters of Fortuin and Kasteleyn
that appear in the high-temperature expansion of the Ising (or more
generally the Potts) model
also suggests itself. Such clusters can also be used
to define the crossing probabilities. They lead to
different measures and to different crossing probabilities, whose universality and conformal
invariance we have not tested.
  
 Finally we point out that the
results for the distributions appear to remain valid at infinite temperature with,
of course, a different value for the parameter appearing in the distribution.

 We are grateful to Michael Aizenman, Christian Mercat, Haru Pinson
and Thomas Spencer for observations and suggestions that have been 
useful to us during the course of
these investigations.\footnote{RPL is grateful to Alois Scharf
for making the bank of computers of the Mathematisches Institut der
Universit\"at Bonn available to him at an important moment in the
experimentation.} 
Our most important debt is to the statistician 
Christian L\'eger whose counsel and
advice were of enormous help in the preparation of Section \ref{dis1}.

 Questions and comments of the referee, to which in particular the
discussion relating statistical dependence and the two-point correlation
function in Paragraph \ref{arbitre} is a response, led to
at least one important
modification of the conclusions in the original version of the paper. 
\section{Distribution of $h$ at the boundary.}\label{dis1}

\def\visible#1{#1}
\def\invisible#1{}

\subsection{The free boson on domains with boundary conditions.}\label{boson}

The partition functions of a free boson $\tilde\phi$, 
with compactification radius
$R$, are familiar objects when the domain is a torus, or 
a rectangle with the field satisfying
Dirichlet boundary conditions, thus equal to $0$ on the boundary. For a general
Riemann surface with boundary and for an arbitrary specification of the
field at the boundary, it may still be possible
to describe the partition functions explicitly (see \cite{CG}). For a cylinder
we use the formula of \cite{L}.
As it suggested
some of the statistical quantities for the Ising model studied in this
paper,
we review this formula.

The cylinder is described as the quotient of the region 
$0\le\Re w \le-\ln q$, $1\ge q>0$, in the complex plane
by the transformations generated by $z\rightarrow z+\omega$,
$\omega=2i\pi$.
If the fundamental domain is chosen to be $0\le\Re w
<-\ln q$ and $0\le \Im w< 2i\pi$, the map $w\rightarrow e^{-w}$ identifies
the cylinder with the annulus of outer radius $1$ and inner radius
$q$. We shall use freely the terminologies associated with the cylinder
and with the annulus. Observe that
$q$ is close to zero for long
cylinders. The angle $\theta$ is used as the parameter on both the inner and the outer
boundary.

The extremal fields $\tilde\phi$ on the domain are real harmonic functions
$$\tilde\phi(z,\bar z) = \phi_0 + a \ln z + b \ln \bar z + 
   \sum_{n\neq 0}(\phi_n z^n + \bar\phi_n\bar z^n).$$
The boundary conditions fix the restriction $\phi$ of
$\tilde\phi$ to the boundary. On the inner circle where $z=qe^{i\theta}$
and $\bar z=qe^{-i\theta}$, this restriction is
$$\phi_{\text{in}}(\theta)=\phi_0+(a+b)\ln q+i\theta(a-b) + 
   \sum_{k\neq 0}a^B_k e^{ik\theta}$$
with the reality condition $a^B_{-k}=\bar a^B_k$ and on the outer circle
$$\phi_{\text{out}}(\theta)=\phi_0+i\theta(a-b) + 
   \sum_{k\neq 0}b_k^B e^{ik\theta}$$
with $b_{-k}^B=\bar b_k^B$. (The superscript stands for boson.)
The compactification condition does not
require $\tilde\phi$ to be periodic but imposes a milder condition:
$\tilde\phi(e^{2i\pi}z,e^{-2i\pi}\bar z)=
\tilde\phi(z,\bar z)-2\pi nR, n\in \mathbb Z$, thus\ $(a-b)=in R, n
\in \mathbb Z$. Since the Lagrangian function (\ref{eq:lagrangien})
does not depend on the term $\phi_0$, this constant can be set to zero.
Therefore only the difference of the constant terms in $\phi_{\text{in}}$
and $\phi_{\text{out}}$ is of significance and we choose to parametrize
it with a variable $x\in[0, 2\pi R)$ and an integer $m\in \mathbb Z$:
\begin{equation}
-(a+b)\ln q=x+2\pi mR.\label{eq:x}
\end{equation}

The partition function on the cylinder with the boundary values of $\tilde
\phi$ specified by $\phi_{\text{in}}$, $\phi_{\text{out}}$, or
equivalently by $x, \{a_k^B\}$ and $\{b_k^B\}$, is a product of three terms [L]
\begin{align}
Z(\phi_{\text{in}}, \phi_{\text{out}}) &= Z(x, \{a_k^B\}, \{b_k^B\})\\
&=\Delta^{-\frac12}Z_1(x)Z_2(\{a_k^B\}, \{b_k^B\})\label{eq:z}
\end{align}
where $\Delta$ is the $\zeta$-regularization of the determinant of
the Laplacian for the annulus. It is given by $\Delta=-i\tau\eta^2(\tau)$
where $q=e^{i\pi\tau}$ and $\eta(\tau)=e^{i\pi\tau/12}\prod_{m=1}^\infty
(1-e^{2im\pi \tau})$ is the Dedekind $\eta$ function. Since this factor is
independent of the boundary data, it will be disregarded.
The crucial terms here are the two other factors $Z_1(x)$ 
\begin{equation}
\sum_{u,v\in\mathbb Z}e^{iux/R}q^{\frac{u^2}{4R^2}+v^2R^2}\label{eq:z1}
\end{equation}
and $Z_2(\{a_k^B\}, \{b_k^B\})$
\begin{equation}
\prod_{k=1}^\infty \exp\left[-2k\left(
(a_k^Ba_{-k}^B+b_k^Bb_{-k}^B)\frac{1+q^{2k}}{1-q^{2k}} -
(a_k^Bb_{-k}^B+b_k^Ba_{-k}^B)\frac{2q^k}{1-q^{2k}}\right)\right].
\label{eq:z2}
\end{equation}

If measurements
are made disregarding the variable $x$, only 
$$\int_0^{2\pi R}Z(x, \{a_k^B\}, \{b_k^B\}) dx$$
is of importance and this gives, after proper normalization, a 
probabilistic measure on the space of boundary data parametrized by
$(\{a_k^B\}, \{b_k^B\})$. The mixing of the boundary data at
both extremities becomes more and more intricate
when $q$ approaches $1$ or, in other
words, when the cylinder becomes a narrow ring. When $q$
is taken to zero, the measure simplifies as it becomes the product
of two terms, each one depending on $\{a_k^B\}$ or $\{b_k^B\}$. Moreover,
in the limit $q=0$, the probabilistic interpretation of $Z$ is simply
that of the gaussian measure in the variables $a_k^B$ and $b_k^B$.

Even though the Coulomb gas provides a description of the minimal models,
we do not know of any similar explicit formula for the partition functions
of these models for general boundary conditions, although Cardy's
paper \cite{C} treats explicitly the case of conformally invariant boundary
conditions. There are indeed only a finite number of these, and one of the difficulties
addressed in this paper is how to introduce continuously varying
conditions. Nonetheless we proceed
boldly using the partition function (\ref{eq:z}) as a guide 
for the Ising model. In contrast to the free-boson model,
the Ising model defined on a graph $\mathcal G$
does not have a field taking its values in the whole real line that we could 
easily identify 
with $\phi$ -- the spin field $\sigma$ takes  
its values in $\{+1, -1\}$. 
Starting from the spin field $\sigma$, defined on the sites of a (finite)
graph $\mathcal G$, one can construct the function $h$
as described in the introduction. It is such that, if $p$ and $q$ are joined, 
then $h(p)-h(q)=\pm \pi$ if $\sigma(p)\neq \sigma(q)$
and $h(p)=h(q)$ otherwise. If the graph $\mathcal G$ is embedded in a surface $D$,
for example, a cylinder or $\mathbb
R^2$, this function $h$ can be extended to a function locally constant on $D$
except on the edges of the dual graph
where it has jumps. The Ising measure $m_I$ on the space of configurations
on the graph $\mathcal G$
of mesh $a$ endows the (finite) set ${\mathfrak H}^a_D$ of possible functions
$h$ with a (discrete) probability measure. (As observed above, this measure
is such that $m^a_D(h)=2 m_I(\sigma)/N_\sigma$ where $N_\sigma$ is the number
of distinct $h$'s that lead to $\sigma$ and $-\sigma$.)

\newcommand{\entrylabel}[1]{\mbox{\emph{#1}}\hfil}
\newenvironment{MesListes}%
{\begin{list}{}%
   {\renewcommand{\makelabel}{\entrylabel}%
    \setlength{\parsep}{2pt}%
    \setlength{\topsep}{2pt}%
    \setlength{\partopsep}{0pt}%
    \setlength{\labelsep}{2pt}%
    \setlength{\itemsep}{4pt}%
    \setlength{\labelwidth}{5mm}%
    \setlength{\leftmargin}{30pt}
   }%
}%
{\end{list}}

Take the graph to be the subset of the lattice $a{\mathbb Z}^2$
of mesh $a=1/LV$ formed by the points
$(am,an)$, $0\leq m<LH$, $0\leq n\leq LV$. We identify upper and lower
edges and regard the graph as a subset of the cylinder: 
$z=m+in\rightarrow\exp(-2\pi z/LV)$. How can we compare ${\mathfrak H}^a_D$
to the field-theoretic measure of the free boson? Using the same letters
$a_k$ and $b_k$ (but without the superscript ``$B$'') for the Fourier
coefficients of the restriction of $h$ to the extremities of a cylinder:
$$h_{\text{in}}(\theta)=\sum_{k\in\mathbb Z} a_k e^{ik\theta}\quad
\text{and}\quad h_{\text{out}}(\theta)=\sum_{k\in\mathbb Z} b_k e^{ik\theta},$$
we study the dependence upon $a_k$ and $b_k$, $k\in\{-N,-N+1,\dots,N-1,N\}$
of the measure $m^a_D$ on ${\mathfrak H}^a_D$, disregarding all other
coefficients. The object obtained this way is a measure $m^{a,N}_D$ 
on ${\mathbb R}^{2(2N+1)}$ concentrated on a finite
set.
Keeping $N$ fixed, we then take the mesh $a$ on 
to zero. If the limit of the measures on
${\mathbb R}^{2(2N+1)}$ exists, presumably as a continuous
distribution, the limit as the number $2N+1$ of 
Fourier coefficients is taken to infinity can be considered.
We name the limiting object
\begin{equation}
m_D=\lim_{N\rightarrow\infty}\ \ \lim_{a\rightarrow 0}\ \ m^{a,N}_D.
\label{eq:lamesure}
\end{equation}
This measure, if it exists, is therefore defined on a space ${\mathfrak H}_I$ with
coordinates $(\{a_k\},\{b_k\})$ and we shall denote the elements of 
this space by $\phi_I$. This measure is to be compared with the
probability measure induced by (\ref{eq:z}) on the space of $\phi/R$. 
(The radius of compactification $R$ appears here because we normalized the
jumps of the function $h$ to be $\pm\pi$, forcing $h$ to change by an integral
multiple of $2\pi$ as $\theta$ winds around one extremity.) The first, and principal,
question is:
\begin{MesListes}
\item[(i)] does the measure $m_D$ exist?
\end{MesListes}
The parallel just suggested can be pushed further. We introduce first the
derivative $H=dh/d\theta$ of the restriction of $h$ to either extremity.
It is clearly a sum of delta functions concentrated half way between those
sites $p$ and $q$ at the boundary such that $\sigma(p)\neq \sigma(q)$.
The mass of each jump is $\pm\pi$. We shall use the letter $A_k$
for its Fourier coefficients,
$$H(\theta)=\sum_{k\in\mathbb Z} A_k e^{ik\theta}.$$
Clearly $A_k=ika_k$. At the other end we use $B_k=ikb_k$. 
We will use $A_k$ equally for
the coordinates parametrizing $\psi_I$, the derivative
$\frac{d\phi_I}{d\theta}$.
For the boson, the probabilistic
interpretation of the partition function $Z$ implies that the $k$-th Fourier
coefficient of the restriction $\phi_{\text{in}}$ is distributed 
(up to a normalizing factor) as
$\exp(-2k|a_k^B|^2)$ in the limit $q=0$. Consequently, if we use
the Fourier coefficients of
$$\frac 1R\frac{d\phi}{d\theta}=\sum_{k\in\mathbb Z}C_k e^{ik\theta}$$
with $ika^B_k=RC_k$, the probability density is $e^{-2R^2|C_k|^2/k}$, 
again up to normalization. For a long cylinder the parallel drawn here
raises the following questions on $m_D$ granted, of course, that the
answer to $\emph{(i)}$ is positive:
\begin{MesListes}
\item[(ii)] are the random variables defined by the
Fourier coefficients $A_k$ of $\psi_I$ distributed
normally as $e^{-\beta_k|A_k|^2}$?
\item[(iii)] is there a constant $R=R_B$ such that the constants $\beta_k$
are simply related by
\begin{equation}
\beta_k=\frac{2R_B^2}k?\label{eq:iii}
\end{equation}
\item[(iv)] is the joint distribution a product of independent single-variable
distributions?
\end{MesListes}

The rest of this section will describe the response to these questions 
provided by numerical simulations. The next will 
provide evidence that this limit measure is both universal and conformally
invariant.

\subsection{The distribution of $h$ at the boundary of a long cylinder.}
\label{cylinder}

\begin{figure}
\begin{center}\leavevmode
\includegraphics[bb = 74 240 534 554,clip,width = 10cm]{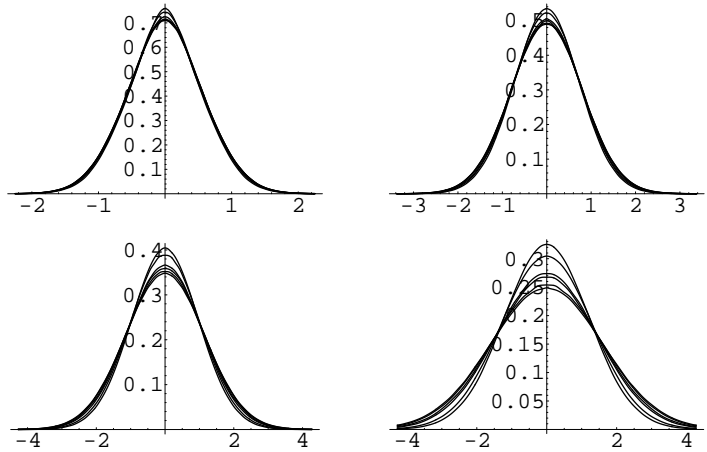}
\end{center}
\caption{The distribution of $\Re A_k$, $k=1,2,4,8$, with respect to the measure
$m_D^a$.
The mesh size $a$ corresponds
to the lattices $59\times 401$, $79\times157$,
$157\times 1067$, $199\times397$, $397\times793$ and $793\times1585$.
The curve $59\times 401$ is at the top when $\Re A_k=0$.
(See text.)\label{fig:2.1}}
\end{figure}

The diagrams of Figure \ref{fig:2.1} are some evidence for the existence of the 
distribution $m_D$ on the space ${\mathfrak H}_I$. They represent the probability
distribution densities of $H$ restricted to one of the extremities of 
various cylinders, in terms of a single variable (either $\Re A_1,
\Re A_2, \Re A_4$ or $\Re A_8$), all others being
disregarded or, thinking in terms of the limit, integrated out.
By rotational symmetry these densities are (almost) identical to those with
respect to the imaginary part of the same coefficients. 
(A small discrepancy could arise from the fact that the numbers of sites
along the circumference were not divisible by 4.)
The square lattices contained $59\times 401$, $79\times157$,
$157\times 1067$, $199\times397$, $397\times793$ and $793\times1585$ sites.
The first number ($LV$) is the number of sites around the circumference and is half
the number of sites along the length 
($LH$) minus one, or less. The Appendix gives some
further technical details on the simulations.
We note at this point that the partition function (\ref{eq:z}) is
obtained by summing over the integer $n$ parametrizing
the linear term $(a-b)=inR$ in both $\phi_{\text{out}}$
and $\phi_{\text{out}}$. The analogue of this term for Ising
configurations is straightforward: a configuration with exactly
two clusters (of opposite signs) extending from one end to the
other of the cylinder will have two longitudinal jump lines.
Depending on the choice of the jump across these lines,
$h$ will increase by $0, 2\pi$ and $-2\pi$ as $\theta$ wraps
around the boundary. Other (even) integral multiples of $\pi$
appear for configurations with more clusters crossing
from one end to the other and the numbers $2n\pi$ can be used
to partition the set of configurations. We have not differentiated
the measure $m_D$ for these various classes. We should add that,
for the cylinders studied in the present section, the configurations $h$
whose linear term is zero are by far the most probable.
The multiples $\pm2\pi$ occurred with a probability about $.0005$;
higher multiples we did not see at all.

Even though the raw data clearly differentiate the curves attached to
smaller cylinders, smoothing helped to separate
the curves between the two largest one ($397\times793$ and $793\times1585$).
This smoothing was done using the kernel method with a gaussian kernel;
the smoothing parameter was chosen according to Eq.\ (3.28) of [Si], in which
$\sigma$ was taken to be the sample standard deviation.

The narrowing of the gaps between the curves as the number of sites is increased
is a good qualitative argument for the existence of the limit 
$m_D=\lim_{N\rightarrow\infty}\lim_{a\rightarrow 0}m^{a,N}_D$. The peaks of
the curves go down systematically as $LV$ and $LH$ increase, except
for the dependence on $A_1$. In this case the center of the curve for
$793\times1585$ lies slightly above the center for $397\times793$ on
the small interval $(-0.05, 0.05)$. Around $\Re A_1 = - 0.05$ the two
curves cross and the curve for $793\times1585$ remains below that for
$397\times793$ until approximately $\Re A_1= + 0.5$. From then on ($|\Re A_1|
>0.5$) the two curves are so close that they cross each other several
times, probably due to the limitation of our samples.
This puzzled us and was checked independently by two of us. We have
no explanation for it. As will be seen below however, the variance of the samples,
a more global indicator, increases systematically over the spectrum
of all the cylinders considered;
in particular that of $793\times1585$ is larger than that of $397\times793$.

\def\er#1#2{\overset{#2}{#1}}

\begin{figure}
\begin{center}\leavevmode
\includegraphics[bb = 72 240 540 554,clip,width = 10cm]{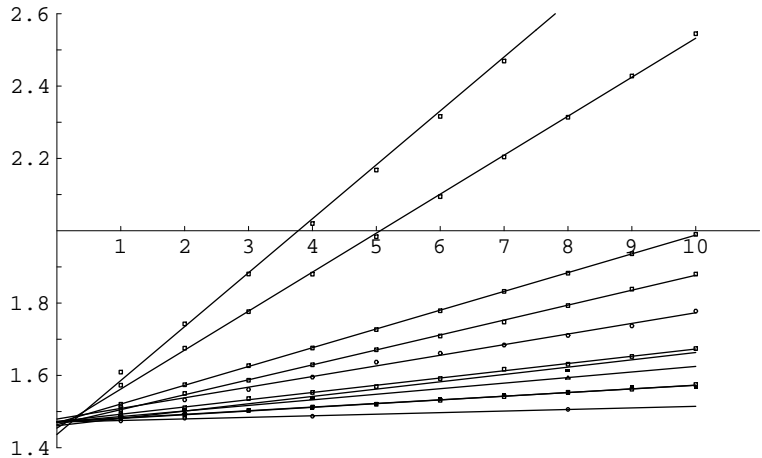}
\end{center}
\caption{The numbers $\hat\omega_k$, $k=1,\dots, 10$ for the
cylinders, the squares and the disk.\label{fig:2.2}}
\end{figure}

\begin{figure}
\begin{center}\leavevmode
\includegraphics[bb = 72 240 540 554,clip,width = 10cm]{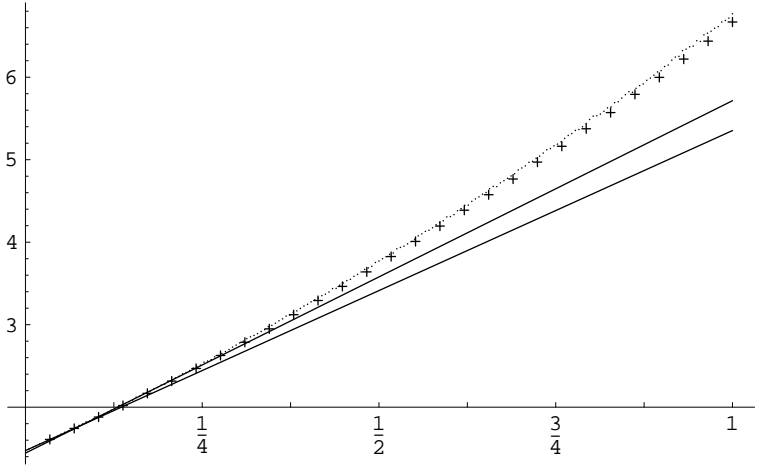}
\end{center}
\caption{The numbers $\hat\omega_k$ for $59\times401$ and $397\times793$
with the horizontal axis scaled proportionately to $1/LV$.\label{fig:2.3}}
\end{figure}

These distribution densities are so similar to normal curves that their
variances are a natural tool for a more qualitative assessment of the finite
size effects. In order to answer questions \emph{(ii)} and \emph{(iii)},
we plot in Figure \ref{fig:2.2}
the numbers 
$$\hat\omega_k^{LV\times LH}=\frac k{2(\hat\Sigma^{LV\times LH}_k)^2}$$
where $\Sigma^{LV\times LH}_k$ is the square root of the variance with respect to the
variable $\Re A_k$ for the cylinder with $LV\times LH$ sites. 
(If both questions were to be answered positively, then the numbers
$k\beta_k= k/2\Sigma_k^2$ for $\Sigma_k=\lim_{LV,LH\rightarrow\infty}
\Sigma^{LV\times LH}_k$ would be a constant. Note that we follow the usual 
statistical convention of distinguishing between the theoretical value $\alpha$ 
of a quantity and
its measured value $\hat \alpha$.)
We plotted
these numbers for $k=1, \dots, 10$ (or, sometimes, $k=1,2,4,8$),
together with a linear fit of these ten points
for every cylinder size on the square lattice considered,
the largest triangular and hexagonal lattices, the anisotropic lattice,
and the $254\times 254$ square and
disk geometries. The latter will be discussed in in Section \ref{dis2}.
The data, read from the top, appear in the order: 
cylinders of size $59\times401$, $79\times 157$, $157\times 1067$, 
$199\times 397$ for the square lattice $\mathcal G_\square$; 
of size $464\times 1069$ for the hexagonal lattice $\mathcal G_{\text{hex}}$;
then of size $397\times 793$ for $\mathcal G_\square$;
of size $312\times 963$ for the anisotropic lattice;
of size $416\times 721$ for the triangular lattice $\mathcal G_\triangle$; 
the cylinder of size $793\times 1585$ for $\mathcal G_\square$
and the square of size $254\times 254$ are superimposed; and 
finally the
disk of radius $r=300.2$ for $\mathcal G_\square$.
The numerical data are also recorded in Table I for $k=1,2,4,8$
together with those for triangular and hexagonal lattices on a 
cylinder and those on an ellipse covered by an anisotropic lattice. The digit
after the vertical bar gives the statistical error on the digit just
before; for example, 
the first element in the table ($1.609|3$) means
that $1/2(\hat\Sigma^{59\times401}_1)^2$ is $1.609$ with the
95\%-confidence interval being $[1.606, 1.612]$. The (statistical) error
bars were not drawn on Fig.\ \ref{fig:2.2} as their length is approximately the size
of the symbols used, or less. All the linear fits meet in a very small neighborhood
on the vertical axis. For the two cylinders with the greatest number
of sites, the disk and the two squares, the ordinates at the origin are all
in the interval $[1.47071, 1.47262]$, while the largest cylinder and the disk
meet at essentially equal values ($1.47071$ and $1.47095$ respectively).
It is likely that, for the two smallest cylinders,
a positive quadratic term would have improved
the fit and narrowed the gap with the intersection of the others.

\begin{figure}
\begin{center}\leavevmode
\includegraphics[bb = 72 240 540 554,clip,width = 10cm]{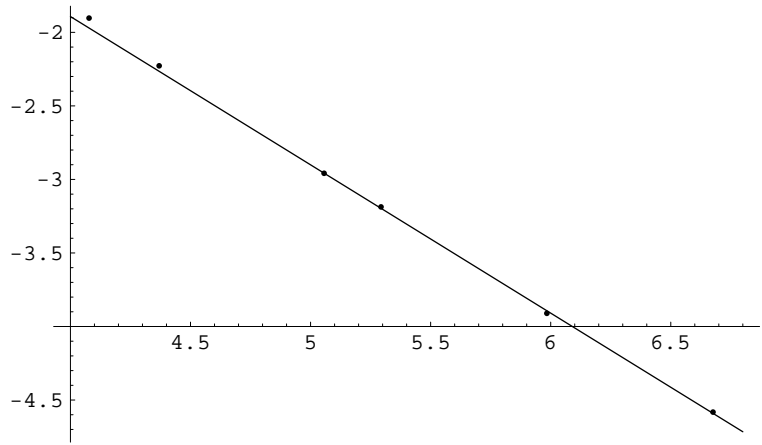}
\end{center}
\caption{Log-Log plot of the slopes of the linear fit of $\hat\omega_k$
as function of $LV$.\label{fig:2.4}}
\end{figure}

Figure
\ref{fig:2.3} reinforces this impression. The numbers $\hat\omega_k^{LV\times LH}$
were drawn for all the linearly independent Fourier modes (but the constant
one) for the cylinders $59\times 401$ and $397\times793$. Since each function
$H^{59\times401}$ is the sum of multiples of the same 59 $\delta$-functions
on the circumference,
it can be identified with a point in ${\mathbb R}^{59}$ that we choose to
parametrize with $A_0, \Re A_1, \Im A_1, \dots, \Re A_{29}, \Im A_{29}$.
Again the distributions with respect to $\Re A_k$ and $\Im A_k$ are identical
and the corresponding samples can be united. The $29$ crosses on the plot
are the data for $59\times401$ and the $198$ dots are those for $397\times793$.
The horizontal axis was scaled differently for the two cylinders: the data
were spread evenly on the interval $[0,1]$, starting at $\frac1{29}$ for
$59\times401$ and at $\frac1{198}$ for $397\times793$. The crosses and the
dots follow almost the same curve when scaled that way. Hence, the change
in the slopes for the various cylinders (Fig.\ \ref{fig:2.2}) can be seen to be the effect
of calculating the slope of a curve at the origin taking 10 values lying in an
interval of length proportional to $1/LV$. This is confirmed by a log-log
plot of these slopes (Fig.\ \ref{fig:2.4}). The six dots can be fitted 
linearly and the slope
is found to be $-1.031$ or, if the two smallest cylinders
are discarded, $-1.008$. These results are indeed very close to $-1$. (It is
this second fit that is drawn of the figure.)
Consequently the numbers $k/2\Sigma_k^2$ are likely to be all equal to one and
the same constant $2R_B^2$ whose four first digits are $1.471$.

This observation together with
the previous data indicates that the distribution $m_D$ quite probably exists
and that the variances $\sigma^2_k$ with respect to the variables $a_k=A_k/k$ are
inversely proportional to $k$:
\begin{equation}
\sigma_k^2=\frac1{k^2}\Sigma_k^2=\frac{\text{cst}}k\label{eq:ici}
\end{equation}
with ${\text{cst}}=0.3399$ close to, but unlikely to be, $\frac13$.
We have not discussed yet whether the distributions are gaussian but
the form (\ref{eq:ici}) is in fact in agreement with the form (\ref{eq:iii}), 
at least for the variances of the distributions with respect to
one of the variables when all the others are integrated. 
The constant $R_B^2$ is
therefore $0.7355$.

\bigskip
\begin{center}
\begin{tabular}{|l|l|c|c|c|c|}
\hline
&&&&& \\
Geometry (lattice) & Size & $\hat\omega_1$ & $\hat\omega_2$ & $\hat\omega_4$ & $\hat\omega_8$\\
&&&&& \\
\hline
&&&&& \\
\multirow{6}{25mm}{Cylinder ($\mathcal G_\square$)} & $59\times 401$& $1.609|3$ & $1.742|3$ & $2.020|4$ & $2.627|5$ \\ 
& $79\times157$    & $1.573|2$ & $1.675|2$ & $1.880|2$ & $2.314|3$ \\ 
& $157\times 1067$ & $1.520|3$ & $1.574|3$ & $1.676|3$ & $1.883|4$ \\ 
& $199\times397 $  & $1.506|6$ & $1.550|6$ & $1.629|7$ & $1.793|8$ \\ 
& $397\times793$   & $1.494|3$ & $1.511|3$ & $1.553|3$ & $1.631|3$ \\ 
& $793\times1585$  & $1.482|3$ & $1.491|3$ & $1.512|3$ & $1.553|3$ \\ 
&&&&& \\
\hline
&&&&& \\
Cylinder ($\mathcal G_\square$, $J_h=2J_v$) & $312\times 963$ & 1.487& 1.507 & 1.540 & 1.614 \\
&&&&& \\
\hline
&&&&& \\
Cylinder ($\mathcal G_\triangle$) & $416\times721$ & 1.491 & 1.496 & 1.536 & 1.593 \\
&&&&& \\
\hline
&&&&& \\
\multirow{3}{25mm}{Cylinder ($\mathcal G_{\text{hex}}$)} & $116\times267$ & 1.599 & 1.719 & 1.946 & 2.418 \\
& $235\times 535$ & 1.535 & 1.601 & 1.717 & 1.952\\
& $464\times1069$ & 1.502 & 1.530 & 1.560 & 1.716\\
&&&&& \\
\hline
&&&&& \\
Disk ($\mathcal G_\square$)& $r=300.2$   & $1.474|3$ & $1.482|3$ & $1.487|3$ & $1.506|4$ \\
&&&&& \\
\hline
&&&&& \\
\multirow{2}{40mm}{Ellipse ($\mathcal G_\square$, $J_h=2J_v$)}& major axis $=749.2$, & 
\multirow{2}{15mm}{\ \ $1.477$} &
\multirow{2}{15mm}{\ \ $1.480$} & 
\multirow{2}{15mm}{\ \ $1.489$} &
\multirow{2}{15mm}{\ \ $1.505$} \\
 & minor axis $=485.2$ & & & & \\
&&&&& \\
\hline
&&&&& \\
\multirow{2}{25mm}{Square ($\mathcal G_\square$)} & $80\times80$& $1.502|4$ & $1.535|4$ & $1.600|4$ & $1.728|5$ \\ 
& $254\times254$     & $1.480|5$ & $1.494|5$ & $1.510|5$ & $1.552|5$ \\
&&&&& \\
\hline
\end{tabular}
\end{center}
\medskip
\noindent Table I: The numbers $\hat\omega_k$, $k=1,2,4,8$ as measured
on the cylinder, the disk, the ellipse and the square. Only the square
lattice on cylinders is discussed in this section. See Section \ref{dis2} for the others.

\bigskip

We turn now to question \emph{(ii)}: are the Fourier coefficients $A_k$ of $\psi_I$
distributed normally as $e^{-\beta_k|A_k|^2}$?  
To address this question we used
three complementary methods that we shall refer to as the graphical
method, the method of moments, and the method 
of goodness-of-fit.

\begin{figure}
\begin{center}\leavevmode
\includegraphics[bb = 72 240 540 554,clip,width = 10cm]{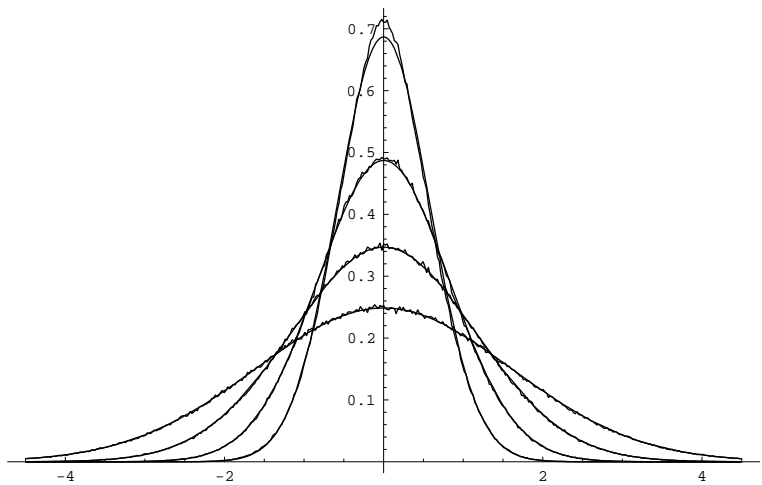}
\end{center}
\caption{Comparison of the empirical distribution $m_D^{793\times1585}$ 
as a function of $\Re A_k$, $k=1,2,4,8$ with gaussian densities whose variance
is the sample variance. The case $k=1$ is at the top.\label{fig:2.5}}
\end{figure}

Graphical methods seem a coarse way to assess whether an empirical distribution
is a given theoretical one. Still they are a  
natural first choice among the arsenal of statistical techniques
designed for this purpose.
Figure \ref{fig:2.5} plots the empirical histograms for $m_D$ as measured on
the cylinder $793\times 1585$ as functions of a single variable ($\Re A_1,
\Re A_2, \Re A_4$ and $\Re A_8$), all other dependence being integrated out.
For these plots we have
joined the data for $\Re A_1, \Im A_1, \Re B_1$ and
$\Im B_1$ which brings the sample to 1424000 configurations.
(The symmetries of $m_D$ insure that these variables are identically
distributed. We are not assuming here that they are statistically
independent. This will be discussed in the next paragraph.)
Besides these four empirical distributions, four normal curves have been
plotted whose variances are those of the data. (These variances can be deduced
from Table I.)
We have left these empirical distributions as they are, in contrast to those
seen on the Figures \ref{fig:2.1}, to distinguish them from the (smooth)
normal curves \emph{and} to give to the reader an idea of the difference
between raw and smoothed data. 
For the dependence on $\Re A_2, \Re A_4$ and $\Re A_8$, the ragged and smooth
curves are essentially identical and, based on this evidence, one is tempted
to claim that $m_D$ is distributed normally with respect to these variables.
The agreement between the two curves for $\Re A_1$ is clearly not so good. The
empirical curve lies above the normal curve at the center, crosses it before
$|\Re A_1| = 0.5$ and remains under it at least till $|\Re A_1| = 1.0$. 
The departure
from normality is statistically significant for the dependence upon $\Re A_1$.
After having observed this fact, one also sees, looking more closely, a gap
at the center of the curves for $\Re A_2$, though on a significantly smaller
scale. (It might not even be visible if Figure \ref{fig:2.5} has been
too compressed.) Since this departure from normality surprised us and, especially,
as it is easily observable only for $\Re A_1$, we
tried to explain it as a 
finite-size effect. The curves for the smaller cylinders
are however similar and the gaps seem similar to the eye. 
(The curves for $\Re A_1$ become wider as the
number of sites is increased, as is seen on Fig.\ \ref{fig:2.1}, but 
the variance of each sample also becomes larger.) 
If one is convinced of conformal invariance, 
discussed in Section \ref{dis2}, one can also use the data from an analogous 
simulation performed on a disk whose boundary contained 2400 sites.
For this geometry, the two curves for $\Re A_1$, empirical and gaussian, show
a similar gap. Thus, on graphical evidence only, we cannot conclude that the gap
seen between $\hat m_D(A_1)$ and the normal curve is a
finite size effect and that it is likely to disappear as $LV, LH\rightarrow \infty$.
The other distributions (for $A_2, A_4$ and $A_8$) are, however, extremely close
to gaussian.

Our first attempt at a more quantitative statement is through
calculation of the moments of the samples. We shall quickly see, however,
the limitations of this approach. We denote by $\mu_{k,i}^L$ the $i$-th moment
of the distribution $m_D^L$ with respect to the variable $\Re A_k$
$$\mu_{k,i}^L=\int (\Re A_k)^i m_D^L(A_0, A_1, \dots) dA_0 \prod_{l=1}^\infty
d\Re A_l\ d\Im A_l.$$
The even moments of the normal distribution are known to be
the mean (the $0$-th moment, in our case $0$ by definition), the variance 
$\sigma^2$ (the second moment, in our case an unknown) 
and $\mu_{2s}=(2s-1)!!\sigma^{2s}$. The first
five non-vanishing moments are therefore $\sigma^2, 3\sigma^4, 15\sigma^6, 
105\sigma^8, 945 \sigma^{10}$. None of the statisticians among our
colleagues suggested the moments as a quantitative tool, probably because of the
enormous errors that these measurements carry. Indeed the variance on a
measurement of $\mu_{k,i}^L$ is $(2i-1)!!\sigma_k^{2i}$ 
if $i$ is odd and $\bigl((2i-1)!!-((i-1)!!)^2\bigr)\sigma_k^{2i}$
if $i$ is even. Consequently the error on $\mu_{k,i}^L$
rapidly grows out of hand as $i$ increases. Nonetheless
the first ten moments were calculated for
the samples for the cylinders $59\times 401$, $79\times157$,
$157\times 1067$, $397\times793$ and $793\times1585$.

Since the distribution $m_D^L$ is, by definition, an even function in all
its variables, all the $i$-th moments, with $i$ odd, are zero. The data only
support this weakly, as more than 10\% of the odd moments lie outside what would
be the 95\% confidence interval if the distributions were gaussian.

\begin{figure}
\begin{center}\leavevmode
\includegraphics[bb = 72 240 540 554,clip,width = 10cm]{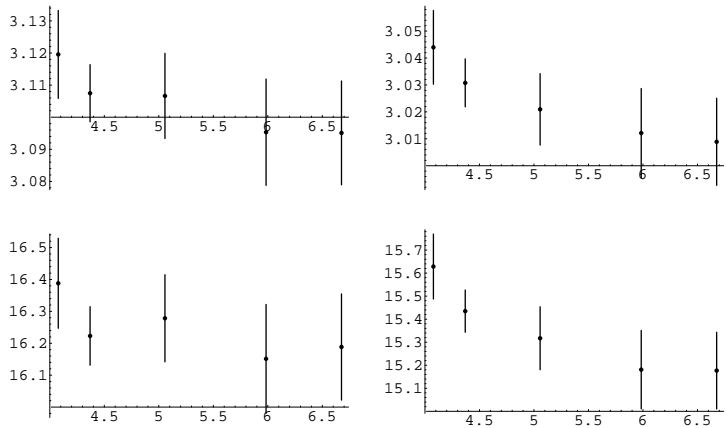}
\end{center}
\caption{The quotients $\hat\mu_{1,i}^L/\hat\sigma_1^{i}$ and
$\hat\mu_{8,i}^L/\hat\sigma_8^{i}$ for $i=4$ and $6$ as functions of
$\log LV$.\label{fig:2.6}}
\end{figure}

Even though the errors on the moments $\mu_{k,i}^L$ are large, 
it is instructive to plot some of the moments as functions of $\log LV$, for $LV=59, 79,
157, 397, 793$. Figure \ref{fig:2.6} shows the quotients
$\hat\mu_{1,i}^L/\hat\sigma_1^{i}$  and $\hat\mu_{8,i}^L/\hat\sigma_8^{i}$, 
for $i=4,6$, that should tend to $3$ and $15$ respectively if the 
limit distributions are gaussian. (The case $i=2$ is the variance and was 
discussed previously.) For the $8$-th Fourier
coefficient, these quotients are monotone decreasing
for both $i=4$ and $6$ and Fig.\ \ref{fig:2.6} repeats in another way
the visual observation made from Fig.\ \ref{fig:2.5} that the distribution
$m_D$ as a function of $\Re A_8$ is very close to a gaussian. The plots
for the first Fourier coefficient are less conclusive: the overall behavior
is decreasing, but not systematically, and the sixth moment is still rather far
from $15$, perhaps an indication that $15$ is not the limit.

The goodness-of-fit technique is our last attempt to quantify the
departure from normality of the dependence on the Fourier coefficients,
particularly of $A_1$. An overview of this technique (or
more precisely this set of techniques) is given in \cite{dAS}.
We are going to concentrate on the random variable 
\begin{equation}
w^2_n=n\int_{-\infty}^\infty \left(F_n(x)-F(x)\right)^2dF(x)\label{eq:w2}
\end{equation}
known as the Cram\'er-von Mises statistic (\cite{dAS}, chap.4). In this expression $n$ is the size 
of the sample and $F(x)$ the cumulative distribution function to
which the data are to be compared, in our case the gaussian whose variance
is that of the sample. If the data $x_1, x_2, \dots, x_n$ are ordered
($x_i\le x_{i+1}$), then the empirical distribution function $F_n(x)=
F_n(x;x_1, x_2, \dots, x_n)$ is a step function defined by
\begin{equation*}
F_n(x)=\begin{cases}
0,& x<x_1\\
\frac in,& x_i\le x\le x_{i+1}, \ \ i=1,\dots, n-1\\
1,& x_n\le x.
\end{cases}
\end{equation*}
The measure of integration $dF(x)$ in $w^2_n$ is equal to $f(x)dx$, 
where $f(x)$ is the probability
distribution corresponding to $F(x)$. The integral therefore gives more
weight to intervals in which the random variable $x$ is more likely to fall.
This is particularly well-suited for our purpose as the gap between empirical
and proposed distributions is precisely where the distribution peaks. 
Note that, if the data $x_i, i=1, \dots, n$, are not distributed according to the
distribution $F$ proposed, the variable $w^2_n$ will grow with the sample
size $n$. 

\begin{figure}
\begin{center}\leavevmode
\includegraphics[bb = 72 240 540 554,clip,width = 10cm]{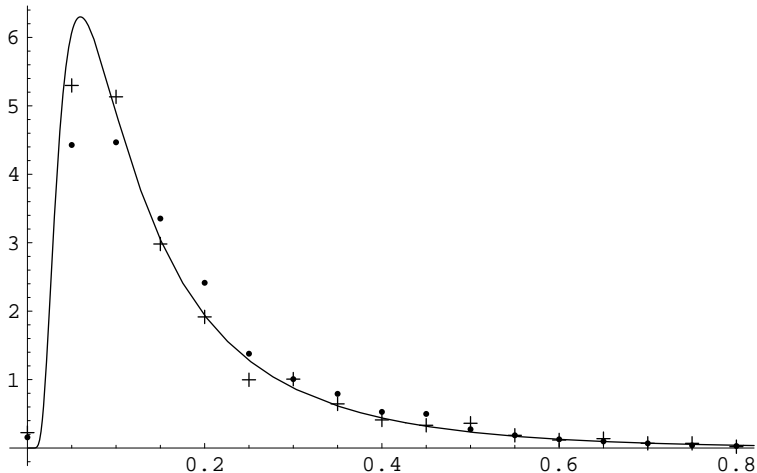}
\end{center}
\caption{The asymptotic distribution $a(w^2)$ together with
the histograms for $A_1$ (dots) and $A_8$ (crosses) for the
cylinder $59\times401$ and $n=1000$.\label{fig:2.7}}
\end{figure}

The null hypothesis $H_0$ is, henceforth, that $F_n(x)$ is a 
measurement of a variable whose distribution is $F$. 
Under the null hypothesis $H_0$, Anderson and Darling \cite{AD} gave an 
analytic expression $a(w^2)$ for the asymptotic probability distribution
of $w^2_n$, that is, the distribution of the variable $w^2_n$ when the sample
size $n$ is taken to infinity. We used their formula (4.35) to plot the curve
of Figure \ref{fig:2.7}.
In Chapter 4 of \cite{dAS}, Stephens indicates
corrections to be applied to $\bar w^2_n$ that allows finite samples to
be compared to the asymptotic distribution. 
For the $n$'s that will be used below these corrections are negligible.
The two first moments of the
distribution for $w^2_n$ are $\frac 16$ (independent of $n$) and 
$\frac{4n-3}{180n}$ \cite{PS}.

We concentrate on two Fourier coefficients, $A_1$ and $A_8$, as our goal
here is to see whether the departure from normality for $A_1$ can be
quantified and whether it decreases with the increase in
the number of sites of the lattice. Again we consider $\Re A_i, \Im A_i,
\Re B_i$ and $\Im B_i$ as independent and following the same statistics.
We can either split the whole available samples into smaller sets 
of $n$ elements or 
measure the variable $w^2_n$ for a very large $n$. With the first method,
a good average $\overline{w^2_n}$
can be calculated if the number $N/n$ of smaller sets is large enough.
The second method
will provide a single number that will, with luck, clearly reject $H_0$
if it has to be rejected. 
We apply both. 

In splitting the large samples into smaller sets, we have to make a careful
choice for $n$. One restriction comes from the actual values of the
$L^2$-integral that we want to measure. Using the data for the cylinder
$397\times 793$ in the first format described in the appendix (that is, grouped
in 401 bins), we can estimate an order of magnitude of 
$w^2_n/n = \int_{-\infty}^\infty (F_n(x)-F(x))^2 dF(x)$ for the whole
sample. (This was done using not the technique suggested by statisticians
\cite{dAS, AD} but using rather the naive Riemann integral over these 401 bins,
$F(x)$ being estimated at the center of these bins. We did not attempt to
evaluate the error in these calculations.) For $A_1$ this integral is
$0.000046$ and approximately 10 times smaller for $A_8$. Even
though there is an important statistical error on these numbers they
give us an idea of the order of magnitude.
We are therefore measuring a very small departure from normality if any.
On the one hand, the strategy of splitting the sample requires
to get an average 
$\widehat{\overline{w^2_n}}$ good enough that, if it is different from
$\frac16$, the difference is unlikely to be of statistical origin and
should instead indicate that $H_0$ needs to be rejected. In other words,
one should break the sample into several smaller samples to get a good average.
On the other hand,
if $H_0$ is false, the quantity $\overline{w^2_n}$ increases with $n$. 
Since the second moment of the distribution
for $w^2_n$ is rather large
($\approx\frac1{45}$), we need to choose $n$ 
large enough that the statistical 
error on $\overline{w^2_n}$ 
be reasonably smaller that the number itself. A rough estimate
of this error is given by $\sqrt{\frac1{45(N/n)}}$ where $N/n$ is the
number of sets obtained by splitting the sample of size $N$ into subsets
of $n$ elements. There is an obvious compromise to be struck and we chose
$n=1000$. 

We measured $\overline{w^2_n}$ for the three cylinders $59\times401$,
$157\times1067$ and $397\times793$ using the methods described in
\cite{dAS, AD}. These cylinders are the three runs whose data were kept
in the second format described in the appendix,
so that the exact values of all the $x_i$ were
available. This format allowed us
to compute again the coefficients $A_i$ (and $B_i$ for $397\times 793$).
The two histograms for $A_1$ (dots) and $A_8$ (crosses) for the
cylinder $59\times401$ are plotted on Figure \ref{fig:2.7} together
with the asymptotic distribution $a(w^2)$. The range $[0, 0.8]$ accounts
for more than 99\% of the observations. Even though the crosses seem
to follow more closely the curve than the dots, a quantitative 
assessment is not inappropriate.
The number $N/n$ of $\widehat{w^2_n}$ is at least $2064$ for each of
the three cylinders and, consequently, the statistical error on the 
resulting $\overline{w^2_n}$ listed in Table II is 
$2\sqrt{1/(45\times2064)}\sim0.0066$. Note that, for
$A_8$, the intervals of confidence around the average $\overline{w^2_n}$
always contain $\frac16$, the predicted mean. Any departure from normality
for $A_8$, if any, cannot be observed from this test.
For $A_1$, the predicted $\frac16$ always falls outside of the 95\%-confidence
interval, though barely so for $397\times793$. This confirms the graphical observation made earlier and forces us to reject
$H_0$.

As described earlier the other strategy is to compute the numbers
$w^2_n$ for a large $n$. We chose $n=250000$. The disadvantage of
doing so is clearly that one has a single measurement of $w_n^2$,
not an average. The results appear also in Table II. The
(single) $\widehat{w^2_n}$ for the dependence on $A_8$
is small for all three cylinders and the hypothesis that 
as the size of the cylinder
(as well as $n$) goes to infinity the distribution of $w^2_n$ approaches
$a(w^2)$ is totally acceptable. However, the values of $\widehat{w^2_n}$ of
$A_1$ indicate that, almost surely, they do not follow these statistics.
The null hypothesis $H_0$ must be rejected for $A_1$.

\bigskip
\begin{center}
\begin{tabular}{|l||c|c||c|c|}
\hline
 & \multicolumn{2}{c||}{$n=1000$} & \multicolumn{2}{c|}{$n=250000$}\\
\cline{2-5}
 & $A_1$ & $A_8$ & $A_1$ & $A_8$ \\
\hline
&&&& \\
$59\times401$    & 0.1818 & 0.1673 & 5.578 & 0.1877 \\
$157\times 1067$ & 0.1802 & 0.1638 & 3.490 & 0.0682 \\
$397\times793$   & 0.1745 & 0.1656 & 2.587 & 0.0947 \\
&&&& \\
\hline
\end{tabular}
\end{center}
\medskip

\centerline{Table II: The means $\overline{w^2_n}$ for $n=1000$ and
the number $\widehat{w^2_n}$ for $n=250000$.}

\bigskip

\begin{figure}
\begin{center}\leavevmode
\includegraphics[bb = 72 240 540 554,clip,width = 10cm]{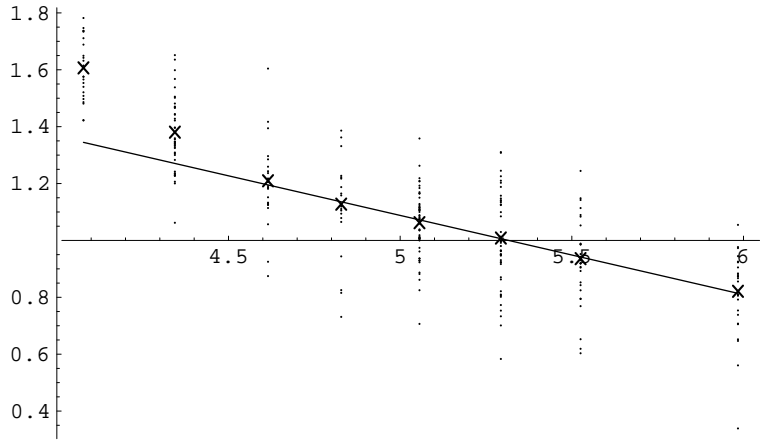}
\end{center}
\caption{Log-log plot of $w^2_n$, $n=250000$,
as a function of $LV$.\label{fig:2.8}}
\end{figure}

The null hypothesis refers, however, to a lattice of a given size
and it is not these with which we are ultimately concerned;
it is rather the limit of the distributions as the lattice
size tends to infinity
that is relevant. One obvious observation from Table II is that the gap between
the empirical data and a gaussian curve is narrowing as the
number of sites increases. Even though $H_0$ has been rejected,
we used the variable $w^2_n$, $n=250000$, to examine
the relationship between the gap and $LV$. We did further runs,
calculating only the value of $w^2_n$ for the dependence on $A_1$.
For each of the lattices $59\times 157, 77\times 155, 101\times 203, 
125\times 251, 157\times 313, 199\times 399, 251\times501$ and
$397\times793$, we obtained between $20$ and $53$ measurements
of the variable $w^2_n$. Since $H_0$ does not hold, we do not
know the distribution of this random variable. On the log-log
plot \ref{fig:2.8}, we drew the average for each lattice ($\times$) together
with the whole sample (dots). The spread in the sample for each lattice
shows that the variance
is very large and thus underlines the difficulty of 
obtaining a reliable mean for $w^2_n$. None the less the function $\bar w^2_n(LV)$
is monotone decreasing and the linear fit of the data (with the first
two excluded) plotted on Figure \ref{fig:2.8}
indicates that a power law, $(w^2_n-\frac16)\propto \alpha LV^\epsilon$
($\alpha\sim 2.48, \epsilon\sim -.278$),
is a reasonable hypothesis. We point out however that, with
our measurements of $\bar w^2_n$, we could hardly choose between 
the above power law or any of the form
$(w^2_n-x)\propto LV^\epsilon$ with $x$ in the interval $[0,1]$.
For this we would need $\alpha LV^\epsilon\ll1/6$ or
$LV\gg10000$. 

So, are the Fourier coefficients $A_k$ distributed normally?
For $k$ large enough (say $k\ge4$), it is impossible with our samples to see or 
calculate any difference between the empirical and the normal distributions.
For small $k$, particularly for $A_1$, the gap is obvious but the goodness-of-fit
technique provides clear evidence that it decreases as the size is increased.
That the gap vanishes as $LV,LH\rightarrow\infty$ is not a claim on
which we care to insist given only the present data.

\subsection{Statistical dependence and the two-point function.}\label{arbitre}

The previous paragraph studied the distribution $m_D$ with respect
to a single $\Re A_k$ or $\Im A_k$, all the others being integrated.
We now turn to the last question raised in Paragraph \ref{boson}, that of
statistical dependence of the variables $A_k$ and $B_k$.

The test for statistical independence that comes first to mind is
the correlation coefficients between the random variables $\Re A_k$, $\Im A_k$,
$\Re B_k$ and $\Im B_k$. These were calculated for the cylinder with
$397\times 793$ sites.
According to Ch.\ 5 of \cite{W} the
correlation coefficient of a pair of independent gaussian
variables is distributed with mean $0$ on $[-1,1]$
as
\begin{equation}
\frac{\Gamma(\frac{N-1}{2})}{\sqrt{\pi}\Gamma(\frac{N-2}{2})}
(1-r^2)^{\frac{N-4}{2}}dr.\label{eq:wi}
\end{equation}
Here $N$ is the size of the sample used to measure 
the correlation coefficient ($N=281000$ for the present
calculation) and no longer the cutoff
$N$ used to measure the distribution $m_D^{a,N}$.
If we set $r=s/\sqrt{N}$
and apply Stirling's formula,
(\ref{eq:wi}) becomes
approximately
$$
\frac{1}{\sqrt{2\pi}}(1-\frac{s^2}{N})^{N/2}ds\sim 
\frac{1}{\sqrt{2\pi}}e^{-s^2/2}ds.
$$
Of the correlation coefficients for {\em all} pairs of distinct variables
in $\Re A_k$, $\Im A_k$,
$\Re B_k$ and $\Im B_k$, $k=1,\dots, 198$, the largest turned out to be
$0.0097$, very small indeed. However this test is (almost) useless! 
The measure $m_D$ is invariant under rotation of the cylinder
around its axis, or at least, $m_D^{397\times 793}$ is invariant under
a finite subgroup. Under a rotation by an angle $\phi$,
the Fourier coefficient $A_k$ picks up a phase $e^{ik\phi}$ and
the expected value $E(A_k A_l)$ must vanish unless $k=-l$.
For
pairs of variables attached to the same extremity, the previous
numerical calculation is not useful. It is meaningful only for the pairs
$(\Re A_k, \Re B_k)$,
$(\Im A_k, \Re B_k)$,
$(\Re A_k, \Im B_k)$
and $(\Im  A_k, \Im B_k)$ of variables at different extremities,
but a more discriminating test of independence is certainly required.

The two-point correlation function of spins along the boundary
turns out to be a striking test for the independence of the variables
at one end of the cylinder. Because of the identification $\sigma(q)=e^{ih(q)}$
introduced in Paragraph \ref{boson}, the measure $m_D$ on the space of functions
$h$, or more precisely on the space ${\mathfrak H}_I$, should allow
for the computation of the correlation function $\langle \sigma(\theta_1)
\sigma(\theta_2)\rangle$ of spins along the extremity. Arguments
have been given in the literature (e.g.\ in \cite{C3, CZ}) that this two-point
function should behave as the inverse of the distance between the
two points, namely the cord length $\sin((\theta_1-\theta_2)/2)$ for
the geometries of the disk and of the cylinder. If we distinguish
between the functions $h$ and the elements $\phi$ of the limiting
space ${\mathfrak H}_I$, the function $\langle \sigma(\theta_1)
\sigma(\theta_2)\rangle$ should be $\langle e^{i 
(\phi(\theta_1)-\phi(\theta_2))}\rangle$
with
$$
\phi(\theta_1)-\phi(\theta_2)=\sum_{k=1}^\infty\{a_k(e^{ik\theta_1}-e^{ik\theta_
2})
+\bar a_k(e^{-ik\theta_1}-e^{-ik\theta_2})\}.
$$
(The relative minus sign between the $\phi$'s removes the irrelevant
constant term.) Now assume that the variables $\Re a_k$ and $\Im a_k$ are
statistically independent and normally distributed with variance $1/(2R_B\sqrt 
k)$.
Gaussian integrations lead to 
$$
\langle e^{i(\phi(\theta_1)-\phi(\theta_2))}\rangle=
\prod_{k=1}^\infty\exp\left(-\frac{|z_k|^2}{2kR_B^2}\right),
$$
with
$$|z_k|^2=|e^{ik\theta_1}-e^{ik\theta_2}|^2=2|1-\cos k\theta|,
\quad\theta=\theta_1-\theta_2. $$
Since $\sum_{k=1}^\infty \cos kx/k=-\ln(2\sin x/2)$, we
obtain up to an (infinite) constant
\begin{equation}
\langle 
e^{i(\phi(\theta_1)-\phi(\theta_2))}\rangle=\frac1{\sin^\alpha(\theta/2)}
\label{eq:deuxpoints}
\end{equation}
with $\alpha=1/R_B^2\approx 1.360$, a number that is not a simple fraction and 
certainly not 1.
Since the small departure from normality discussed in the previous paragraph
is unlikely to change much this result, there is here an obvious conflict 
between
the prediction $\alpha=1$ and this result based on the hypothesis of 
independence of
the $a_k$'s.

\begin{figure}
\begin{center}\leavevmode
\includegraphics[bb = 72 240 540 554,clip,width = 10cm]{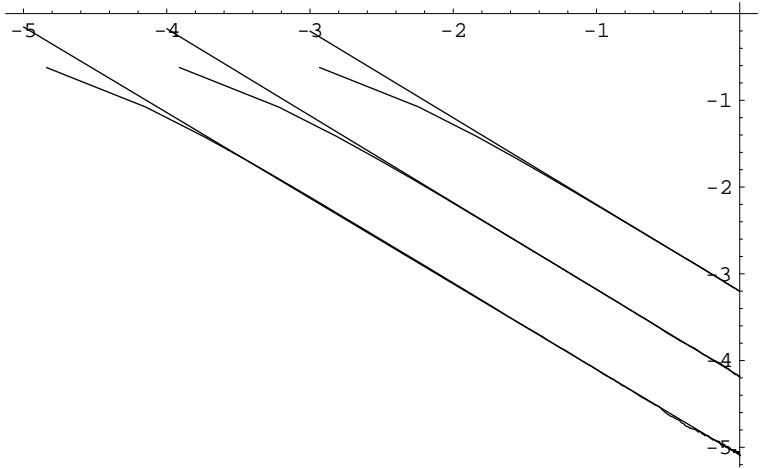}
\end{center}
\caption{Log-Log plot of $\langle \sigma(\theta_1)
\sigma(\theta_2)\rangle$ as a function of $\sin((\theta_1-\theta_2)/2)$
for the cylinders $59\times 401$ (top), $157\times 1067$ and
$397\times 793$ (bottom) together with linear fits.\label{fig:2.9}}
\end{figure}

We do not know if the prediction $\langle \sigma(\theta_1)
\sigma(\theta_2)\rangle\propto 1/\sin((\theta_1-\theta_2)/2)$ has
ever been checked through simulations. However the correlation can be retrieved 
easily
from our data for the cylinders $59\times 401$, $157\times 1067$
and $397\times793$. Figure \ref{fig:2.9} presents the results
together with the linear fits of the data after deletion of the seven first
(short-distance) points. The slopes of these fits are
$0.993$, $1.001$ and $0.988$ for the small, middle and large
cylinders. This prediction requires no further scrutiny.

We are left with the possibility that the variables $A_k$ are statistically
dependent. To show that this is most likely the case, we offer the following two 
data
analyses. We first  study
the conditional distributions of Fourier coefficients.  Namely, we
consider the distribution of
$m(\Re A_k |x_{min} <\Re A_l< x_{max})$, that is, the distribution of $A_k$
when $A_l$ is restricted to values between $x_{min}$ and $x_{max}$ and
all the others variables are integrated. Similar conditional distributions
with the imaginary parts are also considered.
If the Fourier coefficients were independent, every value or interval for the
restricted coefficient would lead to the same distribution. 

\begin{figure} 
\begin{center}
\leavevmode
\includegraphics[bb = 72 240 540 554,clip,width = 10cm]{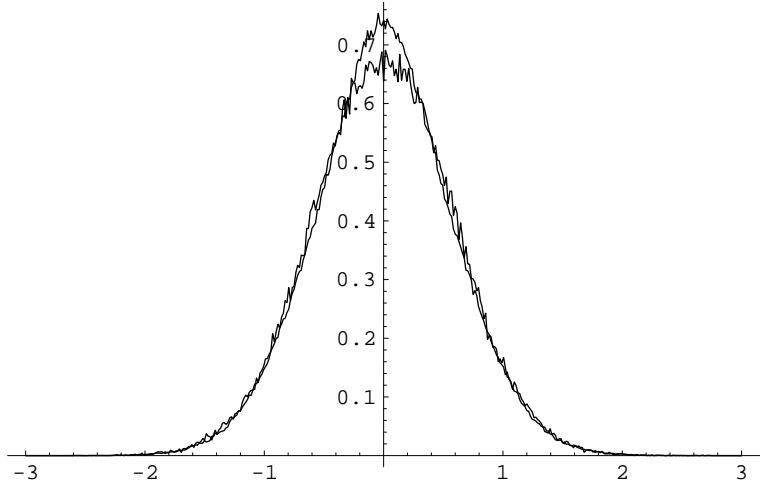}
\end{center} 
\caption{The conditional distribution of $\Re A_1$ on the $157\times 1067$
cylinder.  The top graph contains configurations 
with  $|\Re A_2|>1.125$, while the
lower one contains those with $|\Re A_2|<1.125$.
\label{fig:2.11}}
\end{figure}

In Figure \ref{fig:2.11}, we present the distribution of $\Re A_1$
given two windows on the values of $\Re A_2$, for a $157\times 1067$ cylinder.
The windows were chosen in such a way that both distributions had
similar statistics.  The numerical data clearly show that the two
distributions are different, and thus that these two Fourier
coefficients are correlated.  However, this correlation could be
affected by the finite size of our lattices.  This question of the importance of 
such effects is difficult to address. Since we have easy access 
to only three cylinder 
sizes, we omitted a rigorous study of finite-size effects.

Nevertheless, to acquire a feel for the dependence of $m(\Re A_k |x_{min} <\Re A_l< 
x_{max})$ on the choice of $k$, $l$ and the finite size, we
computed, for several values of $k$ and $l$, the ratio of the
variances of the conditional distributions of $\Re A_k$ when $|\Re
A_l|>1.125$ and $|\Re A_l|<1.125$ (which we will denote $r(\Re A_k,\Re A_l)$).  
We also made the same comparison
for the real part of $A_k$ and the imaginary part of $A_l$.  If the
distributions were independent, all these ratios would be one.  We
studied these ratios for cylinders of size $59\times 401$, $157\times
1067$ and $397\times 793$.  The first observation is that almost all 
these ratios diminish when lattice size increases, so that there is an
finite-size effect.  For example, $r(\Re A_1,\Im A_1)$
goes from $1.19$ for the 
$59\times 401$ cylinder to $1.05$ for the $397\times 793$ one.  Besides this 
finite-size effect, comparing ratios for different values of $k$ and $l$, we 
observed that the statistical dependence of Fourier coefficients $A_k$ and $A_l$ 
diminishes rapidly when $|k-l|$ increases, and is weaker 
for larger $k$ or $l$.  For instance, for the biggest cylinder, $r(\Re A_1,\Re 
A_2)=1.06$, while $r(\Re A_5,\Re A_6)=1.01$ and $r(\Re A_1, \Re A_{12})=0.99$.
These numbers are not conclusive, and further experiments would be
essential were there not another more compelling argument to establish
the dependence of the variables.

\begin{figure}
\begin{center}\leavevmode
\includegraphics[bb = 72 240 540 554,clip,width = 10cm]{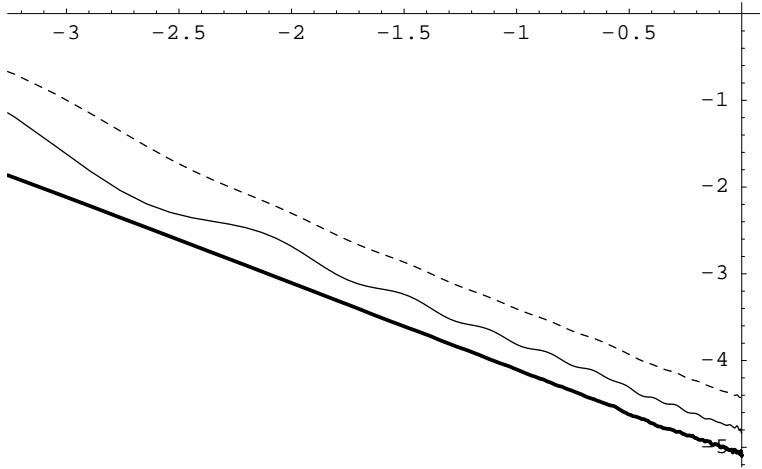}
\end{center}
\caption{Log-Log plot of $\langle \sigma(\theta_1)
\sigma(\theta_2)\rangle$ and of $\langle e^{i 
(\phi(\theta_1)-\phi(\theta_2))\rangle}$
as functions of $\sin((\theta_1-\theta_2)/2)$
for the cylinder
$397\times 793$. (See text.)\label{fig:2.10}}
\end{figure}

As the second analysis we measure the two-point correlation
$\langle e^{i (\phi(\theta_1)-\phi(\theta_2))}\rangle$ using the measure
$m_D$. This is {\em not} the same as directly measuring $\langle 
\sigma(\theta_1)
\sigma(\theta_2)\rangle$ from the configurations as we just did to obtain
Figure \ref{fig:2.9}. Recall
that $m_D$ is obtained by the limit $m_D=\lim_{N\rightarrow\infty}
\lim_{a\rightarrow 0} m^{a,N}_D$ (see eq.\ (\ref{eq:lamesure})). Consequently
we need to set a cut-off $N$ and compute the correlation function
on a sufficiently large cylinder using as an approximation for $\phi$
the truncation of $h$ to its $N$ first Fourier coefficients. If the
cylinder is large enough, the distribution $m_D$ as a function
of $A_k, k=-N,\dots, N$ will be fairly well
approximated by $m_D^{a,N}$. There remains the limit $N\rightarrow \infty$. To a
good approximation this limit may probably be forgotten altogether. 
The previous analysis showed that the dependence between Fourier
coefficients with small indices and those with large ones is significantly
smaller than the dependence amongst the first Fourier coefficients.
If this is so, the gaussian approximation and the independence
hypothesis are good ones for the distribution of $A_k$, $k>N$. If the
function $h$ being approximated is smooth enough, the error around 
$\theta_1-\theta_2=\pi$ for example should be of order $o(\frac1N)$ according
to the computation leading to (\ref{eq:deuxpoints}). By definition
the functions $h$ are piecewise continuous and their smoothness might
be improved by smearing functions as in the usual mathematical
treatment of Green's functions. (See Section \ref{hh2}.) 
We performed the calculation with and without smearing. 
The results for the cylinder
$397\times 793$ are shown on Figure \ref{fig:2.10}.
The thick curve is the log-log plot of $\langle \sigma(\theta_1)
\sigma(\theta_2)\rangle$ as a function of $\sin((\theta_1-\theta_2)/2)$ that
was plotted on Figure \ref{fig:2.9}. The middle, undulating curve has been
obtained by repeating the following two steps over the whole set of 
configurations: first replace the function $\phi$ by the
truncation $\hat h$ of $h$
to the sum of its $30(=N)$ first
Fourier coefficients and, then, add the resulting complex number
$e^{i (\hat h(\theta_1)-\hat h(\theta_2))}$ 
to the sum of the numbers previously
obtained. Only the real part of the average
is plotted as the imaginary one is essentially zero.
The first term neglected by the truncation ($a_{31}$)
is responsible for the wavy characteristic of the curve. The local extrema 
occurs
at every 6 or 7 mesh units in agreement with the half-period ($397/31/2\approx 
6.4$).
A linear fit of this curve (after deletion of the seven
first data) has a slope of $-1.027$. The top curve was obtained in a similar
fashion, except that the two steps were preceded by the smearing of
the function $h$. This smearing was done by convoluting the
functions $h$ with a gaussian whose variance was $2.5$ in mesh units. The wavy
structure is essentially gone. The curve appears above the two others because
the smearing introduces in $h(\theta_1)$ and $h(\theta_2)$
contributions of spins at points between $\theta_1$ and $\theta_2$ and thus
more strongly correlated. The smeared correlation
function is therefore larger than
the two others. A linear fit with the deletion of
the same short-distance data
gives nevertheless a slope of $-1.062$. 

The conclusion is thus that the random variables
$A_k$ (or $a_k$) are {\em statistically dependent} and that the computation
of $\langle e^{i (\phi(\theta_1)-\phi(\theta_2))}\rangle$ using the distribution
$m_D$ leads to the predicted critical exponent $\alpha=1$ for the spin-spin
boundary correlation. A consequence of the statistical dependence is that
we cannot offer as precise a description of the measure $m_D$ as would
have been possible if the answer to question {\em (iv)} had been positive.
This detracts neither from its universality nor from its conformal invariance.

\section{Universality and conformal invariance\\
of the distributions of $h$ on closed loops.}\label{dis2}

\subsection{Two hypotheses.}\label{twohyp}

Various crossing probabilities were measured in \cite{LPS} for several
percolation models at their critical points. Their
fundamental character was stressed by two general hypotheses, 
one of universality,
the other of conformal invariance, that were convincingly demonstrated 
by the simulations. The same two hypotheses will be demonstrated for the 
Ising model at criticality in Section \ref{trav}. In this section, 
we propose similar hypotheses for the 
distribution of the function $h$ introduced above and confront them
with simulations. 

We have considered in the previous section the Ising model 
on the square lattice. Other lattices could be used. The strength of
the coupling could vary from one site to another. Aperiodic lattices
could be considered or even random ones. 
It is, however, easier to be specific and to consider  
two-dimensional planar {\em periodic graphs} $\mathcal G$. We adopt, 
as in \cite{LPS}, the definition used by Kesten \cite{K} in his book 
on percolation: {\em (i)} $\mathcal G$ should not have any loops
(in the graph-theoretical sense), {\em (ii)} $\mathcal G$ is periodic
with respect to translations by elements of a lattice $L$ in $\mathbb R^2$
of rank two, {\em (iii)} the number of bonds attached to a site in $\mathcal G$
is bounded. {\em (iv)} all bonds of $\mathcal G$ have bounded length and every
compact set of $\mathbb R^2$ intersects finitely many bonds in $\mathcal G$
and {\em (v)} $\mathcal G$ is connected. An Ising model is a pair $(
{\mathcal G}, J)$ where $J$ is a positive function defined on bonds, periodic
under $L$. The function $J$ is to be interpreted as the coupling between
the various sites. 
Only some of the models $({\mathcal G}, J)$ will be critical, or, as often expressed,
each model is critical only for certain values of the couplings $J$.
The following discussion is restricted to models at criticality.

Let $D$ be a connected domain of $\mathbb R^2$ whose boundary is 
a regular curve and let $C$ be a parametrized regular 
curve (without self-intersection) in the closure of $D$. 
If $({\mathcal G}, J)$ is an Ising 
model, 
one can measure the distribution $m_{D, C}(\{a_k\}; {\mathcal G}, J)$ 
as we did in the previous section for $m_D$ on the square lattice.
(Although the coordinates $A_k$ will ultimately become our preferred
coordinates, we continue for the moment with the $a_k$.) 
The limit on the mesh can be taken either by dilating 
$C$ and $D$ with the dilation parameter going to infinity
while $\mathcal G$ fixed or by shrinking the planar lattice $\mathcal G$
uniformly while keeping $C$ and $D$ fixed.
As before we shall assume that the limit measure exists for every
regular $C$.
The previous section gave strong support for this supposition when
$C$ is the boundary of
$D$ and $({\mathcal G},J)$ the isotropic Ising model on the
square lattice. We examine the following hypothesis.
\medskip

\noindent{\scshape Hypothesis of universality}: {\itshape
For any pair of Ising models $({\mathcal G}, J)$ and $({\mathcal G}', J')$,
there exists an element $g$ of $GL(2,\mathbb R)$ such that 
for all $D$ and $C$} 
\begin{equation}
m_{D, C}(\{a_k\}; {\mathcal G}, J)=
m_{gD, gC}(\{a^g_k\}; {\mathcal G}', J').\label{eq:univ}
\end{equation}

\noindent The notation $gD$ and $gC$ stands for the images of $D$ and $C$
by $g$. The Fourier coefficients $a^g_k$ are obtained by integrating on $gC$
with respect to $\theta^g$, the image
by the linear map $g$ of the parameter $\theta$ on $C$.
The transformation $g$ \emph{does not} affect the underlying
lattice $\mathcal G$. For example, if ${\mathcal G}'$ 
is the regular square lattice,
it remains the regular square lattice. The domains $D$ and $gD$ are simply
superimposed on $\mathcal G$ and on ${\mathcal G}'$.
For the usual Ising models, those 
defined on other symmetric graphs (the triangular and the hexagonal)
with constant coupling or 
the model with anisotropic coupling on a square lattice,          
the matrix $g$ is diagonal. It is easy to introduce
models for which $g$ would not be diagonal. We have not done so
for the Ising models, but an example for percolation is to be found in \cite{LPS}.

To introduce the hypothesis of conformal invariance of the distributions
$m_{D, C}(\{a_k\}; {\mathcal G}, J)$, it is easier to restrict at first the 
discussion to the Ising model on the square lattice $\mathcal G_\square$
with the constant coupling function $J_\square$. A shorter notation will
be used for this model: $m_{D, C}(\{a_k\})=
m_{D, C}(\{a_k\}; {\mathcal G}_\square, J_\square)$. We endow $\mathbb R^2$ with
the usual complex structure, in other words we identify it with the
space of complex numbers in the usual way.
For this complex structure any holomorphic or antiholomorphic map $\phi$
defines a conformal map, at least locally. Given two domains $D$  and $D_1$ 
we consider 
maps $\phi$ that are bijective from the closure of $D$ 
to the closure of $D_1$ and holomorphic 
(or antiholomorphic) on $D$ itself. Thus $D_1=\phi D$.  Let $\phi C$ be the image
of $C$. 
\medskip

\noindent{\scshape Hypothesis of conformal invariance}: {\itshape
If $\phi$ satisfies the above conditions, then
\begin{equation}
m_{D,C}(\{a_k\})=m_{\phi D, \phi C}(\{a_k^\phi\})\label{eq:conf}
\end{equation}
where the Fourier coefficients $a_k^\phi$ appearing as arguments of 
$m_{\phi D, \phi C}$ are measured with respect to the arc-length parameter
on $\phi C$ in the induced metric, or equivalently as:
\begin{equation}
a_k^\phi=\frac1{2\pi}\int_0^{2\pi}h^{\phi D}\circ \phi_C(\theta) e^{-ik\theta}
d\theta
\end{equation}
where $\phi_C$ is the restriction of $\phi$ to $C$, $h^{\phi D}$
is the function $h$ on the domain $\phi D$ and $\theta$ is the (usual)
arc-length parameter of the original loop $C$.}
\medskip

\noindent Even though we have formulated this hypothesis for the Ising
model on the square lattice with constant coupling, it is clear that it can
be extended to any model $({\mathcal G}, J)$ using the hypothesis of 
universality.

\subsection{Simulations.}\label{simconf}

\begin{figure}
\begin{center}\leavevmode
\includegraphics[bb = 72 300 540 500,clip,width = 12cm]{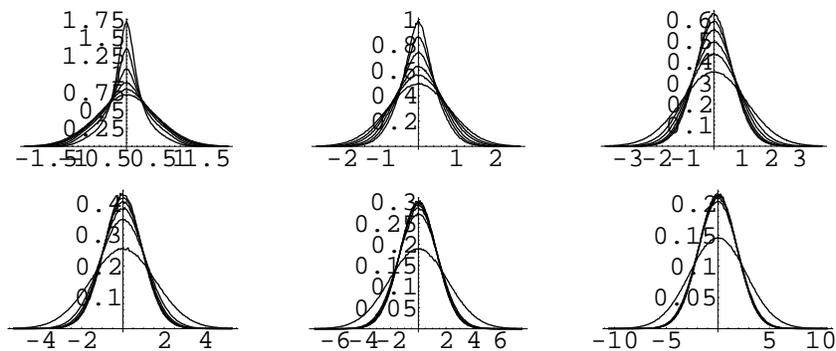}
\end{center}
\caption{Distributions $m_{D,C_i}$ as functions of the real
part of $A_k$ in the natural order: $k=1$,
$k=2$,
$k=4$,
$k=8$,
$k=16$,
$k=32$.\label{fig:3.1}}
\end{figure}

Since the curve $C$ is no longer necessarily an extremity of a cylinder,
our first step is to acquire some intuition about the measure $m_{D,C}$
for curves $C$ inside the domain $D$. To do so, we continue our investigation
of the cylinder 
for $({\mathcal G}_\square, J_\square)$. Thus $D$ remains the cylinder, 
but we select several curves
inside it. On the cylinder $397\times793$ the curves $C_i$ are sections
coinciding with the leftmost column ($C_0$), the $9$-th column ($C_1$),
the $17$-th ($C_2$), the $33$-rd ($C_3$),
the $65$-th ($C_4$) and the middle column ($C_5$). 
These curves are at a distance of
$0, 0.0201, 0.0403, 0.0806, 0.161$ and $0.997$ from the boundary measured
as a fraction of the circumference. 
We have not checked that the measurements on curves and
their mirror images with respect to the middle of the cylinder
are statistically independent. The closest pair 
(the curves on columns $65$ and $729$) are, however, at a distance
of 665 mesh units, that is, more than $\frac56$ of the full length of the
cylinder. So to the distributions on the first five curves
(all but the central one),
we have joined those on their mirror images,
doubling the numbers of configurations studied.
Figure \ref{fig:3.1} presents the measure $m_{D,C_i}$ as functions of the
real part of the Fourier coefficients $A_k$, $k=1,2,4,8,16,32$.
Each graph shows the dependence on a fixed $A_k$ for the six curves.
On each graph the lowest curve at the origin corresponds to $C_0=$ boundary, 
the case studied in Section \ref{dis1}. As the curve $C$ is taken closer
to the center of the cylinder, the distribution becomes sharper at the center.
This is perhaps to be expected as the sites at the boundary are freer to create
clusters of intermediate size than are the sites in the bulk, increasing thereby the values
of the various Fourier coefficients. (See Figures \ref{ising12} and
\ref{cond7A}.) Another natural feature is the gathering
of the distributions for all the interior curves on the plot for $\Re A_{16}$
and $\Re A_{32}$. Indeed the higher Fourier coefficients $A_k$
probe small scale structure, at the approximate scale of $\frac1k$ in 
circumference units. For example, the Fourier coefficient $A_{32}$ 
will be sensitive
mostly to clusters having a ``diameter'' of $\approx 12$ mesh units or less and
these clusters intersecting the curves $C$ at a distance of $32$
or of $64$ mesh units from the extremity
should be distributed more or less the same way. 
In other words the bulk behavior is reached closer to the boundary for
higher Fourier coefficients. One last
observation about these plots is that
the bulk distribution is definitely not a gaussian in $\Re A_1$! It is sharply
peaked at the center but still has a wide tail. (The distribution in
$\Re A_1$ measured along the mid-curve of the cylinder can be better seen
on Figure \ref{fig:3.2} below.)

\begin{figure}
\begin{center}\leavevmode
\includegraphics[bb = 72 290 540 500,clip,width = 10cm]{ising12.ps}
\end{center}
\caption{Two ``typical'' configurations on a disk of radius
200 with free boundary.\label{ising12}}
\end{figure}

To examine the hypotheses of universality
and conformal invariance,
we ran simulations on other pairs $({\mathcal G}, J)$ and
on other geometries.
We discuss both at the same time.
Three other pairs $({\mathcal G}, J)$ were considered:
the regular triangular and hexagonal lattices ${\mathcal G}_\triangle$ 
and ${\mathcal G}_{\text{hex}}$ with
the constant function $J$ and the regular square
lattice ${\mathcal G}_\square$ with a function $J$ that takes
a constant value $J_h$ on the horizontal bonds and another
constant value $J_v$ on the vertical ones with $J_h=2J_v$. 
We shall call this model the anisotropic Ising model.
This choice of $J$ makes the horizontal bonds stronger
than the vertical ones and clusters of identical spins
will have a shape elongated in the horizontal direction
as compared to those of the isotropic model 
$({\mathcal G}_\square, J_\square)$.

The critical couplings are determined by $\sinh 2J_h \sinh 2J_v=1$
(see \cite{B} or \cite{MW}). If the hypothesis of universality
is accepted then it follows from formula (5.9) of XI.5 
of \cite{MW} that the matrix $g$ that appears
in (\ref{eq:univ}) (when $({\mathcal G}, J)$
is the critical model on the square lattice and
$({\mathcal G}', J')$ the anisotropic model)
is\footnote{The point is that because of the anisotropy
the two-point correlation function decays more slowly
in the horizontal direction than in the vertical, behaving at large distance as
$1/(x^2+a^2y^2)^{1/4}$ with $a=\sinh 
2J_h\approx 1.54$. The appropriate conformal structure is
that defined by the ellipse $x^2+a^2y^2=1$.
We are grateful to Christian Mercat for  
this reference. See also his thesis (\cite{Me}) in which the conformal properties of the
Ising model are discussed from quite a different standpoint.}
$$\left(\begin{array}{cc}
1&0\\
0&\sinh 2J_h
\end{array}\right).$$
(The critical value of $J_h$ for which $J_h=2J_v$ is $0.609378\dots$.)
The lattice used for the anisotropic model has $LV=312$ and
$LH=963$. These dimensions correspond to a cylinder on the square lattice
with a horizontal/vertical ratio of $1.999$, very close
to the one used for the square lattice $397\times793$
that has a ratio $1.997$. The lattice used for the triangular
lattice was oriented in such a way that every triangle had one
side along the horizontal axis and the dimensions used were
$LV=416$, that is, the number of horizontal lines containing
sites, and $LH=712$, the number of sites on these lines.
The aspect ratio for a square lattice corresponding to these
numbers is $2.001$. The largest hexagonal lattice used was of size $464\times
1069$. Again $464$ is the number of horizontal lines containing
sites and $LH=1069$ is the length of the cylinder in mesh units.
The corresponding aspect ratio for the square lattice is $1.995$.
We also measured the smaller
hexagonal lattices of sizes $116\times 267$ and $235\times 535$.
The difference between these four ratios
is smaller than the limitation due to finiteness discussed in
\cite{LPPS}. The distances of the curves $C_i$ from the boundary
were chosen as close as possible to those used for the
cylinder on $({\mathcal G}_\square, J_\square)$ and given above. 
(The manner in which the Fourier coefficients of the restriction
of $h$ to these curves were calculated is described in the appendix.)

As evidence for the hypothesis of conformal invariance,
we compared three different geometries, namely the cylinder used in 
Paragraph \ref{cylinder}, a disk, and a square. We identify
the cylinder with the rectangle in the complex plane of height $v$ (its
circumference) and of length $h$. The analytic function $z\rightarrow
e^{-2\pi z/v}$ maps this cylinder onto an annulus. With our choice
of dimensions for the cylinder ($v=397, h=793$), the ratio of the inner
and outer radii is less than $10^{-5}$ and unless the outer diameter of the
annulus is larger than $10^5$, the inner circle contains a single site.
We took the liberty of adding this site to the domain and of identifying
it with a disk. In other words, although
the geometries of the cylinder and of the disk
\emph{are not} conformally equivalent in the sense of the hypothesis, the
finite size realization used here for the disk differs by a single
site from the annulus conformally equivalent to the cylinder. 
The radius of the disk was taken to be $300.2$. The disk can be mapped
onto the square by the Schwarz-Christoffel formula
\begin{equation}
\phi(z) = \int_0^z \frac1{\sqrt{(w^2-e^{i\pi/2})(w^2-e^{-i\pi/2})}} dw  \label{eq:sc-ch}
\end{equation}
which defines a map, with the unit disk as domain, holomorphic except
in the four points $\pm e^{\pm i\pi/4}$. Both maps satisfy our
requirements.
For the square and the disk, 
the distributions were measured at the boundary.
For the disk, they were also measured
on the four circles corresponding
to the inner circles on the cylinder that are not at its center. This latter
circle on the cylinder is mapped, inside the disk of radius $r=300.2$, 
onto a circle of radius $\approx 0.57$, less than one mesh unit. The 
distribution $m_{D,C}$ on this circle
is clearly impossible to measure for this lattice size.

Table I of the previous section has been completed with the data
$\hat\omega_k=k/{2(\hat\Sigma_k)^2}$ for six new experiments:
the three new Ising pairs $({\mathcal G},J)$ (triangular lattice,
hexagonal lattice and 
anisotropic function $J$) on the original cylindrical geometry; 
the two new geometries (disk and square) covered by the square lattice;
an ellipse covered by the square lattice with anisotropic interaction.
For the square, two runs were made on a lattice of $80\times80$ and
$254\times254$ sites. The data for the disk and both squares were
also drawn on Figure \ref{fig:2.2}. As discussed previously,
it can be seen there that their $\hat\omega_k$'s
follow exactly the same pattern as those of the cylinders and that
the ordinate at the origin of their fits falls in the same very small
window $[1.47071, 1.47262]$. 
It is interesting to notice that the small lattice $80\times80$ on
the square geometry leads to $\hat \omega_k$'s that are between those
of the lattices $199\times397$ and $397\times793$ for the cylinder.
Considering that the number of sites in the lattice $199\times397$
is more than twelve-fold that in the small square, this might seem
surprising. The explanation is likely to be that the number of sites
on the boundary where the distribution is measured is the leading
cause of the finite size effect.
The $\hat\omega_k$'s for the triangular
lattice and for the anisotropic model were obtained from the 
401-bin histograms of the empirical distributions. (See the appendix.)
No attempt was made
to provide confidence intervals. 
The linear fits of the $\hat\omega_k,
k=1,2,4,8$, are $1.4723 + 0.0152 k$
(triangular lattice) and $1.4695 + 0.0180 k$ (anisotropic model).
For the largest of the square lattices it was $1.4712+0.0102 k$
and for the disk $1.4710 + 0.0044 k$.
The ordinates at the origin ($1.4723$, $1.4695$ and $1.4712$) are extremely 
close to the narrow window above for the larger cylinders,
the disk and squares, especially striking as
the samples for these experiments (200K) were the smallest of all in
this section and the previous one. The linear fit for the largest of the
hexagonal lattices
is $1.4793+0.0294 k$ and the ordinate at the
origin is not quite so good but the slope remains large compared with the
other fits. 
Indeed the product of the slope and the circumference $LV$ is
in the four cases: $\sim 8.1$ (square); $\sim 6.3$ (anisotropic); $\sim 6.4$ (triangular); $\sim 14.$ (hexagonal).
This suggests that the circumference of the hexagonal lattice must be
twice that of the triangular lattice in order to obtain comparable results, 
perhaps because it contains only half as many bonds per site
as the triangular lattice.
The ordinate at the origin ($1.4770$) is nevertheless close
and this is important because it confirms the suitability of the
construction of the function $h$ that is described in the appendix, a construction less obvious and
more difficult to implement for the hexagonal lattice than for the others.
The anisotropic lattice on an ellipse was included to demonstrate that
the measure $m_{D,C}$ at the boundary is able to select the appropriate
conformal structure even when it is not obvious by symmetry. One 
map between the structure attached to the anisotropic lattice,
thus the square lattice with the indicated asymmetric interaction, and that
attached to the square lattice with symmetric interaction takes an ellipse
$x^2+ay^2\leq1$, $a=1.54369$, to the disk $x^2+y^2\leq1$. As ellipse we 
took one whose major and minor axes were of lengths $749.2$ and 
$485.2$. The usual linear fit of $\hat\omega_k, k=1,\dots, 10$ yielded
$1.4712+.0044 x$.

\begin{figure}
\begin{center}\leavevmode
\includegraphics[bb = 72 300 540 500,clip,width = 12cm]{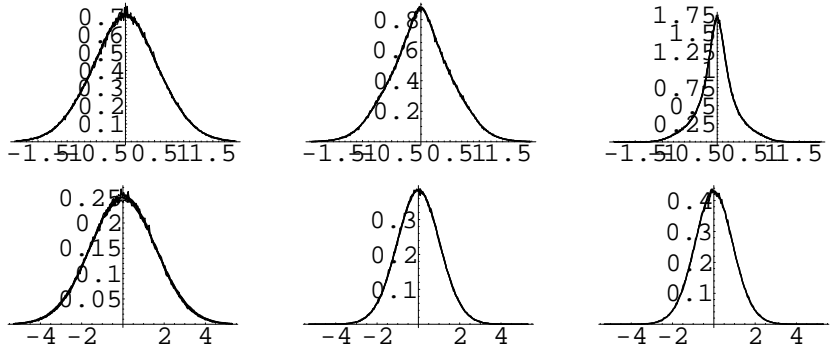}
\end{center}
\caption{Distributions $m_{D,C_i}$ as functions of the real
part of $A_k$ ($k=1$ on the first line, $k=8$ on the second)
on three different curves $C$: the boundary (first column),
the curve at a distance of 16 mesh units on the cylinder 
$397\times793$ and its
conformal images (second column) and the curve at the middle of the
cylinder (last column).\label{fig:3.2}}
\end{figure}

The plots of Figure \ref{fig:3.2} show the measure as a function
of $\Re A_1$ and $\Re A_8$ when $h$ is restricted to three
different curves on the cylinder or to their conformal images on
other geometries: the boundary, the second inner circle
(at a distance $0.0403$ from the boundary measured as a fraction of
the circumference) and the circle in the middle of the cylinder.
For the boundary (first column of Figure \ref{fig:3.2}) five models
have been drawn: the cylinder covered with the square, the triangular
and the anisotropic lattices, the disk and the square both covered
with the square lattice. (The numbers of sites on the various lattices are
those given earlier in this paragraph; only the data for 
the square of side $254$ were drawn here.)
For the second column of the figure, the
same models were used but no measurements were made for the square.
For the curve in the middle (third column), 
only the three lattices on the cylinder were measured,
because the corresponding circle on the disk is too small to 
allow for reliable measurements. (See below.)
To these three lattices
a fourth square lattice, with $199\times2399$ sites, 
was added on the $\Re A_1$ plot.
The agreement is convincing, as it is for
the distributions along the other curves $C_i$ that we measured.

 At first glance no cogent comparison can be made between the
central circle on the cylinder and a circle in the disk. 
A circle in the middle of a short cylinder is
equivalent to a circle in an annulus, but when the cylinder
becomes extremely long, it is more like a  
circle in the plane.
For example, if the cylinders of size $397\times793$ and $199\times2399$
are mapped to an annulus of outer radius 1, the inner radii will be
$4\times 10^{-6}$ and $1\times 10^{-33}$ and the images of
the central circles will have radii $2\times10^{-3}$ and $4\times 10^{-17}$
respectively (too small to make a measurement).
All circles in the plane are, however, conformally equivalent. 
So we can still compare the distributions on the central circle
of a cylinder 
with the distribution on a circle in the plane.
This is easier said than done, because the larger the 
circle the larger the domain needed to make useful
measurements. There is none the less a method, 
so that the distributions measured on
the central circle on both cylinders $397\times 793$ and 
$199\times 2399$ can be considered as distributions in the bulk.

\begin{figure}
\begin{center}\leavevmode
\includegraphics[bb = 72 140 540 650,clip,width = 10cm]{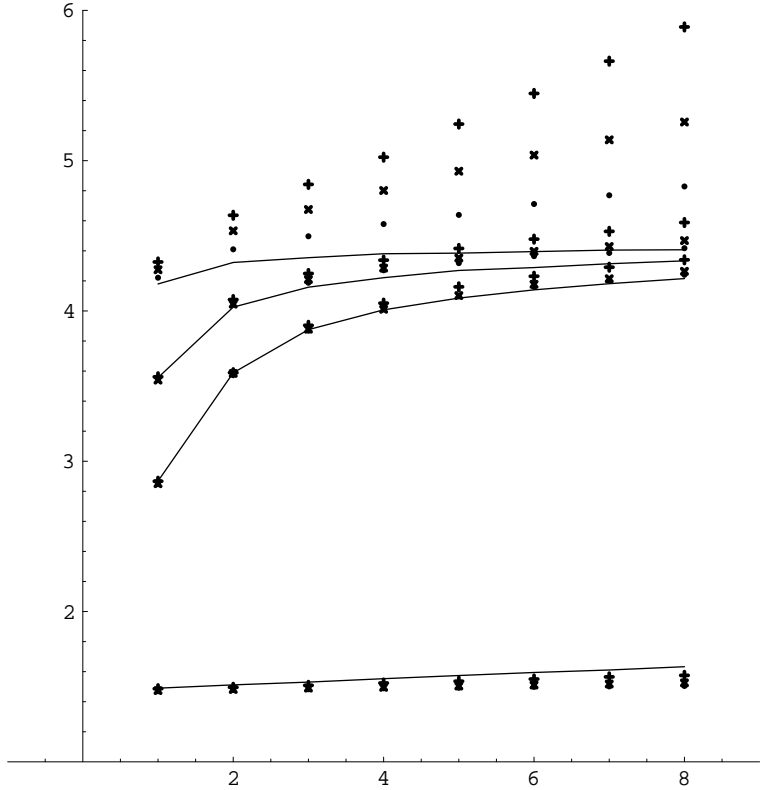}
\end{center}
\caption{The $\hat\omega_k$, $k=1,\dots,8$ on the boundary 
and on three inner circles of the disks of radius $100.2$ ($+$),
$200.2$ ($\times$) and $300.2$ ($\bullet$). The continuous lines
represent the corresponding data for the cylinder $397\times 793$.
\label{fig:3.4}}
\end{figure}

We first calculate the $\omega_k$'s
for progressively smaller circles inside disks and observe that
they do tend toward a limiting distribution.
The main difficulty is again
the finite-size effect revealed in Figure \ref{fig:2.2}. We  
compare corresponding inner circles on disks of radius $100.2, 200.2,
300.2$. On each of these, the distributions were measured on inner circles
of radius $1., 0.5, 0.4, 0.3, 0.2, 0.1319, 0.1$ and $0.04790$ times the
outer radius. The smallest inner circle on the disk of radius $100.2$
has a radius $4.8$ in mesh units. Finite-size effects will be indeed
important! Though we measured the $\omega_k$'s for $k$ up to $32$, the
overall behavior is clear for $k=1,\dots,8$, as presented on Figure
\ref{fig:3.4}. Only the circles of relative radius $1., 0.4, 0.2$ and $0.0479$
were retained for ease of reading. The ``$+$'' are for the disk of
radius $100.2$, the ``$\times$'' for $200.2$, the ``$\bullet$'' for
$300.2$ and the corresponding data for the cylinder $397\times793$
are joined by straight lines. For the boundary, the three disks give
a better approximation of the limiting distributions than the cylinder
but for the inner circles the roles are exchanged. The spread between
the three disks, and between them and the cylinder, is particularly
important for the smallest inner circles but the way it decreases
with the increase of the disk radius supports the hypothesis that a
common distribution for these two geometries exist on each of these
circles.

\begin{figure}
\begin{center}\leavevmode
\includegraphics[bb = 72 140 540 650,clip,width = 8cm]{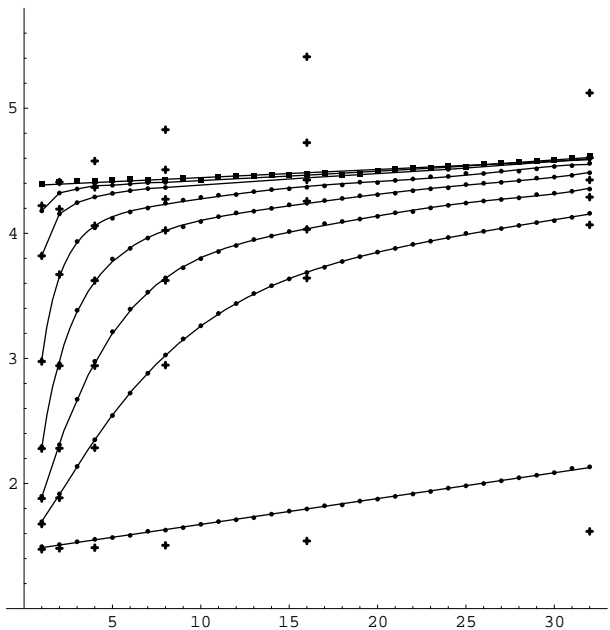}
\end{center}
\caption{The $\hat\omega_k$, $k\in\{1,2,\dots, 32\}$ on a disk ($+$) and on
a cylinder ($\bullet$ and $\blacksquare$) for various
inner circles.\label{fig:3.5}}
\end{figure}

A comparison of the distributions on inner circles for the disk and the
cylinder is therefore possible. Figure \ref{fig:3.5} shows the $\omega_k$'s,
$k\in\{1,2,\dots, 32\}$, for the cylinder $397\times 793$ ($\bullet$)
and the disk of radius $300.2$ ($+$). Curves were added to help
the eye. Seven circles were used. Their distance from the boundary of the
cylinder, in mesh units, and their relative radius for the disk
(in parenthesis) are $0 (1.), 8 (0.881), 16 (0.776), 32 (0.602),
64 (0.363), 128 (0.132)$ and $192 (0.0479)$. The measurements on the
central circle of the cylinder were added ($\blacksquare$). Only for the
two smallest circles of radius $0.132$ and $0.0479$ is the agreement
less convincing but, again, the previous figure showed how the gap
diminishes as the outer radius increases. We shall therefore refer to the
limiting distribution, approached by that on central circles of cylinders and
on very small inner circles of disks, as the {\em bulk
behavior} irrespective of the global geometry.

Another way to check that the bulk behavior is almost reached
in the middle of the cylinder is to compute the spin-spin correlation
along the central parallel. Using the conformal map
from the (infinite) cylinder to a disk one can see that the function
$\langle \sigma(\theta_1)\sigma(\theta_2)\rangle$ should be
proportional to $\sin^{-\alpha_{\text{bulk}}} ((\theta_1-\theta_2)/2)$. The
conformal exponent $\alpha_{\text{bulk}}$ is $\frac14$ (see \cite{MW}).
A log-log fit of $\langle \sigma(\theta_1)\sigma(\theta_2)\rangle $
as a function of $\sin ((\theta_1-\theta_2)/2$
gives a slope of $-0.257$. We can also verify that the measure
$m_D$ on the distributions on this curve allows us to recover this
$\alpha_{\text{bulk}}$ by the measurement of $\langle e^{i(\phi(\theta_1)-
\phi(\theta_2))}\rangle$. As in Paragraph \ref{arbitre} we did this
by first smearing the functions $h$ with a gaussian
and then truncating their Fourier expansion at $N$. (As before
we set $N=30$ and the variance of the gaussian to $2.5$ mesh units.)
The linear fit of the log-log plot leads to an $\alpha_{\text{bulk}}\approx
-0.260$, in fair agreement with the expected value.

Although the random variable $\Re A_1$ is not normal, Figure
\ref{fig:3.2} and, less clearly, Figure \ref{fig:3.1} show that higher Fourier
coefficients are close to normal. In fact, starting around $k=8$,
the histograms of the $\Re A_k$ are graphically undistinguishable
from the normal curves whose variances are those of the samples.
One may ask quite naturally if the distribution of these variables 
is given, at least asymptotically, by the law (\ref{eq:iii})
with, maybe, another constant $R_{\text{bulk}}$. 
The linear fit for the $\hat\omega_k^{\text{bulk}}$ is
$4.380 + 0.0065 k$ and the slope is slightly smaller than that of the boundary,
an indication that finite-size effects are smaller in the bulk. It therefore seems
likely that the variables $A_k$ are asymptotically normally distributed
as in (\ref{eq:iii}) with $(R_{\text{bulk}})^2=2.190$. The ratio 
$(R_{\text{bulk}})^2/R_B^2$ is $2.98$, very close to 3.

The existence of a nontrivial bulk behavior on 
curves in the plane, was 
by no means initially evident and may, in the long run,
be one of the more mathematically significant facts revealed by
our experiments.
One supposes that the distribution of spins in a fixed, bounded region 
for the Ising model on the complete planar lattice at criticality on a lattice 
whose mesh is going to $0$  
is such that they are overwhelmingly of one sign
with substantially smaller islands of opposite signs and that these islands in their
turn are dotted with lakes and so on. This is confirmed by the
two typical states of Figure \ref{ising12} in which the large islands
of opposite spin appear only in regions influenced
by the boundary. Typical states for
the cylinder are similar (Figure \ref{cond7A}) but the conformal geometry is 
such that the bulk state is reached closer to the boundary.
The conclusion is not, apparently, that in an enormous disk, thus
in the plane, the integral of $h$
against any fixed smooth function on a fixed smooth curve
is generally very close to $0$, so
that the distribution of each of the Fourier coefficients $\Re A_k$ and
$\Im A_k$ approaches a $\delta$-function. Rather, 
they are approaching a distribution which is not trivial but is, at least
for $k$ small, clearly not a gaussian.
What we may be seeing is the
effect of the shifting boundaries of the large regions of constant spin.
Once a circle in the plane is fixed, the boundary between even two very large regions of
different spin can, as the configuration is varied, cut it into intervals of quite
different size. 

Indeed the existence of a nontrivial limiting measure on the 
space of distributions on the
boundary was itself not certain beforehand. In spite of the attention we 
gave in Section \ref{dis1} to the possibility of its being gaussian
for the boundary of a circle, the exact form is perhaps of less
mathematical significance than its universality and conformal invariance.

\subsection{Clarification.}\label{clar}

 In order not to encumber the initial discussion with unnecessary abstraction,
we worked with the distributions $m_{D,C}(\{a_k\}, {\mathcal G}, J)$.
A better theoretical formulation would be in terms of a measure 
$m_{D,C}$ on the set of real-valued
distributions in the sense of Schwartz on the oriented smooth curve $C$, or if $C$ were
merely regular (thus sufficiently differentiable)
on some Sobolev space. To be  
more precise,  
the measure is on the set of distributions that annihilate the constant
functions. 
(To be even more  
precise, this is so only if the curve is contractible.
For other curves, such as the circumference of a cylinder,
the set of distributions whose value 
on the constant function $1$ lies in $\{2m\pi\,|\,m \in  \mathbb Z,
 m\neq0\}$,
may have a nonzero measure. Under many circumstances, it
is small enough that it can for numerical purposes be
supposed $0$.)
To introduce the measure $m_{D,C}$ concretely,
we need a basis for the dual space, thus in principle just a basis for the
smooth (or regular) functions on $C$
modulo constants. If this basis is $\{\varphi_k|k=1,\infty\}$ then
$\lambda \rightarrow \{\lambda(\varphi_k)\}$ defines a map of the distributions
into a sequence space and a measure is just a measure on the collection of real infinite
sequences, $\{\mu_1,\mu_2,\dots\}$. It would have to be defined by 
some sort of limiting process
from measures on $\mathbb R^N$. The simplest such measures are product measures.
Given such a measure on the space of real sequences, it defines, at least
intuitively, a measure on
distributions if for almost all sequences $\{\mu_k\}$, the assignment
$\varphi_k\rightarrow\mu_k$ extends to a distribution, thus in particular if it
lies in some Sobolev space. For example, if a parametrization $x(\theta)$,
$0\le\theta\le2\pi$ of the curve has been fixed then one possible
choice of the basis is the collection 
$$
\{x(\theta)\rightarrow\Re e^{ik\theta},x(\theta)\rightarrow\Im e^{ik\theta}|k>0\}
$$
Then it is better to put the $\mu_k$ together in pairs and to
use sequences $\{A_k\}$ (or $a_k=A_k/ik$) of complex numbers.
This has been the point of view of this paragraph. Starting with a given
parametrization, we examined the joint distributions of the complex random
variables $a_k$.

 The parametrization also allows us to introduce the measure $d\theta/2\pi$ and
thus to identify functions with distributions. In particular, in order
for a measure on sequences to yield a measure on distributions it is necessary,
and presumably usually sufficient, that the sum
$$
  \sum_{k=-\infty}^{\infty} a_ke^{ik\theta}
$$
converges as a distribution for almost all sequences $a_k$. 
For example, if the measure is a gaussian defined by 
\begin{equation}
\exp(-\alpha\sum_{k=1}^\infty k|a_k|^2) \label{eq:ss}
\end{equation}
then 
the expectation
$\text{E}(|a_k|^2)$ is $1/(2\alpha k)$,
so that $\sum |a_k|^2/k$ converges almost everywhere.
As a result, the sum (\ref{eq:ss}) converges almost 
everywhere as a distribution.
This conclusion remains valid provided only that the
expectations $\langle a_ka_l\rangle$ are those of the gaussian (\ref{eq:ss}),
a property that according to the results of Section \ref{dis1} the measure
$m_D$ has a good chance of possessing. Therefore, if 
$\lambda=\sum_k\lambda_ka_k$
is any random variable that is a linear function of the $a_k$ then the
expectation $E(\lambda^2)$ is calculated as though the measure were gaussian. 

 Our method is numerical, so that we approximate the measure on sequences from
a large, finite scattering of functions $h$, or rather of their
derivatives, because the derivative $H$ is well defined as a distribution,
although $h$ itself is not. The distribution $H$
is a sum of $\delta$-functions, with mass $\pm\pi$ at each point
where the curve crosses a contour line of the function $H$. 
Since its value on the constant function $1$ is 
the sum of those jumps, this value is $0$, as noted, whenever that sum necessarily
vanishes, either because the curve is contractible or because the cylinder
is extremely long. 

 Although our construction required 
a specific parametrization, the resulting measure on
distributions may be independent of the parametrization and, more generally, even
of the choice of basis. We did not attempt to verify this. 
It may be useful, however, to describe an example.

\def\Q{{\mathfrak Q}}
\def\D{{\mathfrak D}}

 When $D$ is a disk of radius $1$ with the boundary $C$ parametrized 
in the usual way by arc length $\theta$,
the function $h$ can be recovered by integration
with respect to $d\theta$ from the
distribution $H$.
The measure on distributions on $C$
is, as we discovered, not equal to the
gaussian measure associated to a constant, $2R_B^2$, times
the Dirichlet form $\Q (H)$, but, if we ignore the reservation
expressed at the end of Paragraph \ref{boson}, the variance of linear functions
of the Fourier coefficients can be calculated
as though it were. We recall that to calculate $\Q (H)$, or $\Q(H,H)$
if we want to stress that it is a quadratic form, we extend the function
$h$ as a harmonic function to the interior and then 
$$
  \Q(H)=\D (h)=\D (h,h)=\frac{1}{4\pi}\int\{(\frac{\partial h}{\partial x})^2+
         (\frac{\partial h}{\partial y})^2\}dxdy.
$$
or, extending it to an hermitian form,
\begin{equation}
  \Q(H)=\D (h)=\D (h,h)=\frac{1}{4\pi}\int\left\{\left|\frac{\partial h}{\partial x}\right|^2+
         \left|\frac{\partial h}{\partial y}\right|^2\right\}dxdy
\end{equation}
if we use again the symbol $h$ for the harmonic function inside $D$.
If we identify formally distributions with functions by means of the bilinear form
$$
\frac{1}{2\pi}\int_0^{2\pi}h_1(\theta)h_2(\theta)d\theta=\langle h_1,h_2\rangle,
$$
or, in complex terms,
\begin{equation}
\frac{1}{2\pi}\int_0^{2\pi}h_1(\theta)\overline{h_2(\theta)}d\theta=\langle h_1,h_2\rangle,
\end{equation}
and regard therefore $\Q$ and $\D$ as operators, so that $\Q(H)=\langle \Q
H,H\rangle$
and $\D (h)=\langle \D h,h\rangle$, then, as a simple
calculation with the functions $e^{ik\theta}$ shows, $\D $ has $0$ as an eigenvalue
of multiplicity one, eigenvalues $\frac12, \frac22,\frac32,\dots$,
each with multiplicity two, $\Q $ has eigenvalues $\frac12,\frac14,\frac16,\dots$, each with
multiplicity two and on the domain of $\Q $,
the orthogonal complement of the constant functions, $4\D =\Q ^{-1}$. More precisely, and
this is the best form for our purposes, if the Fourier expansion of $h$ is
$\sum_k a_ke^{ik\theta}$ then 
$$
  \D (h)=\frac12\sum_{k\neq 0}|k a_k|^2,
$$
or if $h$ is real,
$$
  \sum_{k>0}k|a_k|^2.
$$

 Suppose now that $D'$ is any domain, $C'$ its boundary, and $\varphi'$ 
any smooth function on
$C'$. The function $\varphi'$ 
defines a linear form 
\begin{equation}
\lambda\rightarrow \lambda(\varphi') \label{eq:rv}
\end{equation}
on distributions. By conformal invariance,
the measure on distributions on $C'$ is 
obtained by transport of the measure on distributions on
$C$ using any conformal transformation $\phi$
from $D$ to $D'$. If the measure on the distributions is in fact well-defined,
independently of any choice of basis, then the characteristic function of
(\ref{eq:rv}) is formally calculated as
$$
\int\exp\left(-2R_B^2(\Q\lambda,\lambda)+i\alpha\lambda(\varphi)\right)
/\int\exp\left(-2R_B^2(\Q\lambda,\lambda)\right)
$$
which is
$$
  \exp(-\alpha^2\Q(\lambda_\phi)/8R_B^2)
$$
if $\varphi=\varphi'\circ\phi$ 
and $\lambda_\varphi$ is the distribution such that 
$\lambda(\varphi)=\Q (\lambda,\lambda_\varphi)$.
Consequently, the probability distribution of the
random variable defined by
(\ref{eq:rv}) will be gaussian with variance $\Sigma^2$ given by
$$
   1/2\Sigma^2=2R_B^2/\Q (\lambda_\varphi).
$$
But $\lambda_\varphi=\Q ^{-1}\varphi$ so that
\begin{equation}
  1/2\Sigma^2=R_B^2/2\D (\varphi).\label{eq:sigma}
\end{equation}

\begin{figure}
\begin{center}\leavevmode
\includegraphics[bb = 72 240 540 554,clip,width = 10cm]{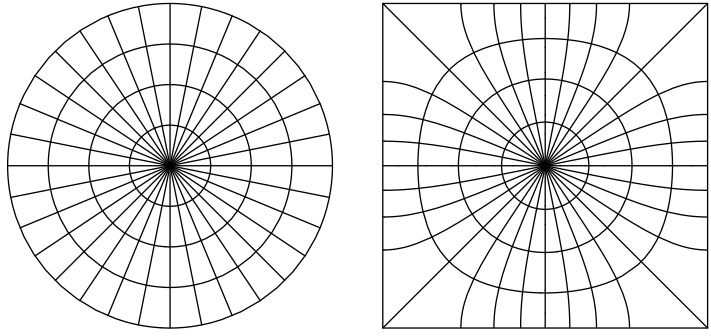}
\end{center}
\caption{The Schwarz-Christoffel map from the disk to the
square. The images of the rays of the disk are the curves intersecting
at the center of the square.\label{fig:3.3}}
\end{figure}

 Suppose for example that $D'$ is a square of side 
$\frac{\pi}2$ and that we parametrize its
boundary by arc length: $s=s(t)$, $s(0)$ being one of the vertices. 
We write $t=t(s)$ for the inverse function. 
The Schwarz-Christoffel map $\phi$ of the disk onto the square
is depicted in Figure \ref{fig:3.3}
where the curves intersecting at the center of the
square are the image of the rays on the disk. Although arcs of equal
length on the circular boundary
are mapped to intervals of different lengths on the edge of the
square, 
this effect is important only close to the vertices.
Thus if $\varphi'_k$ is the function $s(t)\rightarrow \cos(kt)$ 
and $\varphi_k=\varphi'_k\circ\phi$ then the distribution
of the random variable defined by $\varphi_k$ should be gaussian with variance
$\D (\varphi_k)/R_B^2$. 
Moreover $\D (\varphi_k)$ is obtained from the Fourier coefficients of
$\varphi_k$ and they are calculated by observing that, apart from a constant factor,
which is unimportant, the Schwarz-Christoffel transformation (\ref{eq:sc-ch})
restricted to the boundary $w=e^{i\theta}$ of the unit disk can be expressed
in terms of the elliptic integral of the first kind,
$$
  F(t|2)=\int_0^t\frac{d\psi}{\sqrt{1-2\sin^2\psi}}.
$$
With our choice of $s(0)$, the function $s(t)$, for $t\in[0,\frac{\pi}2]$,
is 
$$s(t)=\frac{\frac{\pi}4}{F(\frac{\pi}4|2)}F(t-
{\textstyle{\frac{\pi}4}}|2) +\frac{\pi}4.$$
The graph of $(t,s(t))$ on $[-\pi,\pi]$ is obtained from that on
$[0,\frac{\pi}2]$ by translation by $(\frac{\pi}2,\frac{\pi}2)$.
The function $s=s(t)$ is odd and composition of odd or even functions
with $s$ preserves their parity. If the basis $\{s(t)\rightarrow
\cos lt, s(t)\rightarrow \sin lt, l>0\}$ is chosen and $\varphi_l(t)
=\cos(ls(t))$ (or $\sin(ls(t))$) written as
\begin{equation}
\frac{C_{l0}}2+\sum_{k\ge1}(C_{lk}\cos kt+S_{lk}\sin kt),\label{eq:devfou}
\end{equation}
then the Dirichlet form is
\begin{equation}
\D(\varphi_l)=\frac14\sum_{k\ge 1}k(C_{lk}^2+S_{lk}^2).\label{eq:diri}
\end{equation}
For a given $l$ the random variables
$\Re A_l$ and $\Im A_l$ are identically distributed on the
disk, at least when the number of sites at the boundary is
a multiple of 4. On the square they were
shown to be also identically distributed, at least
in the limit of the simulations, when
these variables are measured with respect to the induced parameter.
However,
if the arc-length parameter $s$ is used on the square, the two variables
$\Re A_l^s$ and $\Im A_l^s$ are identically distributed only when $l$ is
odd. The graphs of two functions $\cos(l\phi(t))$ and $\sin(l\phi(t))$ are
translations of each other when $l$ is odd but not when $l$ is even.
The variances of the random variables $\Re A_l^s$ and  $\Im A_l^s$ must
then be distinguished and they are given by
\begin{equation}
(\Sigma_{\Re A_l}^s)^2=\frac1{4R_B^2}\sum_{k\ge1}kC_{lk}^2
\quad \text{and}\quad 
(\Sigma_{\Im A_l}^s)^2=\frac1{4R_B^2}\sum_{k\ge1}kS_{lk}^2.\label{eq:vars}
\end{equation}
Using these formulas we shall compute the numbers 
$$\omega_{\Re A_l}^s=\frac l{2(\Sigma_{\Re A_l}^s)^2}
\quad \text{and}\quad 
\omega_{\Im A_l}^s=\frac l{2(\Sigma_{\Im A_l}^s)^2}$$ 
introduced in Section \ref{dis1}.

The coefficients
$$C_{lk}=\frac1\pi \int^\pi_{-\pi}\cos ls(t) \cos kt\ dt
\quad \text{and} \quad
S_{lk}=\frac1\pi \int^\pi_{-\pi}\sin ls(t) \sin kt\ dt$$
are therefore needed. They can be calculated numerically. The
convergence rate of (\ref{eq:devfou}) is however slow.
The elliptic integral $F(t|2)=\int_0^2(1-2\sin^2\psi)^{-
\frac12}d\psi$ behaves like $(t-\frac{\pi}4)^{\frac12}$ as $t\rightarrow
\frac{\pi}4^-$. Consequently, the function $s$ has a similar
behavior at integer multiples of $\frac{\pi}2$ and the absolute values 
$|C_{lk}|$ and $|S_{lk}|$ decrease approximately as 
$M/k^{\frac32}$. 
Even with the 250 first
Fourier coefficients $C_{lk}, k=1,\dots,250$,
the Parseval
identity for the function $\cos l\phi(t)$
is satisfied to only five decimal digits. We decided
nonetheless to restrict the sums (\ref{eq:vars}) to these $250$ first
coefficients. Since all the terms in $\D (\varphi_l)$
are positive, the truncated sums
will lead to larger estimates of the $\omega$'s than the true sums.

Since we wish to compare $\omega_{\Re A_l}^s$ and 
$\omega_{\Im A_l}^s$ with those measured with the simulations
done on the square of side $254$, it is appropriate to modify
slightly the Dirichlet form (\ref{eq:diri}) to take into account
finite-size effects. Paragraph \ref{cylinder} showed that
the quantity $k/(2\Sigma_k^2)$ is not strictly constant on
a finite lattice but grows slowly. We found that 
$$\frac k{2\Sigma_k^2} = 2R_B^2(1+\epsilon k)$$
was a good approximation. (See Figure \ref{fig:2.3}.) Since
the ratio $k/4R_B^2$ in (\ref{eq:vars}) plays the role of 
the variance, we decided to replace it by
$$\frac k{4R_B^2(1+\epsilon k)}.$$
The slope $\epsilon$ is that of the linear fit appearing
in Figure \ref{fig:2.2} for the square with $254\times254$ sites.

Table III lists the values of $\hat\omega_l^s$, that is
$\hat\omega_{\Re A_l}^s$
and $\hat\omega_{\Im A_l}^s,l=1,2,3,4,5$, as measured
by the simulations and the values $\omega_{\Re A_l}^s$
and $\omega_{\Im A_l}^s$
obtained using the (truncated) sums (\ref{eq:vars}) with finite-size
effects introduced as discussed.
The original values $\hat\omega_l$
have been added to give an idea
of the discrepancy that the use of the arc-length parameter introduces.
The values $\hat\omega_l^s$ and $\omega_l^s$ are close to
one another and the latter are always greater than the former, probably
because of the truncation.

\begin{center}
\begin{tabular}{|l|c|c|c|}
\hline
&&&\\
$l$ & $\hat\omega_l$ & $\hat\omega_l^s$ & $\omega_l^s$\\
&&& \\
\hline
1&1.480 & 1.380 & 1.388 \\
\hline
\multirow{2}{1cm}{2}&\multirow{2}{1.cm}{1.494} & 2.241 & 2.251 \\
& & 0.963 & 0.974 \\
\hline
3&1.505  & 1.365 & 1.386 \\
\hline
\multirow{2}{1cm}{4}&\multirow{2}{1.cm}{1.510} & 1.921 & 1.958 \\
 & & 1.146 & 1.181\\
\hline
5&1.520 & 1.451  &  1.499 \\
\hline
\end{tabular}
\end{center}

\noindent Table III: The numbers $\hat\omega_l$,  $\hat\omega_l^s$ and
$\omega_l^s$ for $l=1,2,3,4,5$ for the square of side $254$.
When they differ, the $\omega$ for $\Re A_l$ is placed above the
$\omega$ for $\Im A_l$.

\subsection{Conditional probabilities.}\label{condprob}

Suppose that the curve $C$ of the previous section is the disjoint union of two curves
$C_1$ and $C_2$. Then the space of distributions on $C$ is the product
of the space on $C_1$ and the space on $C_2$. We fix the model to be the
Ising model at criticality on the square lattice and denote a distribution by
$\psi$ and the measure whose meaning was clarified in the previous paragraph
by $m_{D,C}(\psi)$. Then, {\itshape in principle}, the conditional probability
$m_{D,C}(\psi_1|\psi_2)$ on the set of distributions on $C_1$ is defined 
for each distribution $\psi_2$ on $C_2$. Whether this is so is not so 
easy to test experimentally. 
To approximate the conditional probability
numerically with our methods we have to choose a neighborhood $U$ of 
$\psi_2$ and proceed as before, eliminating from the sample all distributions
$\psi'=(\psi'_1,\psi'_2)$ for which $\psi'_2$ does not fall in $U$.
We recall that $\psi'_i$ is a distribution given by a sum of $\delta$-functions
on $C_i$. First of all, the neighborhood $U$ is a neighborhood in an
infinite-dimensional space, so that it is going to be, in any case, very large.
Secondly, we cannot eliminate too many distributions for then the samples
would be far too small. Thus $U$ is going to have to be enormous.
The notion seems nevertheless to be workable even at a coarse
experimental level.

There are two properties that one might expect. We can introduce
and study experimentally the measure on the distributions on $C_1$
obtained when the spins on $C_2$, or in a small neighborhood of
it, are all taken to be $+1$. This of
course presupposes some kind of compatibility of $C_2$ with the lattice structure,
as in the examples studied where $C_2$ passes through a row of sites,
or some way, either theoretical or practical, of specifying the neighborhood,
but granted this, we consider the measure $m_{D,C_1}(\psi_1|C_2,+)$
obtained from this familiar condition.
It is defined quite differently than the conditional probability
$m_{D,C}(\psi_1|0)$ for $\psi_2\equiv0$. (See Paragraph \ref{5point2}.)
None the less, one could hope that they
were equal. The experiments to be described are too coarse to establish
this with any degree of certainty, but do render the expectation plausible.

The second property is the markovian property. Suppose that $C_1$ is the disjoint
union of $C_3$ and $C_4$, so that $\psi_1$ is a pair $(\psi_3,\psi_4)$.
Suppose moreover -- this is the essential condition -- that $C_2$ separates
$C_3$ from $C_4$. Then one can hope that conditioning the measure
$m_{D,C}(\psi_1|\psi_2)$ on $\psi_4$ leads to a measure 
$m_{D,C}(\psi_3|\psi_4|\psi_2)$ that is equal to $m_{D,C}(\psi_3|\psi_2)$,
thus the measure on the distributions on $C_3$ when the distributions on
$C_2$ and $C_4$ are given is independent of the distribution on $C_4$.
The influence of the distribution $\psi_4$ is not propagated across $C_2$
when the distribution on $C_2$ is fixed.

\begin{figure}
\begin{center}\leavevmode
\includegraphics[bb = 76 296 536 496,clip,width = 12cm]{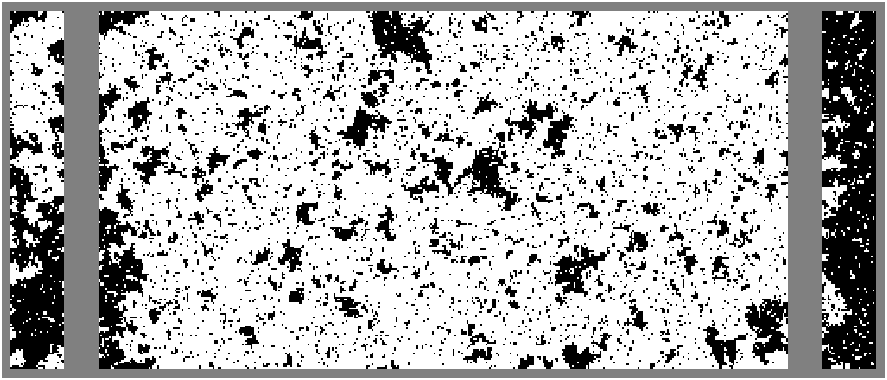}
\end{center}
\caption{Configurations on cylinders.\label{cond7A}}
\end{figure}

We begin by examining Figure \ref{cond7A} in which three typical states are shown,
from left to right: free boundary conditions on a cylinder of circumference 199
and length 31; free boundary conditions on a cylinder of circumference 199
and length 399; and free boundary conditions on the left but $+$ boundary conditions
on the right on a cylinder of circumference 199 and length 31. The picture in the
center is the familiar one: towards the middle there is a tendency to form very
large clusters of constant sign, indeed there is only one very large
(white) cluster but at the boundary the clusters are smaller.
Recall as well that for a cylinder there is conformal distortion. In
Figure \ref{ising12} the phenomenon is  illustrated without distortion:
there is one large (white) cluster on the left and one large (black) one
on the right.
In the picture on the left of Figure \ref{cond7A}, 
the freedom to form smaller clusters is reinforced by
the proximity of the two boundaries. There is almost no bulk behavior at all.
On the other hand, in the picture on the right, the boundary condition is forcing 
a single large cluster on the right and this cluster is attempting to envelop
the left boundary as well.

\begin{figure}
\begin{center}\leavevmode
\includegraphics[bb = 76 256 536 536,clip,width = 12cm]{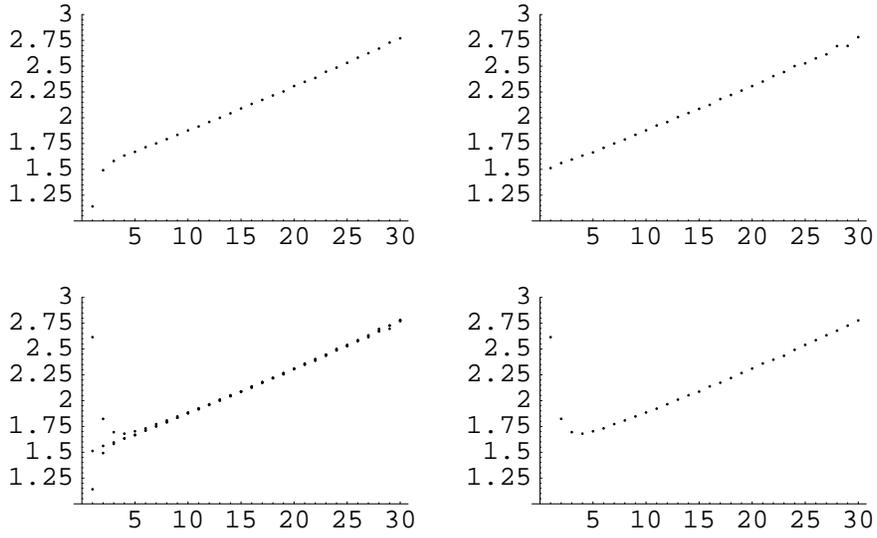}
\end{center}
\caption{The numbers $\hat \omega_k$ associated with the conditional
distributions. (See text.)\label{cond7B}}
\end{figure}

\begin{figure}
\begin{center}\leavevmode
\includegraphics[bb = 76 230 540 560,clip,width = 10cm]{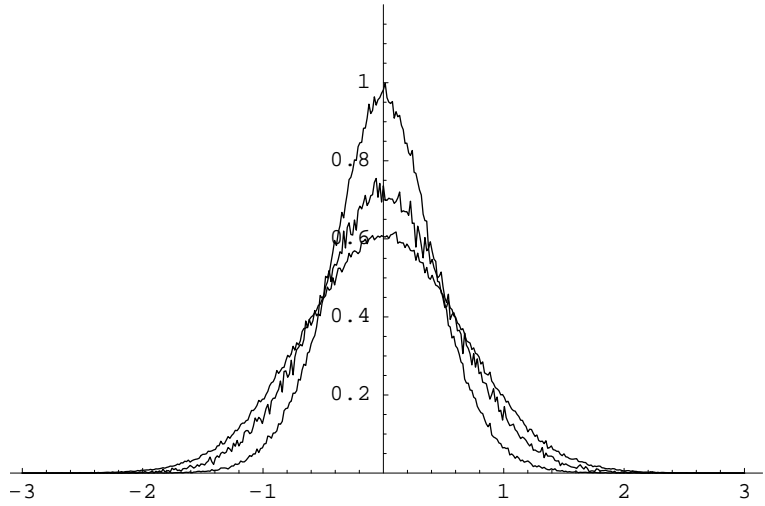}
\end{center}
\caption{The distribution of $\Re A_1$ for the three pairs (domain, boundary
conditions) described in the text. At the center the curves are in the order,
from top to bottom: short cylinder with constant spins on the right, long
cylinder, short cylinder with both sides free.\label{cond7C4}}
\end{figure}

This qualitative description is confirmed by a calculation,
for the measures on the distributions on the left boundaries, of the 
numbers $\hat \omega_k$ introduced in Paragraph \ref{cylinder}. The results 
are plotted in the diagrams of Figure \ref{cond7B} for
the measure associated to the left boundary in the three cases. In clockwise
order from the upper left, they are:
free boundary conditions on a cylinder of size $199\times31$;
free boundary conditions on a cylinder of size $199\times399$;
boundary conditions on the left free, those on the right
constant, and size $199\times31$.  In the diagram on the lower
left, they are superposed.
The graph in the upper right
is like those of Figure \ref{fig:2.2}, except that we have used new statistics
with a smaller sample, so that the graph is somewhat irregular. All graphs
are pretty much the same except for the first four or five values of $k$.
As far as the higher values of $k$ are concerned the two boundaries are effectively
at an infinite distance from each other. For $k=1$, there is a pronounced 
difference between the graphs so that the distribution of $\Re A_1$ on the
short cylinder is flatter than on the long cylinder. On the other hand,
when the boundary condition is imposed the value of $\hat\omega_k$
increases and the distribution of $\Re A_1$ is peaked. The superposition
of the three curves is shown in Figure \ref{cond7C4}.

\begin{figure}
\begin{center}\leavevmode
\includegraphics[bb = 76 240 540 560,clip,width = 8cm]{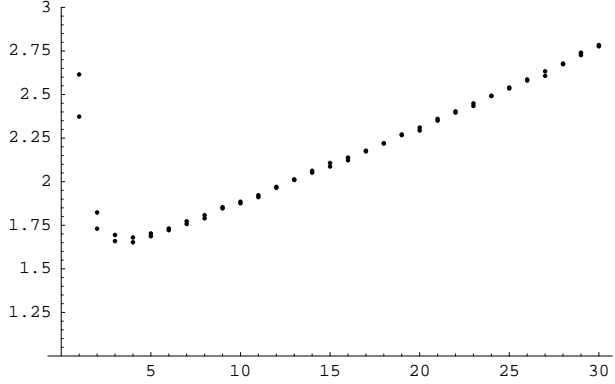}
\end{center}
\caption{The numbers $\hat\omega_k$ for $m_{D,C_1}(\psi_1|C_2,+)$ and
$m_{D,C}(\psi_1|0)$.\label{cond7D}}
\end{figure}

\begin{figure}
\begin{center}\leavevmode
\includegraphics[bb = 76 240 540 560,clip,width = 8cm]{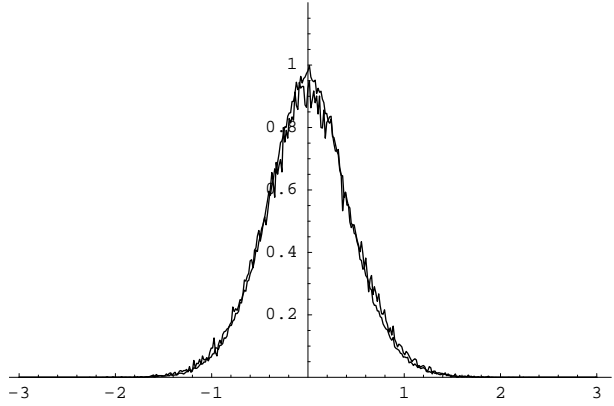}
\end{center}
\caption{The distribution of $\Re A_1$ for $m_{D,C_1}(\psi_1|C_2,+)$ and
$m_{D,C}(\psi_1|0)$.\label{cond7D1}}
\end{figure}

We now take $C_1$ to be the left boundary of a cylinder of aspect ratio $199/31$
and $C_2$ to be the right boundary. To test the assertion that
$m_{D,C_1}(\psi_1|C_2,+)$ is the conditional probability $m_{D,C}(\psi_1|0)$,
we thermalize for free boundary conditions at both ends
of a cylinder of size $199\times31$ but only
keep those samples for which
\begin{equation}
\begin{alignat}{3}
|\Re A_1|&<.125, & \qquad |\Re A_2|&<.2\sqrt2, & \qquad  |\Re A_3|&<.35\sqrt3, \\
|\Im A_1|&<.125, & \qquad |\Im A_2|&<.2\sqrt2, & \qquad  |\Im A_3|&<.35\sqrt3.
\end{alignat}
\end{equation}
About 3 out of every 10,000 states satisfy this condition. So our crude
experiments will not permit a substantially smaller neighborhood of $0$.
In Figure \ref{cond7D}, we plot the resulting collection of $\hat\omega_k$
together with those obtained from the previous experiment with $+$ boundary
conditions on the right side. We see that in spite of the large size
of the neighborhood, the two graphs are quite close. It is the values
of $\hat\omega_1$ and $\hat\omega_2$ that tell.
The graphs of the distributions of $\Re A_1$ are compared in Figure \ref{cond7D1}
to ensure that not only are the variances close but also the probability 
measures themselves. 
Without being at all conclusive, the experiment encourages the belief that
$$
m_{D,C}(\psi_1|0)=m_{D,C_1}(\psi_1|C_2,+).
$$

In order to test whether the probabilities are markovian we
considered on the one hand a cylinder of size $199\times31$ 
on which we thermalized, keeping only the distributions that on the right
boundary satisfied the conditions
\begin{equation}
\begin{alignat}{3}
|\Re A_1| <&.125,  & \qquad |\Re A_2|&<.2\sqrt2, & \qquad |\Re A_3|&<.35\sqrt 3,\notag\\
\noalign{\vskip-5pt}
&&&&&\label{skewcond}\\
\noalign{\vskip-5pt}
.3\le \Im A_1&\le1,& \qquad |\Im A_2|&<.2\sqrt2, & \qquad |\Im A_3|&<.35\sqrt 3.\notag
\end{alignat}
\end{equation}
On the other hand we considered a cylinder of size $199\times61$ on which we
thermalized with spin $+$ as the boundary condition on the right 
and then selected only those states satisfying
the conditions (\ref{skewcond}) on the distributions for the 
central meridian. 
We then 
examined the resulting measure on the distributions on the left boundary,
in particular the distribution of $\Re A_1$ and $\Im A_1$. The markovian
hypothesis asserts that, when we fix the distribution
on the center, the measure on the distributions on the left
boundary is completely shielded from the boundary conditions on the right,
although once again we are prevented by the necessity of allowing the rather
large open neighborhood (\ref{skewcond}) from actually fixing the distribution on
the center. We can only impose very crude constraints on the first few
Fourier coefficients. For the experiments on the smaller cylinder about 3
samples in 10,000 are kept; on the larger, curiously enough, about 1 in 1,000.

\begin{figure}
\begin{center}\leavevmode
\includegraphics[bb = 76 300 530 480,clip,width = 14cm]{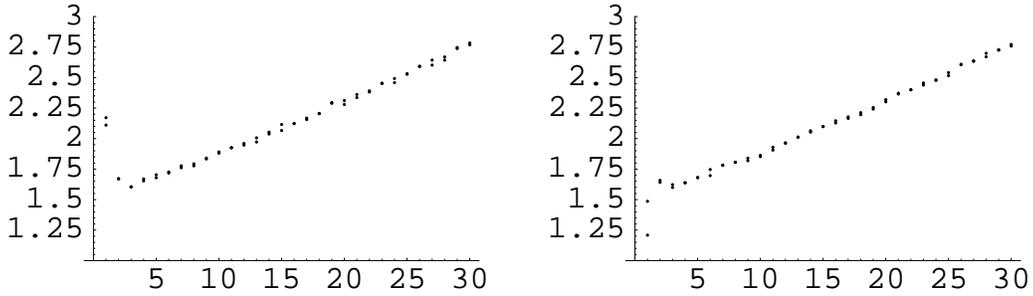}
\end{center}
\caption{Test of the markovian hypothesis.
The numbers $\hat \omega_k$ for the two cylinders with $\Re A_k$
on the left and $\Im A_k$ on the right.\label{cond7E}}
\end{figure}

\begin{figure}
\begin{center}\leavevmode
\includegraphics[bb = 72 300 540 500,clip,width = 10cm]{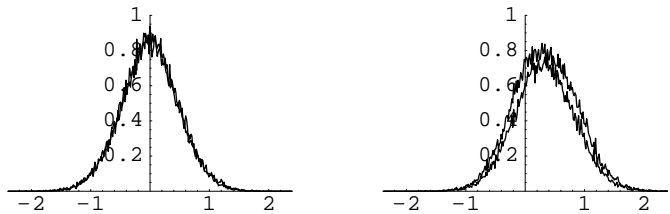}
\end{center}
\caption{The distributions on the left boundary as a function of
$\Re A_1$ (on the left) and of $\Im A_1$ (on the right).\label{cond7F}}
\end{figure}

In Figure \ref{cond7E} the $\hat\omega_k$ are plotted 
and compared once again, on the left those
for $\Re A_k$, on
the right those for $\Im A_k$. For $\Re A_1$ the value
of $\hat\omega_1$ is slightly larger for the broader
cylinder; the other values are very close. For $\Im A_1$
the value is smaller for the narrower cylinder, and the
other values are again very close. In Figure \ref{cond7F}
a similar comparison is made of the distributions
of $\Re A_1$ on the left and of $\Im A_1$ on the right.
As is to be expected from conditions \ref{skewcond},
the distribution of $\Im A_1$ is shifted to the right. 
It is more shifted for
the narrow cylinder than for the broad. The results
encourage the belief in the markovian hypothesis, even though
it is hard to imagine that experiments as
coarse as these could ever successfully refute the hypothesis
because some shielding is inevitable. The question is rather how much.

\section{Cylinders of variable length and the phase.}\label{huit}

We have seen in Paragraphs \ref{arbitre} and
\ref{simconf} that the measures $m_{D,C}$ can be
used to recover the conformal exponent associated to the spin-spin 
correlation at the boundary and in the interior.
Various formulas in the theory of free fields suggest
that critical exponents might also be obtained from
the analogue for
the field $h$ of the variable $x$ defined in Paragraph
\ref{boson} for the free boson $\tilde\phi$. We refer to this variable
as the phase, and our examination in this section, although brief, 
indicates clearly that it also can be used
to reproduce exponents of the classical Ising model.

The variable $x$ for the boson field measures the difference
between the constant terms in $\phi_1$
and $\phi_2$, the restrictions of $\tilde\phi$ to the
two boundaries of the cylinder. It takes its values in the 
interval $[0,2\pi R)$ where $R$ is the radius of compactification.
(See Paragraph \ref{boson}.) An analogue for the Ising model on the
cylindrical $LV\times LH$ square lattice $\mathcal G_\square$ is
defined using
$$x'=\frac1{LV}\sum_p(h(p+\delta)-h(p))$$
where $\delta$ is the unit vector in the {\it horizontal}
direction and the sum runs over all sites $p$ in the lattice
that have a right neighbor. Because the jumps of $h$ are chosen
at random between $\pm \pi$, it is natural to study the distribution of
$$x=x'\mod 2\pi$$
instead of $x'$.
The normalization of $x$ is such that a closed curve of discontinuity
in $h$ that wraps around the cylinder, in other words that is
noncontractible, gives a contribution of $\pm\pi$ to $x$. Clusters
intersecting the boundary contribute $\pi\Delta/LV$ to $x$ where
$\Delta$ is the numbers of boundary sites inside the cluster.
However contractible curves surrounding clusters of constant spins
not intersecting the boundary do not contribute.

In Section \ref{dis1} we introduced, for the cylinder $D$, the measure
$$m_D(\{a_k\},\{b_k\})=\lim_{N\rightarrow\infty}\lim_{a\rightarrow 0}
m_D^{a,N}(\{a_k\},\{b_k\})$$
defined on the space ${\mathfrak H}_I$ with coordinates $(\{a_k\},\{b_k\})$,
$k\in\mathbb Z\setminus\{0\}$. 
As we observed in Paragraph \ref{clar},
this can also be regarded as a measure $m_D(\psi_1,\psi_2)$
on a space of distributions, one $\psi_1$ on the circle at one end of the
cylinder and one $\psi_2$ on the circle at the other end. We 
could as well have defined
$$
m_D(\psi_1,\psi_2,x)=m_D(\{a_k\},\{b_k\},x)=\lim_{N\rightarrow\infty}\lim_{a\rightarrow 0}
m_D^{a,N}(\{a_k\},\{b_k\},x)$$
taking the variable $x$ into account. The probability
$m_D(\psi_1,\psi_2)$ is a conditional probability, thus --
speaking imprecisely -- we have integrated over the variable $x$.
Writing all measures informally as measures absolutely continuous with respect to
a Lebesgue measure on the underlying spaces, we express this as
$$
dm_D(\{a_k\},\{b_k\})=dm_D(\psi_1,\psi_2)=Z_D(\psi_1,\psi_2)\,d\psi_1\,d\psi_2
$$
with
$$
 Z_D(\psi_1,\psi_2)=\int_0^{2\pi} Z_D(\psi_1,\psi_2,x)\, dx.
$$
This is a convenient notation that 
avoids technical explanations about conditional probabilities
and also reminds us of the connection between the
measures and partition functions.
 
\subsection{The measure $m_q(x)$.}

We first consider $dm_q(x)=dm_{D}(x)=Z_q(x)dx$, $D=D(q)$,
with
$$
Z_q(x)=\int Z_D(\{a_k\},\{b_k\},x)\prod_k da_k\,db_k=
\int Z_D(\psi_1,\psi_2,x)\,d\psi_1\,d\psi_2,
$$
the choice between the three notations $(\{a_k\},\{b_k\})$,
($\{A_k=ika_k\}, \{B_k=ikb_k\}$) and
$(\psi_1,\psi_2)$ being a matter of convenience.
We shall parametrize by the variable $q$ the cylinder $D$ in the plane
of length $lA$, $l=\ln(1/q)$, and circumference $2\pi A$, 
with $A$ arbitrary. It is
mapped to an annulus the ratio of whose inner and outer
radii is $q$ by $z\rightarrow \exp(z/A)$. The measure is normalized
$$\int_0^{2\pi}Z_q(x)\, dx=1$$
and its Fourier expansion is
$$Z_q(x)=\frac1{2\pi}+\sum_{k\neq0}\nu_k(q)e^{ikx}.$$
We can try to expand each coefficient in a series of powers of $q$
$$\nu_k(q)=\sum_{j=0}^\infty c_k(\alpha_j) q^{\alpha_j}.$$
We expect from the original calculations on the Ising model
or from arguments of conformal field theory that $\alpha_0=0$, although
we admit $c_k(\alpha_0)=0$, and that $\alpha_1=\frac18$. The 
remaining $\alpha_j$ should be at least $\frac58$. 
(The usual argument of conformal field theory would select the exponents
$0, \frac18$ and $1$, and all those differing from these by two
positive units, but it requires unitarity. It is not yet clear to us
to what extent unitarity is 
pertinent in the present context. The 
whole Kac spectrum could intervene -- at least our experiments are
not fine enough to rule out
$\alpha=\frac58$ which is smaller than  $\alpha=1$.)

\begin{figure}
\begin{center}\leavevmode
\includegraphics[bb = 80 320 530 470,clip,width = 16cm]{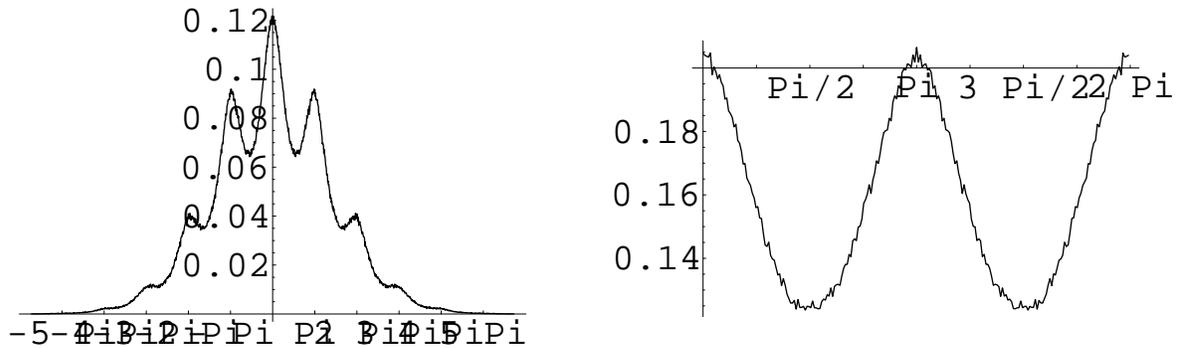}
\end{center}
\caption{The distributions of the variables $x'$ and $x$ for
the cylinder $117\times 801$.\label{fig:5.2}}
\end{figure}

We have run two sets of experiments to measure the smallest exponent
in $\nu_1(q)$, one for $LV=59$, the other for $LV=117$. As $q\rightarrow
0$, that is for long cylinders, the graph of $Z_q$ is practically
of period $\pi$, instead of $2\pi$, and the odd Fourier coefficients
$c_{2k+1}(q)\underset{q\rightarrow0}{\longrightarrow}0$. The physical
reason for this behavior is that, for very long cylinders, several
noncontractible curves of jumps in $h$ are likely to occur and
configurations with an even or an odd number of these curves will
arise in approximately the same numbers. Figure \ref{fig:5.2} 
shows, for the long cylinder of size
$117\times 801$, the 
distribution of the variable $x'$ (before the identification 
$x'\sim x'+2\pi$) and of the variable $x$. The peaks for $x'$ are centered on the integer
multiples of $\pi$, clearly underlining the role of noncontractible
curves of jumps. The figure shows configurations with $n$ curves,
$|n|=0,1,2,3,4$, and the data also indicate that $|n|=5$ and $6$ were
obtained in the sample of $1.6\times10^6$ configurations. Even
for $|n|=4$ the probability is fairly large. It should be remembered
that only $\frac1{16}$ of the configurations with 4 noncontractible
curves will contribute to the peak around $4\pi$. The distribution
$m_q(x)$ is, for this cylinder, almost perfectly periodic of period
$\pi$.

\begin{figure}
\begin{center}\leavevmode
\includegraphics[bb = 70 240 540 550,clip,width = 10cm]{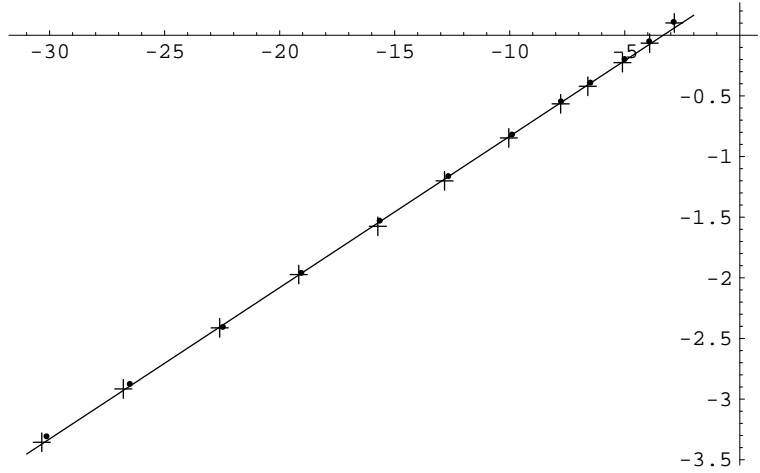}
\end{center}
\caption{Log-log plot of the Fourier coefficient $\hat \nu_1$
as a function of $q$.\label{fig:5.3}}
\end{figure}

Figure \ref{fig:5.3} is a log-log plot of $\nu_1$ as a function of
$q$. The data for the cylinders with $LV=59$ are marked by ``$\bullet$''
and those with $LV=117$ by ``$+$''. The shortest cylinders were
$59\times 27$ and $117\times 53$. We measured several other longer
cylinders for both $LV$'s. We decided to discard for both the figure
and the fits the measurements of $\hat \nu_1$ whose 95\% confidence
interval was more than 5\% of the measurement itself.\footnote{The
Fourier coefficients are given by $\nu_k=\sum_{i=1}^{2LV}c_i p_i$
where $c_i=\cos((i-\frac12)2\pi k/2LV)$ and $p_i$ are the frequencies
for the $2 LV$ bins in which the data are distributed. We use $n_i$
for the number of data in the $i$-th bin and $N$ for the sample size.
Hence $\hat p_i=n_i/N$. Since the distribution of the $n_i$ is a 
multinomial MULT$(N;n_1,n_2, \dots, n_{2LV-1})$, the first
moments are $\langle n_i\rangle=Np_i$ and $\langle n_in_j\rangle=
N(N-1)p_ip_j+Np_i\delta_{ij}$. Therefore Var$(\nu_k)=\frac1{2N}
\sum_{i\neq j}p_i p_j(c_i-c_j)^2$. For the cylinder $117\times 801$
discussed above ($q\approx -43.0$), the measured $\hat \nu_1$ with
the 95\% confidence interval is $0.00468\pm 0.00110$ even though
the sample was larger than $1.6\times 10^6$. It was not used for
the fit.} The linear fits of the log-log pairs give a slope of
$0.12506$ for $LV=59$ and of $0.12478$ for $LV=117$. The line on
the figure is the latter fit. The value $\alpha_1=\frac18$ appears
clearly. We did not check its universality but there is no reason
to doubt it.

\subsection{The ratio $Z_{+-}(q)/Z_{++}(q)$.}\label{5point2}

Let $Z_{++}(q)$ and $Z_{+-}(q)$ be the relative probabilities
that with constant boundary conditions on a cylinder of parameter
$q$ the spins are equal at opposite ends or unequal.
 There is a well-known formula due to Cardy \cite{C},
\begin{equation}
 \frac{Z_{+-}(q)}{Z_{++}(q)}=
\frac{\chi_1(q)-\sqrt2\chi_2(q)}{\chi_1(q)+\sqrt2\chi_2(q)}\label{eq:5.3}
\end{equation}
with
\begin{align*}
  \chi_1(q)=\prod_{m>0,\ m {\text{ odd}}} (1+q^m),\\
  \chi_2(q)=q^{1/8}\prod_{m>0,\ m {\text{ even}}} (1+q^m).
\end{align*}

We could, in experiments, fix the spins along one or both
of the two ends of the cylinder to be constant. 
This leads to alternative
measures $m_q(\{b_k\},x)$,
in which the spins at the left
end are taken to be $+1$, and $(Z_{+-}(q),Z_{++}(q))$. 
The question arises whether
\begin{equation}
  m_q(\{b_k\},x)=m_q(\{a_k=0\},\{b_k\},x)\label{eq:5.1}
\end{equation}
and whether
\begin{equation}
Z_{++}(q)\delta_0+Z_{+-}(q)\delta_\pi =m_q(\{a_k=0\},\{b_k=0\},x).\label{eq:5.2}
\end{equation}
These two equations require some explanation. The measure 
$m_q(\{a_k=0\},\{b_k=0\},x)$
is understood, in so far as it can be assumed to exist, to be
the conditional probability defined by the probability measure
$m_q(\{a_k\},\{b_k\},x)$, the conditions being $a_k=b_k=0$,
or equivalently $A_k=B_k=0$, $\forall k
\in \mathbb Z\setminus\{0\}$.
Experimentally this means that it is a distribution that we
approximate just as we approximate $m_q(\{a_k\},\{b_k\},x)$
itself except that we discard all samples for which the restrictions
$h_1$ and $h_2$ at the ends of the cylinder
do not lie in a suitably chosen neighborhood
of $0$. The neighborhood is thus to be as small as possible but
large enough that we do not reject so many samples that the
number of useful samples becomes impossibly small. We define
$m_q(\{a_k=0\},\{b_k\},x)$ in the same manner, but the condition is now
that $a_k=0$, $\forall k \in \mathbb Z\setminus\{0\}$.

If (\ref{eq:5.2}) is valid the distribution defined by
\begin{equation}
\int_{|A_k|<c_k}\int_{|B_k|<c_k}\ 
Z_q(\{A_k\},\{B_k\},x)\prod_k dA_k dB_k\label{intconst}
\end{equation}
with sufficiently small $c_k$'s should 
be approximately $a(q) \delta_0 + b(q) \delta_\pi$,
thus a sum of two $\delta$-functions with coefficients
whose ratio $b/a$ is given by (\ref{eq:5.3}).  Similarly
the distribution
\begin{equation}
\int_{|A_k|<c_k}\ Z_q(\{A_k\},x)\prod_k dA_k\label{intconst2}
\end{equation}
provides another ratio $b/a$ to be compared with (\ref{eq:5.3}).

Measuring these two ratios $b/a$ is difficult.
The ratio $Z_{+-}/Z_{++}$ decreases from $1$ at $q=0$ to
$0$ at $q=1$. Large ratios $Z_{+-}/Z_{++}$, those easier
to measure, correspond therefore to long cylinders. For
these the variables $A_k$ and $B_k$ are independent and
their distributions are known from previous sections. 
The effect of the constraints can therefore be estimated by using
$r_k=\text{Prob}_{q=0}(|A_k|<c_k)$.
Even by imposing restrictions $|A_k|<c_k$ and $|B_k|<c_k$
only for $k=1,2,3$, leaving the other variables free, 
a choice of $r_1=r_2=r_3\sim 0.1$ cuts the number of admissible configurations
by a factor of one million for the measurement of (\ref{intconst})
and the measurement is impracticable.
For shorter cylinders ($q\rightarrow 1$), the ratio  $Z_{+-}/Z_{++}$
drops quickly. For a circumference four times the length,
the ratio is less than $\frac2{1000}$, again difficult to measure.
We limited ourselves to a small window of $r=l/2\pi$, choosing six values
corresponding to values of $q$ increasing by a factor of approximately
$4$ at each step. Table IV gives the values of $r$, $q$, the (rather
small) lattices we used and Cardy's prediction.
The ratios $b/a$ were measured for the constraints:
\begin{equation}
c_1\sim 0.377\qquad c_2\sim 0.653 \qquad c_3\sim 0.929,
\label{constraints}
\end{equation}
the others being infinite. These numbers correspond to the
following probabilities 
$$\text{Prob}(|A_1|<c_1)= 0.2\qquad
\text{Prob}(|A_2|<c_2)= 0.3\qquad
\text{Prob}(|A_3|<c_3)= 0.4$$
if the cylinder were of size $79\times157$ like the one used in Section \ref{dis1}.
For this long cylinder and these constraints applied at both extremities,
only a fraction $(0.2\times0.3\times0.4)^2\sim 0.0006$
of the configurations would be used. We observed that for the shorter
cylinders of Table IV more configurations passed the test.
The difficulty of getting proper samples for the measurement
of (\ref{intconst2}) is of course less acute.

Three sets of measurements were taken. For the first set the constraints
given by (\ref{constraints}) were applied at both extremities of the
cylinders and is thus of the form
(\ref{intconst}). In Table IV it is refered to as {\it const/const} for
``constrained''.
For the second they were applied at one extremity while the spins
at the other were forced to be the same though they were allowed to
flip simultaneously during the Swendsen-Wang upgrades. This corresponds
to (\ref{intconst2}) and is refered to as {\it const/fixed}. The last
set is the measurement of the ratio $Z_{+-}/Z_{++}$, that is the case
{\it fixed/fixed}.
For each lattice enough configurations ($> 20$ million in each case)
were generated so that at least 30000 contributed to the integral
(\ref{intconst}). Far larger samples were obtained for the two other
sets.

\bigskip

\begin{center}
\begin{tabular}{|c|l||c|c|c|c|c|c|}
\hline
&$LV\times LH$ & $79\times122$ & $79\times104$ & $79\times86$ & $79\times68$ & $79\times52$ & $79\times34$ \\
\hline
&$r=LH/LV$ & 1.544 & 1.316 & 1.089 & 0.861 & 0.658 & 0.430 \\
\hline
&$q$       & 0.0000611 & 0.000256 & 0.00107 & 0.00448 & 0.0160 & 0.669 \\
\hline
&$Z_{+-}/Z_{++}$ & 0.408 & 0.331 & 0.249 & 0.165 & 0.0927 & 0.0260\\
\hline
\hline
\multirow{3}{12mm}{\hfil$b/a$\hfil} & {\itshape const/const} & 0.419 & 0.341 & 0.276 & 0.193 & 0.117 & 0.0419 \\
& {\itshape const/fixed} & 0.411 & 0.338 & 0.259 & 0.179 & 0.101 & 0.0301 \\
& {\itshape fixed/fixed}& 0.4071 & 0.3289 & 0.2494 & 0.1640 & 0.0916 & 0.02539
 \\
\hline
\end{tabular}
\end{center}
\medskip

\centerline{Table IV. Ratios $b/a$ measured for several cylinders.}

\bigskip

Because of the small sample, especially in the case (\ref{intconst}),
large statistical
variations are expected between neighboring bins and smoothing provides
an efficient tool to identify the two local maxima around $x=\pi$ and
$x=0$ whose ratio was used as a measurement of $b/a$. These measurements
appear in the last lines of Table IV. (Smoothing was done as in Paragraph
\ref{cylinder}. The smoothing parameter was chosen as if the distribution of
$x$ were approximately the sum of two gaussians centered at $\theta=0$
and $\theta=\pi$. The ratios $b/a$ did not seem to be very sensitive
to the exact choice of the smoothing parameter. Of course the case
{\it fixed/fixed} does not require any smoothing since the distribution
is actually of the form $a(q) \delta_0 + b(q) \delta_\pi$.)
The measurements for {\it constrained/constrained} and
{\it constrained/fixed} are systematically larger than the predicted values
though they are very close, in fact
closer for longer cylinders than for shorter ones.

\begin{figure}
\begin{center}\leavevmode
\includegraphics[bb = 72 240 540 554,clip,width = 10cm]{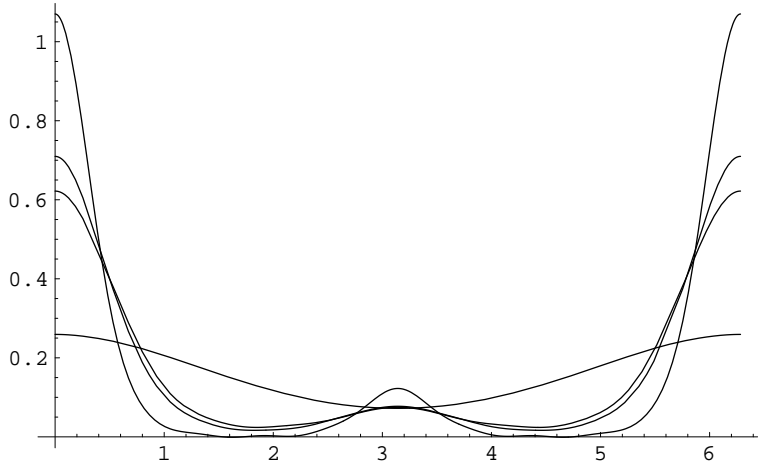}
\end{center}
\caption{The distribution of $x$ for four different sets of 
constraints $\{c_k\}$.\label{fig:5.1}}
\end{figure}

It is useful to see how the choice of constraints changes the measured ratios
$b/a$ and whether the distribution of the variable $x$ is at all similar
to the proposed sum $a \delta_0 + b \delta_\pi$. For the cylinder $79
\times 52$ we compared four sets
of constraints for the measurement of (\ref{intconst}). The first set
consisted of no constraint at all, that is all the $c_k$'s were infinite.
The second was the one used before and the finite $c_k$'s for the third
and fourth sets were
$$c_1\sim 0.259\qquad c_2\sim 0.653 \qquad c_3\sim 0.929$$
and
$$c_1\sim 0.259\qquad c_2\sim 0.441 \qquad c_3\sim 0.614 \qquad c_4\sim 0.782.$$
These $c_k$'s correspond to $r_1=0.1, r_2=0.3, r_3=0.4$ and
$r_1=0.1, r_2=0.15, r_3=0.2, r_4=0.25$.
For the fourth set
only $3152$ configurations were admissible out of the 200 millions
generated and they were distributed in the $2 LV=632$ bins. Errors
are large in this case. Instead of smoothing as before
we compared the four sets by
expanding their histograms in Fourier series keeping only the
first ten terms. The ratios $b/a$ are sensitive to the number
of terms kept. Only the first two digits of the ratios given below, 
at the end of this paragraph,
are reliable. 
The smoothed distributions are shown on Figure \ref{fig:5.1}. If the
distribution goes to $a \delta_0 + b \delta_\pi$
as the constraints become more
stringent then the peaks at $0$ and $\pi$
should be narrower and the
distribution around $\frac{\pi}2$ and $\frac{3\pi}2$ should
go to zero as one goes to the first to the fourth set. This is what 
happens with the four curves. At $\frac{\pi}2$ the top curve is
that with no constraint and the one closer to zero corresponds to
the fourth set of constraints. Even though the values of $a$ and
$b$ for the three last sets are quite different, as they should be, 
their ratios are strikingly close: $0.120, 0.109$ and $0.114$.  

\begin{figure}
\begin{center}\leavevmode
\includegraphics[bb = 72 240 540 554,clip,width = 10cm]{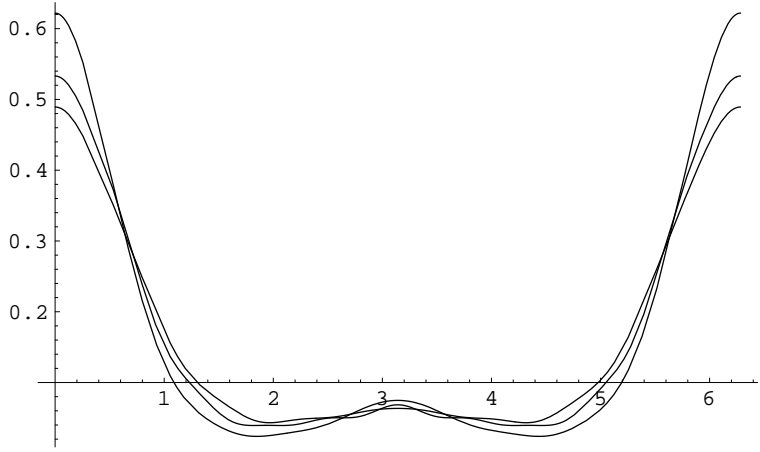}
\end{center}
\caption{The distribution of $x$ for the constraints (\ref{constraints}) 
on the three lattices $79\times 52$, $158\times 104$ and $316\times
208$.\label{fig:5.4}}
\end{figure}

Finally we compared the ratios
$b/a$ for the three lattices $79\times 52$, $158\times 104$ and $316\times
208$ using always the constraints (\ref{constraints}). The numbers of admissible
configurations were 40409, 9931 and 8816 and the ratios $b/a$, obtained
again after truncation of their Fourier series, are
$0.120, 0.129$ and $0.130$. These numbers are the same within
the statistical errors though the values of $a$ and $b$ are again different.
Figure \ref{fig:5.4} shows the three distributions, the sharper peaks
being for the smaller lattices. It seems that smaller $c_k$'s are
necessary for finer lattices if the peaks are to be as sharp
as for the coarse lattice.

It is not clear whether the above measurement technique can
reproduce accurately the ratios $Z_{+-}/Z_{++}$
with a proper choice of the $c_k$'s and
the size of the lattice. The very superficial analysis we have done
does not indicate any decrease in the small gap appearing in Table
IV for the short cylinders. Still the measurements and the predictions
are very close.

\subsection{The measure $m_q(\{a_k\},\{b_k\},x)$ for long 
cylinders.}\label{sec53}

 Some identities are suggested by the previous experiments. 
For infinitely long cylinders the following
hypothesis seems natural
\begin{equation}
Z_{q=0}(\{a_k\},\{b_k\},x)=
\int_0^{2\pi} Z_{q=0}(\{a_{-k}\},y-x)Z_{q=0}(\{b_k\},y) \, dy.\label{eq:old}
\end{equation}
As evidence, integrate with respect to the $a_k$ and $b_k$.
On the left we obtain
$$
\sum_k\nu_k\exp(ikx)
$$
and on the right 
$$
2\pi\sum_k|\mu_k|^2\exp(ikx),
$$
if
$$
  \int Z_0(\{b_k\},x)\prod db_k=\sum_k\mu_k\exp(ikx).
$$
We have, by definition, $\nu_0=\mu_0=\frac1{2\pi}$ and $\nu_k=\mu_k=0$ if $k$ is odd.
Experiments on a cylinder with $59\times 401$ sites yield
\begin{alignat}{3}
\nu_2&\sim .00273&\qquad \mu_2&\sim .0208&\qquad 2\pi\mu_2^2&\sim .00271,\\
\nu_4&\sim .0000267&\qquad \mu_4&\sim .00279&\qquad 2\pi\mu_4^2&\sim .0000488.
\end{alignat}
Unfortunately only the first line carries any conviction.
It may not be possible to measure $\nu_4$ with
any accuracy.

 The measure $dm_0(\{a_k\},x)=Z_0(\{a_k\},x)dx$ may be of some interest, but we
cannot offer any precise hypotheses. It can be expanded in
a Fourier series.
$$
Z_0(\{a_k\},x)=\sum_j\mu_j(\{a_k\})\exp(ijx),
$$
in which $\mu_0(\{a_k\})\equiv1$ and $\mu_j(\{a_k\})\equiv0$
for $j$ odd. Then, for example, $\mu_2(\{a_k\})$
is a function of $\{a_k\}$, or equivalently, of
$\{A_k\}$, but,
in spite of considerable effort, we have no idea
what this function might be.

\begin{figure}
\begin{center}\leavevmode
\includegraphics[bb = 72 240 540 540,clip,width = 12cm]{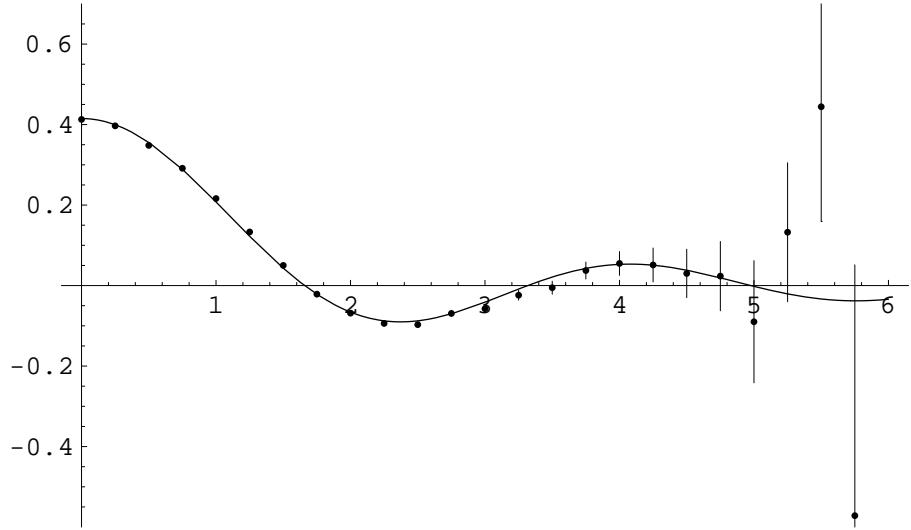}
\end{center}
\caption{The second Fourier coefficient $\mu_2$ as a function of 
$|A_1|$.\label{fig:5.5}}
\end{figure}

 A simpler function is
\begin{equation}
\int \mu_2(\{a_k\})\prod_{k\ge2} da_k =f(|A_1|).\label{eq:5.b}
\end{equation}
The experiments indicate that
\begin{equation}
  f(x)\sim a\frac{\sin(bx\pi)}{(bx\pi)}\label{eq:5.d}
\end{equation}
with $a\sim 0.415$ and $b\sim 0.603$, but this can be no more than
an approximation, as Figure \ref{fig:5.5} indicates. (It was
obtained for the cylinder $157\times 1067$ with a sample of
more than a million configurations. The error bars are
indicated.) 

The functions $\mu_j(\{a_k\})$ possess little symmetry. They are
invariant under a rotation, thus under a simultaneous transformation
of all variables $a_k\rightarrow e^{ik\theta}a_k$,
$\theta$ arbitrary, but not obviously under anything else, so that
for example,
$$
\int \mu_2(\{a_k\})\prod_{k\ge3} da_k
$$
is a function of three variables, $|A_1|$, $|A_2|$
{\it and} $\arg(A_1^2/A_2)$. The functions
$\mu_j(\{a_k\})$ are intriguing, and we would have very
much liked to discover more about them.

\section{Crossings.}\label{trav}

\subsection{Events and the two hypotheses.}

Crossings are one of the main order parameters for percolation
models. Consider, for example, 
a rectangle covered by a regular lattice. A configuration is fixed 
when each vertex has been declared open or
closed and this configuration has a crossing
if it is possible to move on open sites joined by lattice bonds from
the left side of the rectangle to the right one. A probability is
usually defined on the set of all configurations by fixing the
probability $p$ that a site is open, so that a site is then, of course,
closed with probability $1-p$. 
The probabilities for each site
are independent but equal. In the limit
of mesh length zero, the probability of such horizontal
crossing is known (rigorously) to have a singular behavior
as a function of $p$, being $0$ for $p<p_c$ and $1$ for $p>p_c$,
for a certain constant $p_c\in(0,1)$ that depends on the lattice.
This definition can be extended readily to the Ising model by 
replacing crossings on open sites by crossings on spins of a given
sign, say, for example, of positive sign. The probability of crossings
on clusters of $+$ spins is not a familiar order parameter for
the Ising model, and it is not even clear that
it is not trivial, thus identically $1$ or identically $0$.
We examined it, at first, only out of idle curiosity, following
a suggestion of Haru Pinson and were somewhat astonished
to discover that it is far from trivial. With hindsight,
it does have some immediately appealing features
and has been studied before although not with the
same goals \cite{KSC}.
It is related to spontaneous magnetization and 
to the geometry of the main cluster.
It even turns out to share striking
properties of the percolation crossings: universality and conformal
invariance \cite{LPS}. Whether a formula for it
analogous to that of Cardy \cite{C2}
for percolation remains an open question. We recall
the definitions.

Let $D$ be a domain and $D'$ a closed subset of $D$. Let
$\alpha=\{(\alpha_1, \alpha_2), \dots, (\alpha_{2n-1},\alpha_{2n})\}$
and $\beta=\{(\beta_1,\beta_2),\dots, (\beta_{2m-1},\beta_{2m})\}$
be sets of $n$ and $m$ pairs of intervals in the boundary of
$D'$ such that the $2(m+n)$
intervals are pairwise disjoint. (In fact, the intervals
need not be in the boundary of $D'$ but these are the only cases we
treated.) Let a lattice $\mathcal G$ be
superimposed upon the domain $D$. 
Let $\Gamma$ be a configuration for the Ising model $(\mathcal G,J)$ on $D$
and $\Gamma'$ its restriction to $D'$.
We shall say that the event $E$
specified by the data $(D,D', \alpha, \beta)$ occurs for
the configuration $\Gamma$   

\newcommand{\encorelabel}[1]{\mbox{{\itshape (#1)}}\hfil}
\newenvironment{MesNouListes}%
{\begin{list}{}%
   {\renewcommand{\makelabel}{\encorelabel}%
    \setlength{\parsep}{0pt}%
    \setlength{\topsep}{0pt}%
    \setlength{\partopsep}{0pt}%
    \setlength{\labelsep}{0pt}%
    \setlength{\itemsep}{0pt}%
    \setlength{\labelwidth}{0.33in}%
    \setlength{\leftmargin}{\labelwidth}%
   }%
}%
{\end{list}}

\begin{MesNouListes}
\item[i] if for every pair $(\alpha_{2i-1},\alpha_{2i})$, $i=1,\dots,n$,
there is a connected cluster of $+$ spins for $\Gamma'$
that intersects both $\alpha_{2i-1}$ and $\alpha_{2i}$
\item[ii] and if for no pair $(\beta_{2j-1},\beta_{2j})$, $j=1,\dots, m$,
is there a connected cluster of $+$ spins for $\Gamma'$ 
that intersects both $\beta_{2j-1}$ and $\beta_{2j}$.
\end{MesNouListes}
(For percolation the definition of an event is simpler as the introduction
of the larger domain $D$ is superfluous, so that the measure 
on the configurations on $D'$ is independent of 
the choice of $D$. Thus one takes $D=D'$.) 
Let ${\mathcal G}_a$ be the lattice $\mathcal G$ shrunk
by the factor $a$ and let $\pi_E^{(\mathcal G,J),a}$
be the probability of the event $E=(D,D', \alpha,\beta)$
for the Ising model $({\mathcal G}_a,J)$ at its critical point, 
then $\pi_E^{(\mathcal G,J)}$ will be defined
as
$$\pi_E^{(\mathcal G,J)}=\lim_{a\rightarrow 0}\pi_E^{(\mathcal G,J),a}$$
if the limit exists. The two hypotheses of universality and conformal
invariance are then identical to those proposed in \cite{LPS} for
percolation. 

\noindent{\scshape Hypothesis of universality}: {\itshape
For any pair of Ising models $({\mathcal G}, J)$ and $({\mathcal G}', J')$,
there exists an element $g$ of $GL(2,\mathbb R)$ such that} 
\begin{equation}
\pi_E^{(\mathcal G,J)}=\pi_{gE}^{({\mathcal G}',J')},\qquad
\text{\itshape for all events\ }E.\label{eq:univcros}
\end{equation}

\noindent{\scshape Hypothesis of conformal invariance}: {\itshape
Let $({\mathcal G}_\square, J_\square)$ be the Ising model on the
square lattice with critical coupling $J_\square$. 
Let $\phi$ be a map satisfying the
same requirements as in the hypothesis of conformal invariance
of Section \ref{dis2}. Then}
\begin{equation}
\pi_E^{({\mathcal G}_\square,J_\square)}=
\pi_{\phi E}^{({\mathcal G}_\square,J_\square)},\qquad
\text{\itshape for all events\ }E.\label{eq:confcros}
\end{equation}

It is best to observe explicitly that the map $\phi$
acts on both $D$ and $D'$, so that if $D$ is the
whole plane there are very few admissible $\phi$. 
The following two paragraphs describe simulations done to examine these
hypotheses when $D=D'$ (Paragraph \ref{trav2}) or $D'\varsubsetneq D$
(Paragraph \ref{trav3}).

\subsection{$D=D'$.}\label{trav2}

For the first events to be considered we take $D=D'$.
Their description is simple when the geometry of $D$
is that of a rectangle. We introduce the notation $\pi_h(r)$ and $\pi_v(r)$,
instead of $\pi_E$, for events $E$ occuring on $D$, a rectangle
with aspect ratio $r=\text{width}/\text{height}$, with a single
pair $(\alpha_1,\alpha_2)$ and an empty $\beta$. For the probability
of horizontal crossings $\pi_h$ the two intervals $\alpha_1$ and $\alpha_2$
are the left and right sides and for the probability $\pi_v$
of vertical crossings, the top and bottom. The probability $\pi_{hv}(r)$
will give an example of an event with two pairs $\alpha=\{
(\alpha_1,\alpha_2),(\alpha_3,\alpha_4)\}$. It is the probability
of having simultaneously horizontal and vertical crossings in a rectangle
$D$ of aspect ratio $r$. Note that the number $\pi_h(r)-\pi_{hv}(r)$
is the probability to have a horizontal crossing without having a
vertical one. It thus provides an example of event $E$ with one pair
$\alpha$ and one pair $\beta$.
Finally we introduce $\pi_h^A(r)$ and $\pi_v^A(r)$
whose corresponding events have a single pair $(\alpha_1,\alpha_2)$.
For $\pi_h^A$, $\alpha_1$ is the vertical segment splitting the rectangle
in two parts of equal areas and $\alpha_2$ the right side.
For $\pi_v^A$, $\alpha_1$ is the horizontal segment in the middle of
the rectangle and $\alpha_2$ the top side. For these two probabilities, we
could also have taken $D'$ to be the half-rectangle bounded by $\{\alpha_1,\alpha_2\}$
because a path joining $\alpha_2$ to $\alpha_1$ reaches $\alpha_1$
before it leaves this half-rectangle, so that the
sites outside the half-rectangle
are superfluous.

Two difficulties limit the precision of the numerical measurements.
The first one is the limitation due to a choice of convention and
was discussed at length in \cite{LPPS}. Since $\pi_E^{({\mathcal G},J)}$
are approximated by measurements on finite lattices, the exact
position of the domain $D$ with respect to the lattice must be
specified by convention; or, equivalently, a prescription must
be given for calculating $r$ for a rectangle with $LH$ sites in the
horizontal direction and $LV$ in the vertical one. 
To examine the sensitivity
to convention consider an extreme case. Suppose that in convention I the width is
that of the narrowest rectangle containing the $LH$ horizontal sites
and that in convention II, the width is that of the widest. For the
square lattice oriented so that its bonds are parallel to the sides of
the rectangles, the difference between the two widths is 2 mesh units.
If both conventions measure the height in the same way, the discrepancy for
$\pi_h$ between the two conventions is 
$$\frac2{LV}|\pi'_h(r)|,$$
the prime denoting a derivative. 
These numbers
can be estimated from the data of Table VII. 
Table V gives an order of magnitude for this limitation
on precision
for the two probabilities $\pi_h$ and $\pi_h^A$
at the center ($r=1$) and at the extremities ($r=0.1361$ and $7.353$)
of the range of the aspect ratio we measured.  
Our conventions are given in the 
appendix; 
whatever they are, the above limitation is unavoidable.

\begin{center}
\begin{tabular}{|c||c|c|c||c|c|c|}
\hline
&&&&&& \\
$r$ & $\pi_h$ & $\frac2{LV}|\pi_h'|$ & Statistical error 
& $\pi_h^A$ & $\frac2{LV}|\pi_h^{A\prime}|$ & Statistical error \\
&&&&&& \\
\hline
\hline
&&&&&& \\
7.3 & 0.02 & $3\times 10^{-4}$ & $6\times 10^{-4}$ 
  & 0.12 & $1\times 10^{-4}$ & $1\times 10^{-3}$ \\
1.0  & 0.50  & $3\times 10^{-3}$ & $2\times 10^{-3}$ 
  & 0.66 & $3\times 10^{-3}$ & $2\times 10^{-3}$ \\
0.14& 0.98 & $2\times 10^{-3}$ & $6\times 10^{-4}$ 
  & 0.99 & $5\times 10^{-3}$ & $3\times 10^{-4}$ \\
&&&&&& \\
\hline
\end{tabular}
\end{center}
\medskip

\noindent Table V: Sensitivity to convention and statistical
errors for a sample of 200000 for 3 values of $r$ on a lattice containing
$\approx 40000$ sites.
\bigskip

To confirm the conformal invariance we also measure all these
probabilities for comparable geometries on the disk and the cylinder. The Schwarz-Christoffel
map can be chosen so that the four vertices of the rectangle of
aspect ratio $r$ correspond to the four points $\pm e^{\pm i\theta}$
for some $\theta\in [0,\frac\pi 2]$, on the unit circle. Notice that
$r=0$ corresponds to $\theta=\frac{\pi}2$, $r=1$ to $\theta=\frac{\pi}4$
and $r=\infty$ to $\theta=0$. The slope of the function
$\theta(r)$ at $r=0$ is zero. This means that
the sensitivity to convention is magnified for values of $\theta$
close to $\pi/2$. For 
example we measured the probabilities $\pi$ in the rectangular
geometry for five different
values of $r$ in the range $[0.1361,0.1647]$ . The corresponding range
of $\theta$ is $[1.57051,1.57076]$ and, 
on the disk of radius $r=300.2$ mesh units
that we used, at most one site can be 
contained along the boundary in this interval.
This is even worse for the corresponding geometry
on the cylinder of size $397\times 793$ where the
$\pi$'s have also been measured. Such measurements are too imprecise
to be useful and we measured the probabilities, on the disk 
and the cylinder, only
for the $\theta$'s corresponding to the forty-one values in the middle
of the eighty-one we used for the rectangular geometry. The arc between the
two smallest as well as the two largest
$\theta$'s among these 41 values is about $3.7$ mesh units.
Since we have taken the sites in the angles $(\pi-\theta,\pi+\theta)$,
$(-\theta,\theta)$ to define the pair of intervals $(\alpha_1,\alpha_2)$,
it is clear that a rather large systematic error is to be expected.

\begin{figure}
\begin{center}\leavevmode
\includegraphics[bb = 72 240 540 554,clip,width = 12cm]{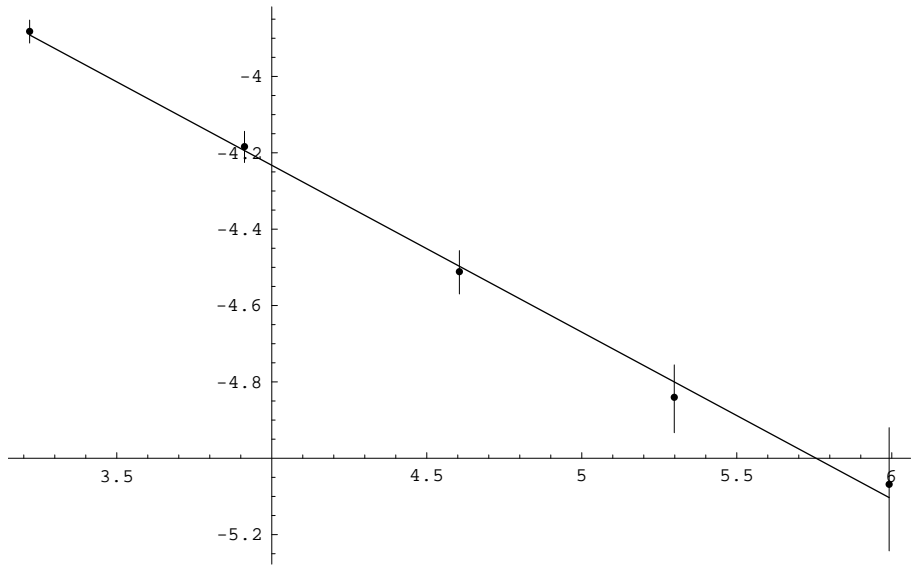}
\end{center}
\caption{Log-log plot of $1-\pi_h-\pi_v$ measured
on a square as a function of the linear number of sites.\label{fig:4.1}}
\end{figure}

Finite-size effects are the origin of the second difficulty. Fortunately
the relation $\pi_h(r)+\pi_v(r)=1$ is verified for the triangular
lattice, even for finite ones. This is a well-known identity for percolation
and the argument for its validity here is the same. For the other
pairs $({\mathcal G},J)$, this relation is not verified for finite
lattices, that is
$\pi_h^{(\mathcal G,J),a}+\pi_v^{(\mathcal G,J),a}\neq 1$. 
Nevertheless, if universality holds, it should be satisfied for the other
pairs $({\mathcal G},J)$ in the limit of zero mesh. 
Departure from zero of the quantity $|1-\pi_h^{(LH,LV)}-\pi_v^{(LH,LV)}|$
for $r=\frac{LH}{LV}$ is therefore a measure of finite-size effects. 
Interpreted differently, this quantity is a measure of the error made on 
$\pi_h(r)$ when the number $\pi_h^{(LH,LV)}$ is used in its stead.
A verification for a square
domain covered by the square lattice indicates that $\pi_h^{({\mathcal
G}_\square,J_\square)}(r)+\pi_v^{({\mathcal G}_\square,J_\square)}(r)=1$ 
is likely
to hold when the number of sites goes to infinity. The log-log plot of
Figure \ref{fig:4.1}
shows that $(1-\pi_h-\pi_v)$ and $LH$
are related by a power law. (The five
points correspond to squares with $25, 50, 100, 200$ and $400$ sites along
their edges.) The slope is $0.437$ and unlikely
to be universal. The crossing probabilities $\pi_h$ and $\pi_v$ on the
square lattice
for the square ($r=1$) of size $200\times200$
were measured to be $0.4963$ and $0.4964$. 
The gap is of order of $3.5\times10^{-3}$, comparable
to the value of $2\pi'_h(r)/{LV}$ at this point. 
Note finally that, even though $\pi_h(r)+\pi_v(r)=1$ holds
for finite rectangular subsets of the triangular lattice, it does not
follow that $\pi_h(r)$ is equal to $\pi_h^{(LH,LV)}$ as finite-size
effects could alter both $\pi_h^{(LH,LV)}$ and $\pi_v^{(LH,LV)}$
while keeping their sum equal to $1$.

\begin{figure}
\begin{center}\leavevmode
\includegraphics[bb = 72 330 540 500,clip,width = 15cm]{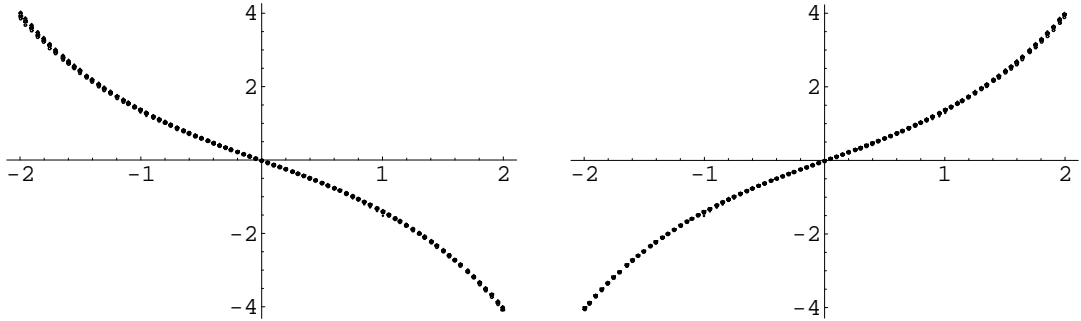}
\end{center}
\caption{$\log \pi_h/(1-\pi_h)$ and $\log \pi_v/(1-\pi_v)$ 
as a function of $\log r$.\label{fig:4.h}}
\end{figure}

\begin{figure}
\begin{center}\leavevmode
\includegraphics[bb = 72 240 540 554,clip,width = 12cm]{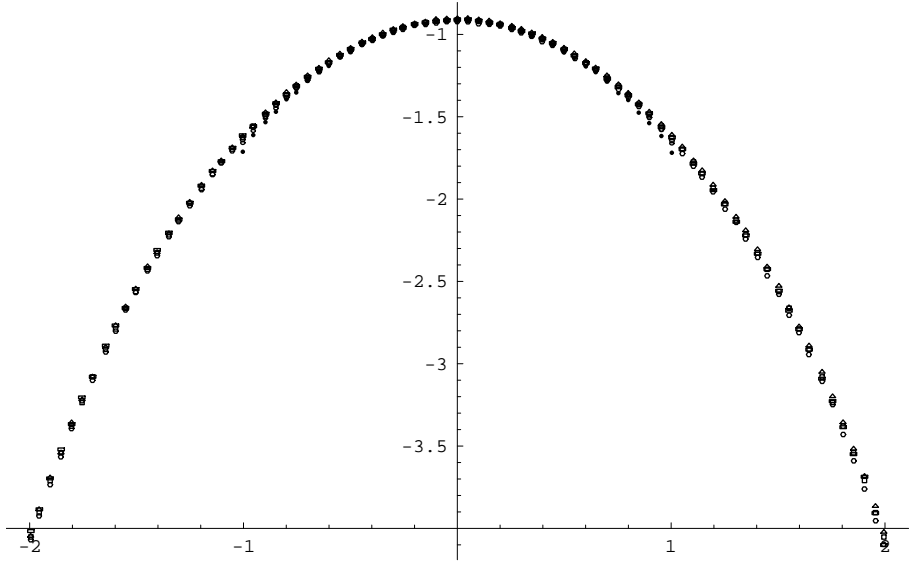}
\end{center}
\caption{$\log \pi_{hv}$ as a function of $\log r$.\label{fig:4.hv}}
\end{figure}

\begin{figure}
\begin{center}\leavevmode
\includegraphics[bb = 72 330 540 500,clip,width = 15cm]{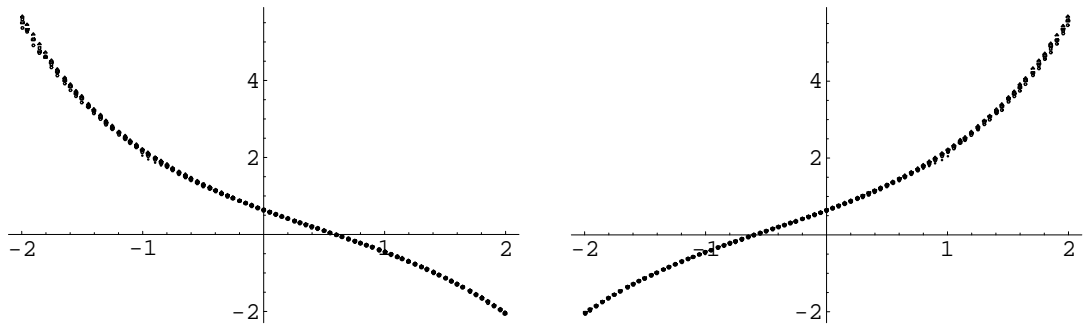}
\end{center}
\caption{$\log \pi_h^A/(1-\pi_h^A)$ and 
$\log \pi_v^A/(1-\pi_v^A)$ as functions of $\log r$.\label{fig:4.hA}}
\end{figure}

The five plots in Figures \ref{fig:4.h} to \ref{fig:4.hA} show all the data 
available: the probabilities for
81 values of the aspect ratio for the rectangles and 41
for the disk and the cylinder. For the rectangles, 4 different Ising
models were studied: the three regular lattices with isotropic coupling
and the square lattice with the anisotropic coupling used in Section
\ref{dis2}. Each figure contains therefore six sets of measurements,
four for the rectangles, one for the disk and one for the cylinder.
The cylinder is treated as though it were infinitely long and
the crossings are from an interval on one end to another disjoint interval of the
same length
on the same end,
the intervals being chosen so that their position on
the cylinder is conformally equivalent to
that of two opposite sides on a rectangle.
Because of the large amount of information on these figures, 
the error bars were not drawn. Some of the $95$\% confidence
intervals for the measurements of $\pi_h$ were listed on Table V
and the difference of the extreme values of these intervals is equal to
$0.07$ at $r=0.136$, $0.02$ 
at $r=1.000$ and $0.07$ at $r=7.351$
for the variable $\log \pi_h/(1-\pi_h)$ that appears in Figure \ref{fig:4.h}.
(For the square lattice, the confidence intervals on the probabilities
are a factor $\frac1{\sqrt{5}}$ smaller since the sample was $5$ times
larger.) The vertical dimension of the dots on this figure
is approximately $0.065$ and thus comparable to
the statistical errors or larger
than them. 

In all the figures, one sees clearly
some spreading of the data at the two extremities of the range of $r$.
The data for the disk and for the cylinder also fall 
slightly beside those for the rectangles around the extreme
values of their range ($\log r \sim \pm 1$). These small discrepancies
can all be explained by the above two limitations.
First, for all the pairs $({\mathcal G}, J)$ but the isotropic
triangular lattice, the quantity $\pi_h+\pi_v$ is {\em less}
than one. It is thus likely that finite-size effects tend to decrease both 
$\pi_h$ and $\pi_v$. Since for $\log r \sim \pm 2$, one of the linear
dimensions of the rectangle is half what it is around $\log r\sim0$, the
values of $\pi_h$ and $\pi_v$ should be spread more at the extremities
than at the center of the range of $r$; and $\pi_h^{(\triangle)}$
should be
the largest of all measurements. This is what is observed though the
spread is noticeable only when the small linear dimension is in the
direction of the crossing. Second, by keeping the sites inside the
sector $(-\theta(r),\theta(r))$ or
$(\pi-\theta(r),\pi+\theta(r))$, the number of sites (necessarily
integral) is underestimated, leading to probabilities
lower than what universality would predict. This is again what is 
observed. But these discrepancies are rather small. As can be seen
from the figures the agreement is remarkable.

Only for the isotropic Ising model on the square lattice is $\pi_h^{(LH,LV)}$
strictly equal to $\pi_v^{(LV,LH)}$. It is then sufficient to measure
the five probabilities $\pi_h, \pi_v, \allowbreak
\pi_{hv},\allowbreak\pi_h^A,\pi_v^A$ for 41 values of $r$ to cover the
same range. We profited from this coincidence and substantially increased the
sample in order to measure the
probabilities with very high accuracy. In this case 
each sample contained at least one million configurations.
For the other models we used samples of at least $200000$ configurations.
As can be seen from Table V, even the smaller sample size yields statistical
errors at worst of the same order of magnitude as the sensitivity
to conventions.
Table VII lists the crossing probabilities
$\pi_h, \pi_v, 
\pi_{hv},\pi_h^A,\pi_v^A$ for the isotropic Ising model on the square lattice;
Table VIII lists them for the triangular lattice.
This table gives an idea of both the difference between the various
probabilities as measured for two different Ising models and
the isotropy of the probabilities: the pairs $(\pi_h,\pi_v)$ and $(\pi_h^A,\pi_v^A)$
are approximately symmetric under the exchange of $r\leftrightarrow r^{-1}$ even though
the lattice is not invariant under a rotation of $\frac{\pi}2$.

\begin{figure}
\begin{center}\leavevmode
\includegraphics[bb = 72 320 540 460,clip,width = 17cm]{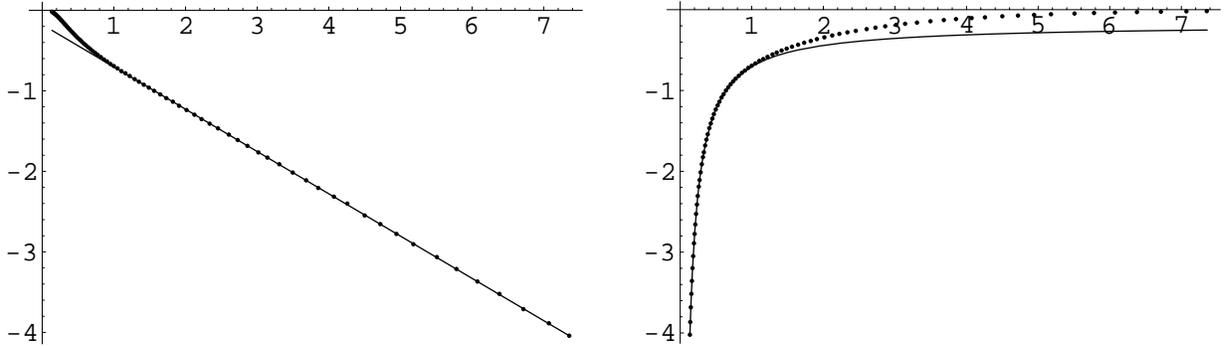}
\end{center}
\caption{Fits of the asymptotic behavior of $\pi_h$:
(a) $\log\pi_h(r)$ and (b) $\log(1-\pi_h(r))$.\label{fig:4.2}}
\end{figure}

For percolation, Cardy's formula predicts the following
asymptotic behavior
$$\log \pi_h^{\text{perco}}(r)\quad
\underset{r\rightarrow\infty}{\longrightarrow}\quad
-\frac{\pi}3 r+\text{constant},$$
or equivalently
$$\log (1-\pi_h^{\text{perco}}(r))
\quad\underset{r\rightarrow0}{\longrightarrow}\quad
-\frac{\pi}{3 r}+\text{constant}.$$
The data
for the Ising model behave similarly.
We used those for the triangular lattice since they respect closely
the relation $\pi_h(r)+\pi_v(r)=1$. We rejected the ten points at
both extremities of the spectrum of $r$ because they carry the
largest finite-size effect. The 
30 remaining points with largest $r$ were fitted to $\log\pi_h(r)
\approx a+br$ and the 30 with smallest $r$ were fitted to
$\log(1-\pi_h(r))\approx c+d/r$. The fits appear in Figure \ref{fig:4.2}.
The constants $b$ and $d$ turned out to be $-0.1672\pi$ and $-0.1664\pi$.
A natural guess for both constants is $-\pi/6$.

\subsection{$D'\varsubsetneq D$.}\label{trav3}

We measured the crossing probabilities
from one curve $C_i$ to another one $C_j$ 
on the cylinder, $0\le i<j\le4$, and
for the corresponding configurations on the disk. (The curves
$C_i$ have been introduced in Section \ref{dis2}.)
The simulations were done on the cylinder with 
$397\times 793$ sites and on the disk of radius
$300.2$ mesh units.
The results are
tabulated in Table VI. In each cell the number on top
is the probability for the disk, the one on the bottom
that for the cylinder and, again, the vertical bar ``$|$''
is used as in Section \ref{dis1} to give the statistical
errors. The agreement is convincing even though the
probabilities for the disk are systematically larger
than those for the cylinders. Again the
geometries of the disk and the cylinder are
not quite conformally equivalent.  Only if the cylinder is infinitely
long can one hope to have perfect agreement. 
Since the relative gap increases
as the two curves $C_i$ and $C_j$ move closer
to the middle of the cylinder, the shortness
of the cylinder is a likely explanation for the
discrepancy.  


Since the numbers of Table VI are all close to
$1.$, one more example of crossing probability was measured.
The event $E$ for the cylinder ($=D$) is given by the following data: 
the domain $D'$ is
delimited by the curve $C_2$ and 
the right-hand side of the cylinder and $\alpha_1$
and $\alpha_2$ are the two intervals on $C_2$ that correspond
to the forty-seventh value of the aspect ratio $r$ considered in the
previous paragraph ($r=1.35$). The data for the disk are
the conformal images of those of the cylinder.
For the disk and the cylinders the numbers 
$\pi_E$ are $0.412|1$ and $0.4096|6$ respectively.

\begin{center}
\begin{tabular}{|c||c|c|c|c|}
\hline
&&&& \\
 & $C_1$ & $C_2$ & $C_3$ & $C_4$ \\
&&&& \\
\hline
\hline
&&&& \\
$C_0$ & $0.99998|1$ & $0.9976|1$ & $0.9634|4$ & $0.8465|9$ \\
      &$0.999979|5$ & $0.99741|6$ & $0.9631|2$ & $0.8456|4$ \\
&&&& \\
\hline
&&&& \\
$C_1$ &             & $0.99958|5$ & $0.9727|4$ & $0.8541|8$ \\
      &             & $0.99932|3$ & $0.9699|2$ & $0.8510|4$ \\
&&&& \\
\hline
&&&& \\
$C_2$ &            &            & $0.9848|3$ & $0.8643|8$ \\
      &            &             & $0.9827|2$ & $0.8614|4$ \\
&&&& \\
\hline
&&&& \\
$C_3$ &            &            &            & $0.8995|7$ \\
      &            &             &            & $0.8973|4$ \\
&&&& \\
\hline
\end{tabular}
\end{center}

\medskip\noindent Table VI: 
Crossing probabilities from one curve $C_i$
to another $C_j$ for the disk and the cylinder.

\bigskip

Another interesting choice is $D'\varsubsetneq D=\mathbb R^2$. That
means measuring crossings on domains $D'$
in the bulk.
We have seen that the $\omega_k^{\text{bulk}}$ are larger than those
at the boundary by approximately a factor of 3. The corresponding variances
$\Sigma_k^2$ are consequently smaller and the number of large clusters
intersecting the central meridians of the cylinder is also smaller. 
Are there enough of them to break crossings? Or is $\pi_h^{\text{bulk}}(r)$
a trivial function, namely equal to $\frac 12$ for all $r$?

Such a measurement would amount, in an ideal situation, to thermalizing an
infinite lattice $\mathbb Z^2$ and then measuring crossings on finite $D'$ inside
this lattice. Only the usual limitations (convention and finite size of $D'$)
would then have to be dealt with. To do the actual simulations, the first idea
is to truncate $D$ to a finite though large lattice and to choose $D'$ as
the largest domain possible inside a region in which the behavior of
the spins is as close
as possible to the bulk behavior. With our present computers, a lattice size
of practical use contains about $10^6$ sites.
If $\mathbb R^2$ is approximated by a square lattice, then it
would be of size $1000\times 1000$. The domains $D'$ used in 
Paragraph \ref{trav2} contained
around 40000 sites and the domain $D'$ with $r=1$ was therefore 
$200\times 200$. If we compare these sizes with disks, as we are interested
only in orders of magnitude, the boundary of $D'$ would correspond to a 
circle of radius one fifth that of $D$. The distribution $m_{D,\partial D'}$
on the boundary of $D'$ is approximately equivalent to that of a circle at a 
distance of $100$ mesh units from the boundary of the cylinder $397\times 793$. 
Figure \ref{fig:3.5} (Paragraph \ref{simconf}) shows that the first four Fourier
coefficients are still far from their bulk distribution. These coefficients are
precisely those measuring the large clusters responsible for creating
crossings or for breaking them.

But as we have seen (Paragraph \ref{simconf}),
the middle of a long cylinder provides a better approximation to bulk
behavior. 
So we confine our experiments to cylinders.
If a square $D'$ of size $200\times 200$ is located in the
middle of a cylinder of size about $397\times 793$ as before, its distance from 
the boundary is about 300 mesh units and its spins behave essentially
as in the bulk as can be seen in Figure \ref{fig:3.5}. This choice has 
one possible drawback. It spoils the 
symmetry between horizontal and vertical directions. 
The mean width of the largest cluster is surely not
equal to its mean height on a long cylinder. Fortunately a simple quantity,
$\pi_h(r)-\pi_v(\frac1r)$, can be used to quantify this symmetry breaking.

To enforce the relation $\pi_h(r)+\pi_v(r)=1$, we took the measurements
on triangular lattices with $426\times 737$ and $852\times 1475$ sites, the $737$ 
and $1475$ sites being
in the longitudinal direction. On these lattices, the crossings $\pi_h, \pi_v$
and $\pi_{hv}$ were measured on rectangles with the 81 aspect ratios
$r$ used before. To keep the rectangles safely in the bulk, we used domains
$D'$ with approximately 10000 sites. (We used the same domains on both cylinders.
See below.) The longest rectangle ($r\approx 7.3$)
has $40\times 253$ sites and its distance from the boundary, for the
cylinder of size $426\times 737$, is similar to
that of the square of size $200\times 200$ square in a cylinder
of size $397\times 793$ discussed
above. The highest rectangle ($r\approx 0.13$) has $293\times 34$ sites and its
height takes up more than $\frac23$ of the circumference of the
smaller lattice, possibly too large
a fraction if the symmetry breaking is important. The larger lattice 
helps to address this question. We also measure the 
crossings $\pi_h, \pi_v$
and $\pi_{hv}$ inside a disk of radius $100.2$ whose center is within one mesh
unit from the central
meridian of the cylinders. Note that the hypothesis of conformal invariance stated
above does not relate the crossings in the bulk on the rectangles and on the disk.
As emphasized, the map $\phi$ must act on both $D$ and $D'$ and there is no
conformal map from the plane ($D$) to the plane taking a rectangle ($D'$) to a
disk.

\begin{figure}
\begin{center}\leavevmode
\includegraphics[bb = 72 330 540 500,clip,width = 15cm]{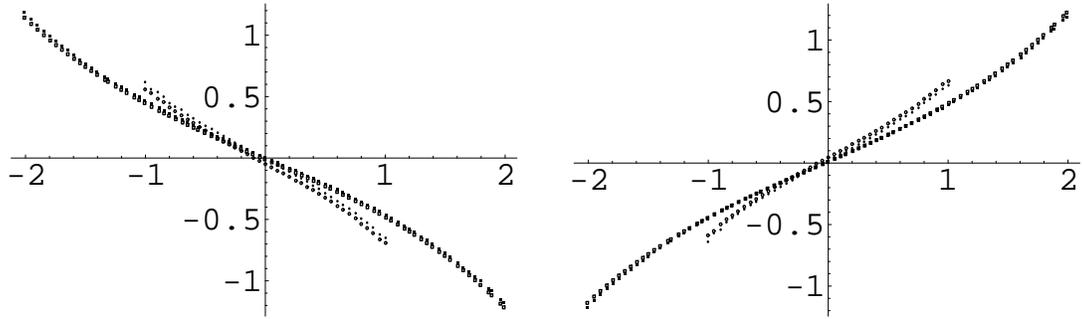}
\end{center}
\caption{$\log \pi_h/(1-\pi_h)$ and $\log \pi_v/(1-\pi_v)$ 
as a function of $\log r$.\label{fig:4.10hetv}}
\end{figure}

\begin{figure}
\begin{center}\leavevmode
\includegraphics[bb = 72 240 540 554,clip,width = 12cm]{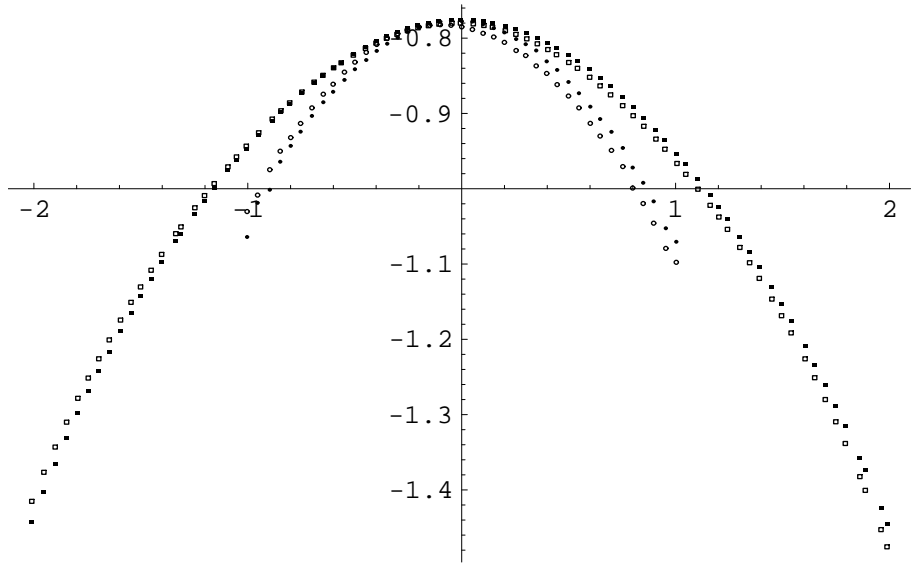}
\end{center}
\caption{$\log \pi_{hv}$ as a function of $\log r$.\label{fig:4.10hv}}
\end{figure}

Figures \ref{fig:4.10hetv} and \ref{fig:4.10hv} present the results. Squares ($\square$)
were used for the crossings on rectangles 
and circles ($\circ$) for those on the 
disk. White symbols are for the $426\times 737$ lattice and black for the $852\times 1475$.
The two samples were $895000$ for the $426\times 737$ cylinder and
and $227000$ for the $852\times 1475$.
Even though these data look almost identical to those presented in Paragraph
\ref{trav2} (Figures \ref{fig:4.h} and \ref{fig:4.hv}), the vertical scale is 
different. When $D=D'$, $\pi_h$ ranges from $0.02$ to $0.98$ as $r$ decreases
from $7.3$ to $0.14$. Here, in the bulk, $\pi_h$ goes from $0.23$ to $0.76$ for the
same interval of $r$. 

There is a definite breaking of the horizontal-vertical symmetry.
The graph of $\pi_{hv}$ for the cylinder with $426\times 737$ sites
is clearly asymmetrical. For the rectangles the quantities
$\pi_h(r)$ and $\pi_v(\frac1r)$ that should be equal if the symmetry was present
differ by about 6\% for $r$ large or small and by 1\% for $r\approx 1$. For the
measurements on the disk their departure from symmetry
varies from 3\% to 7\%. For both cases,
rectangles and disk, the vertical crossings are {\em always} larger than the
corresponding horizontal ones. Large clusters wrapping around the circumference
are more likely than clusters having about the same number of sites but that fail
to surround the cylinder simply because the former have fewer peripheral sites
than the latter. This difference seems to play a role here. 
If this is so,
a better measurement of the $\pi$'s would
therefore be obtained by, say, doubling the linear dimensions of the cylinder
while keeping the number of sites in the domains $D'$ unchanged. This is
why we studied the larger $852\times 1475$ cylinder. For this new experiment,
the asymmetry is essentially gone. For example most of the quantities $\hat\pi_h(r)$
and $\hat\pi_v(\frac1r)$ differ now by less than 0.5\%. Still the data for
the two lattices remain very close and experiments with smaller cylinders show
that the curves in Figures \ref{fig:4.10hetv} and \ref{fig:4.10hv} do not change
much with lattice size, so that we can assert with
some confidence that the crossing probabilities in the bulk
are well defined as the mesh goes to zero,
in other words, the crossing probabilities are defined
even when $D$ is the whole plane.  
The data, especially those for the $852\times 1475$
cylinder, must represent a very good approximation 
to the crossing probabilities
in the bulk for the rectangles and the disk.

\begin{figure}
\begin{center}\leavevmode
\includegraphics[bb = 72 285 540 500,clip,width = 15cm]{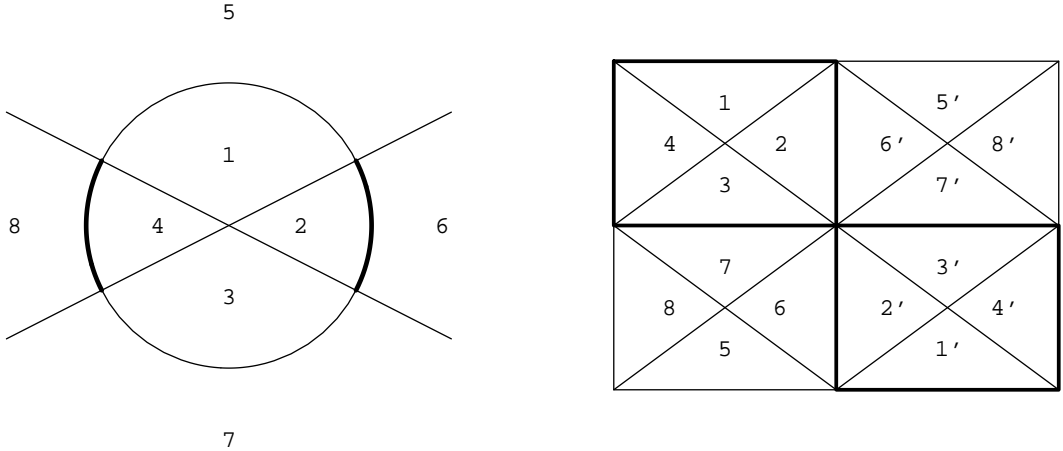}
\end{center}
\caption{The Riemann surfaces $R_1$ and $R_2$ and their corresponding
neighborhoods. Only one sheet of $R_1$ is presented.\label{fig:4.11}}
\end{figure}

In particular, the curves for the crossings on the rectangles 
and on the disk are now distinct
and their difference does not seem to be due to the limitation of the experiments. There was in fact no reason at all to compare
them or to use the parameter $r$ to describe the arcs on the
disk, for it pertains to a conformal transformation $\phi_r$
from the disk to the rectangle of aspect ratio $r$ that is no
longer pertinent. Nevertheless, it does appear that
$\pi_v^{\text{rectangle}}(r)<
\pi_v^{\text{disk}}(r)$ for $r>1$ and that
$\pi_v^{\text{rectangle}}(r)>
\pi_v^{\text{disk}}(r)$ for $r<1$, inequalities for
which we have no very persuasive 
explanation. For each $r$,
the map $\phi_r$ extends to a conformal equivalence $\phi_r$ of a
double covering $R_1(r)$ of the plane, or rather of the Riemann sphere,
ramified at four points with a torus $R_2(r)$. 
Figure \ref{fig:4.11} represents $R_1$ and
$R_2$ and their corresponding neighborhoods. Only one sheet
of $R_1$ is depicted here; the other is identical, all data
being primed ($2\rightarrow 2'$, etc.).
$R_1$ is a double covering of $\mathbb C$ with cuts tying
the four singular points on the unit circle $\omega, \bar\omega,
-\omega$ and $-\bar\omega$. The cuts were drawn along the unit
circle. Consequently the neighborhood
to the left of the domain $2$ on the first sheet is the domain $6'$
on the second. The thicker lines on $R_2$ are not cuts but circumscribe
the images of the disks on $R_1$. The top and bottom sides of the whole
rectangle are identified
as are the left and the right. Both
$R_1$ and $R_2$ are tori. 
The conformal class of $R_1$ and $R_2$
depends on $r$. The hypothesis of conformal invariance
does apply to
$\pi_v^{\text{disk}\subset R_1(r)}$
and $\pi_v^{\text{rectangle}\subset R_2(r)}$. They are expected to be 
equal. We do not know what relations might subsist between
$\pi_v^{\text{rectangle}}(r)$ and $\pi_v^{\text{rectangle}\subset R_2(r)}$
or between $\pi_v^{\text{disk}\subset R_1(r)}$
and $\pi_v^{\text{disk}}(r)$.

\newpage
    
\null\vskip.50 true in

$$
\vbox{\null\hskip1.75 true cm
{\tenpoint\tolerance=1500{
\begin{tabular}{|r|r|c|c|c|c|c|c|c|}
\hline
\hfil$LH$\hfil & $LV$ & $r$ & $r^{-1}$ & $\pi_h$ & $\pi_v$ & $\pi_{hv}$ & $\pi_h^A$ & $\pi_v^A$\\
\hline
200 & 200 & 1.000 &  1.000 &  0.4963 &  0.4964 &  0.4022 &  0.6553 &  0.6554 \\ 
205 & 195 & 1.051 &  0.951 &  0.4811 &  0.5107 &  0.4005 &  0.6403 &  0.6667 \\ 
210 & 190 & 1.105 &  0.905 &  0.4671 &  0.5250 &  0.3989 &  0.6286 &  0.6811 \\ 
216 & 186 & 1.161 &  0.861 &  0.4527 &  0.5396 &  0.3956 &  0.6153 &  0.6937 \\ 
221 & 181 & 1.221 &  0.819 &  0.4389 &  0.5553 &  0.3910 &  0.6023 &  0.7058 \\ 
227 & 176 & 1.290 &  0.775 &  0.4220 &  0.5711 &  0.3839 &  0.5870 &  0.7212 \\ 
232 & 172 & 1.349 &  0.741 &  0.4083 &  0.5837 &  0.3764 &  0.5766 &  0.7324 \\ 
238 & 168 & 1.417 &  0.706 &  0.3922 &  0.5963 &  0.3665 &  0.5627 &  0.7428 \\ 
264 & 164 & 1.488 &  0.672 &  0.3791 &  0.6133 &  0.3582 &  0.5521 &  0.7582 \\ 
250 & 160 & 1.562 &  0.640 &  0.3649 &  0.6288 &  0.3484 &  0.5398 &  0.7714 \\ 
257 & 156 & 1.647 &  0.607 &  0.3482 &  0.6449 &  0.3359 &  0.5242 &  0.7854 \\ 
263 & 152 & 1.730 &  0.578 &  0.3324 &  0.6592 &  0.3228 &  0.5111 &  0.7972 \\ 
270 & 148 & 1.824 &  0.548 &  0.3169 &  0.6753 &  0.3097 &  0.4982 &  0.8102 \\ 
277 & 145 & 1.910 &  0.524 &  0.3028 &  0.6888 &  0.2974 &  0.4847 &  0.8195 \\ 
284 & 141 & 2.014 &  0.497 &  0.2875 &  0.7062 &  0.2836 &  0.4727 &  0.8334 \\ 
291 & 137 & 2.124 &  0.471 &  0.2717 &  0.7208 &  0.2688 &  0.4573 &  0.8447 \\ 
298 & 134 & 2.224 &  0.450 &  0.2571 &  0.7354 &  0.2550 &  0.4456 &  0.8569 \\ 
306 & 131 & 2.336 &  0.428 &  0.2424 &  0.7503 &  0.2408 &  0.4324 &  0.8666 \\ 
314 & 128 & 2.453 &  0.408 &  0.2275 &  0.7651 &  0.2265 &  0.4183 &  0.8786 \\ 
322 & 124 & 2.597 &  0.385 &  0.2115 &  0.7823 &  0.2108 &  0.4033 &  0.8903 \\ 
330 & 121 & 2.727 &  0.367 &  0.1971 &  0.7963 &  0.1966 &  0.3893 &  0.8987 \\ 
338 & 118 & 2.864 &  0.349 &  0.1836 &  0.8101 &  0.1833 &  0.3748 &  0.9094 \\ 
347 & 115 & 3.017 &  0.331 &  0.1697 &  0.8242 &  0.1695 &  0.3597 &  0.9173 \\ 
355 & 113 & 3.142 &  0.318 &  0.1581 &  0.8340 &  0.1579 &  0.3475 &  0.9237 \\ 
364 & 110 & 3.309 &  0.302 &  0.1447 &  0.8486 &  0.1446 &  0.3330 &  0.9338 \\
374 & 107 & 3.495 &  0.286 &  0.1318 &  0.8626 &  0.1318 &  0.3178 &  0.9409 \\ 
383 & 104 & 3.683 &  0.272 &  0.1192 &  0.8749 &  0.1192 &  0.3013 &  0.9490 \\ 
393 & 102 & 3.853 &  0.260 &  0.1089 &  0.8858 &  0.1089 &  0.2880 &  0.9553 \\ 
403 &  99 & 4.071 &  0.246 &  0.09758 &  0.8976 &  0.09757 &  0.2723 &  0.9606 \\ 
413 &  97 & 4.258 &  0.235 &  0.08836 &  0.9069 &  0.08836 &  0.2589 &  0.9656 \\ 
423 &  94 & 4.500 &  0.222 &  0.07719 &  0.9178 &  0.07719 &  0.2428 &  0.9714 \\ 
434 &  92 & 4.717 &  0.212 &  0.06971 &  0.9265 &  0.06971 &  0.2313 &  0.9755 \\ 
445 &  90 & 4.944 &  0.202 &  0.06150 &  0.9343 &  0.06150 &  0.2160 &  0.9792 \\ 
456 &  88 & 5.182 &  0.193 &  0.05432 &  0.9425 &  0.05432 &  0.2035 &  0.9825 \\ 
468 &  85 & 5.506 &  0.182 &  0.04596 &  0.9509 &  0.04596 &  0.1874 &  0.9854 \\ 
480 &  83 & 5.783 &  0.173 &  0.03933 &  0.9573 &  0.03933 &  0.1735 &  0.9881 \\ 
492 &  81 & 6.074 &  0.165 &  0.03407 &  0.9631 &  0.03407 &  0.1612 &  0.9902 \\ 
504 &  79 & 6.380 &  0.157 &  0.02899 &  0.9687 &  0.02899 &  0.1488 &  0.9922 \\ 
517 &  77 & 6.714 &  0.149 &  0.02450 &  0.9738 &  0.02450 &  0.1355 &  0.9938 \\ 
530 &  75 & 7.067 &  0.142 &  0.02015 &  0.9778 &  0.02015 &  0.1246 &  0.9951 \\ 
544 &  74 & 7.351 &  0.136 &  0.01738 &  0.9813 &  0.01738 &  0.1153 &  0.9963 \\
\hline
\end{tabular}
}}\hfill }
$$

\bigskip
\centerline{Table VII.}

\newpage

\null\vskip.75 true in

\vbox{\hsize=8.0 true in
{\null\hskip-2.0 true cm\tenpoint\tolerance=1500{
\begin{tabular}{|c|c||c|c||c|c||c|c||c|c||c|c|}
\hline
$r$ & $r^{-1}$ & $\pi_h(r) $ & $\pi_h(r^{-1}) $ & $\pi_v(r)$ & $\pi_v(r^{-1})$ & $\pi_{hv}(r)$ & $\pi_{hv}(r^{-1})$ & $\pi_h^A(r)$ & $\pi_h^A(r^{-1})$ & $\pi_v^A(r)$
& $\pi_v^A(r^{-1})$ \\
\hline
 1.000 & 1.000  & 0.4997 & 0.4997 & 0.4996 & 0.4996 & 0.4056 & 0.4056 & 0.6586 & 0.6586 & 0.6586 & 0.6586  \\ 
 1.050 & 0.9516 & 0.4868 & 0.5142 & 0.5146 & 0.4852 & 0.4061 & 0.4048 & 0.6456 & 0.6706 & 0.6717 & 0.6445  \\ 
 1.105 & 0.9049 & 0.4707 & 0.5305 & 0.5282 & 0.4710 & 0.4027 & 0.4037 & 0.6316 & 0.6849 & 0.6840 & 0.6330  \\ 
 1.160 & 0.8614 & 0.4574 & 0.5426 & 0.5434 & 0.4544 & 0.3998 & 0.3974 & 0.6190 & 0.6968 & 0.6980 & 0.6191  \\ 
 1.222 & 0.8190 & 0.4411 & 0.5579 & 0.5587 & 0.4406 & 0.3935 & 0.3931 & 0.6060 & 0.7087 & 0.7109 & 0.6059  \\ 
 1.290 & 0.7752 & 0.4258 & 0.5747 & 0.5747 & 0.4246 & 0.3874 & 0.3868 & 0.5919 & 0.7255 & 0.7250 & 0.5915  \\ 
 1.349 & 0.7412 & 0.4120 & 0.5883 & 0.5871 & 0.4119 & 0.3803 & 0.3799 & 0.5800 & 0.7365 & 0.7354 & 0.5814  \\ 
 1.416 & 0.7056 & 0.3979 & 0.6033 & 0.6020 & 0.3969 & 0.3720 & 0.3718 & 0.5678 & 0.7498 & 0.7489 & 0.5671  \\ 
 1.487 & 0.6720 & 0.3827 & 0.6174 & 0.6164 & 0.3818 & 0.3622 & 0.3616 & 0.5545 & 0.7612 & 0.7619 & 0.5547  \\ 
 1.563 & 0.6401 & 0.3679 & 0.6323 & 0.6320 & 0.3685 & 0.3514 & 0.3519 & 0.5431 & 0.7758 & 0.7750 & 0.5427  \\ 
 1.647 & 0.6071 & 0.3516 & 0.6482 & 0.6468 & 0.3511 & 0.3390 & 0.3387 & 0.5295 & 0.7875 & 0.7876 & 0.5279  \\ 
 1.728 & 0.5774 & 0.3375 & 0.6642 & 0.6624 & 0.3365 & 0.3277 & 0.3272 & 0.5152 & 0.8009 & 0.8003 & 0.5165  \\ 
 1.824 & 0.5483 & 0.3202 & 0.6788 & 0.6782 & 0.3219 & 0.3128 & 0.3144 & 0.5009 & 0.8137 & 0.8133 & 0.5034  \\ 
 1.909 & 0.5235 & 0.3060 & 0.6946 & 0.6923 & 0.3055 & 0.3006 & 0.3002 & 0.4876 & 0.8269 & 0.8250 & 0.4908  \\ 
 2.013 & 0.4964 & 0.2908 & 0.7103 & 0.7088 & 0.2906 & 0.2868 & 0.2867 & 0.4758 & 0.8370 & 0.8372 & 0.4760  \\ 
 2.124 & 0.4708 & 0.2745 & 0.7267 & 0.7264 & 0.2739 & 0.2716 & 0.2712 & 0.4603 & 0.8499 & 0.8500 & 0.4606  \\ 
 2.224 & 0.4497 & 0.2599 & 0.7412 & 0.7403 & 0.2612 & 0.2578 & 0.2592 & 0.4488 & 0.8616 & 0.8608 & 0.4505  \\ 
 2.337 & 0.4280 & 0.2446 & 0.7555 & 0.7553 & 0.2448 & 0.2432 & 0.2434 & 0.4348 & 0.8719 & 0.8716 & 0.4356  \\ 
 2.454 & 0.4077 & 0.2310 & 0.7693 & 0.7702 & 0.2301 & 0.2300 & 0.2291 & 0.4229 & 0.8813 & 0.8821 & 0.4218  \\ 
 2.596 & 0.3839 & 0.2141 & 0.7864 & 0.7855 & 0.2108 & 0.2134 & 0.2102 & 0.4067 & 0.8938 & 0.8935 & 0.4033  \\ 
 2.727 & 0.3666 & 0.2004 & 0.8003 & 0.8001 & 0.1986 & 0.1999 & 0.1982 & 0.3940 & 0.9020 & 0.9024 & 0.3916  \\ 
 2.863 & 0.3492 & 0.1864 & 0.8144 & 0.8135 & 0.1854 & 0.1861 & 0.1852 & 0.3770 & 0.9123 & 0.9118 & 0.3781  \\ 
 3.018 & 0.3315 & 0.1712 & 0.8288 & 0.8287 & 0.1712 & 0.1710 & 0.1711 & 0.3637 & 0.9204 & 0.9222 & 0.3626  \\ 
 3.142 & 0.3182 & 0.1613 & 0.8395 & 0.8383 & 0.1611 & 0.1612 & 0.1610 & 0.3530 & 0.9285 & 0.9274 & 0.3523  \\ 
 3.308 & 0.3023 & 0.1477 & 0.8522 & 0.8517 & 0.1476 & 0.1477 & 0.1475 & 0.3370 & 0.9353 & 0.9361 & 0.3368  \\ 
 3.496 & 0.2861 & 0.1337 & 0.8667 & 0.8666 & 0.1331 & 0.1336 & 0.1330 & 0.3209 & 0.9441 & 0.9446 & 0.3204  \\ 
 3.683 & 0.2715 & 0.1216 & 0.8790 & 0.8781 & 0.1212 & 0.1215 & 0.1212 & 0.3060 & 0.9507 & 0.9514 & 0.3050  \\
 3.853 & 0.2596 & 0.1120 & 0.8899 & 0.8888 & 0.1099 & 0.1119 & 0.1099 & 0.2930 & 0.9573 & 0.9569 & 0.2911  \\ 
 4.071 & 0.2456 & 0.09977 & 0.9012 & 0.9010 & 0.09799 & 0.09976 & 0.09798 & 0.2774 & 0.963 & 0.9634 & 0.2743  \\ 
 4.258 & 0.2349 & 0.08975 & 0.9093 & 0.9100 & 0.08995 & 0.08975 & 0.08994 & 0.2625 & 0.9675 & 0.9678 & 0.2611  \\  
 4.500 & 0.2222 & 0.07988 & 0.9218 & 0.9210 & 0.07853 & 0.07988 & 0.07853 & 0.2468 & 0.9733 & 0.9730 & 0.2458  \\
 4.717 & 0.2119 & 0.07010 & 0.9297 & 0.9287 & 0.07042 & 0.07010 & 0.07042 & 0.2328 & 0.9765 & 0.9766 & 0.2334  \\ 
 4.944 & 0.2023 & 0.06240 & 0.9378 & 0.9371 & 0.06288 & 0.06240 & 0.06288 & 0.2196 & 0.9802 & 0.9806 & 0.2196  \\ 
 5.181 & 0.1931 & 0.05558 & 0.9453 & 0.9444 & 0.05514 & 0.05558 & 0.05514 & 0.2058 & 0.9833 & 0.9834 & 0.2066  \\ 
 5.506 & 0.1816 & 0.04739 & 0.9533 & 0.9533 & 0.04602 & 0.04739 & 0.04602 & 0.1908 & 0.9865 & 0.9871 & 0.1874  \\ 
 5.784 & 0.1730 & 0.04084 & 0.9598 & 0.9597 & 0.04010 & 0.04084 & 0.04010 & 0.1762 & 0.9893 & 0.9894 & 0.1760  \\
 6.074 & 0.1647 & 0.03488 & 0.9655 & 0.9648 & 0.03486 & 0.03488 & 0.03486 & 0.1649 & 0.9913 & 0.9912 & 0.1638  \\ 
 6.381 & 0.1567 & 0.02971 & 0.9705 & 0.9704 & 0.02902 & 0.02971 & 0.02902 & 0.1511 & 0.9930 & 0.9932 & 0.1497  \\ 
 6.715 & 0.1489 & 0.02517 & 0.9756 & 0.9746 & 0.02494 & 0.02517 & 0.02494 & 0.1384 & 0.9945 & 0.9945 & 0.1378  \\ 
 7.067 & 0.1414 & 0.02097 & 0.9795 & 0.9793 & 0.02057 & 0.02097 & 0.02057 & 0.1264 & 0.9957 & 0.9957 & 0.1266  \\ 
 7.352 & 0.1360 & 0.01793 & 0.9824 & 0.9818 & 0.01750 & 0.01793 & 0.01750 & 0.1172 & 0.9965 & 0.9966 & 0.1162 \\
\hline
\end{tabular}
}}\hfill }

\bigskip\bigskip
\centerline{Table VIII.}

\newpage

\def\vrulesub#1#2{\hbox{\,\vrule height13pt depth9pt\,}_{#1}^{#2}}
\section{Comparison with free fields.}\label{hh2}

\def\visible#1{#1}
\def\invisible#1{}

If we take $\eta=\tilde\phi R$ the interaction for
the free field on a square lattice is
$$
        \frac{g}{8\pi}\sum\left(\eta(p)-\eta(q)\right)^2,\qquad g=2R^2,
$$
the sum running over all pairs of nearest neighbors. In the continuum limit this
becomes formally
\begin{equation}
      \frac{g}{4\pi}\int\left\{
\left(\frac{\partial \eta}{\partial x}\right)^2+\left(\frac{\partial \eta}{\partial y}\right)^2\right\}dxdy.\label{eq:4.0}
\end{equation}
We observe that there is an inconsistency in \cite{L} between the discrete 
and continuous hamiltonians. For consistency the denominator in (4.3) of
that paper has to be replaced by $4\pi$. We have used the formulas based
on the continuous hamiltonian.
 
 There are at least two properties of free fields that appear again in other models.
Either might be chosen as a basis of comparison and a means of studying these
models. The property commonly chosen is the
asymptotic behavior of correlation functions. In particular, in the plane,
\begin{equation}
     \langle(\eta(p)-\eta(0))^2\rangle\sim \frac2g \ln|p|, \label{eq:4.1}
\end{equation}
where $|p|$ is the distance between $x$ and $0$; and on a cylinder of
circumference $2\pi$
\begin{equation}
     \langle(\eta(p)-\eta(0))^2\rangle \sim a+\frac1g |p|,\label{eq:4.2}
\end{equation}
if $p$ and the origin $0$ lie on the same generator and $|p|$ is the distance between
$p$ and the origin with respect to the metric that yields a circumference of $2\pi$.
We shall briefly recall below the pertinent calculations. In the formula $a$ is
a constant that depends on the mesh. It could approach infinity as the mesh
approaches zero.

 Another property is described in \cite{L}. Consider the partition function 
$Z(\phi)$ with
boundary conditions, either on a disk so that $\phi$ is a function on the circle,
defined however only modulo constants, thus for 
simplicity with constant term $0$, or on a cylinder, taken to be
infinitely long, so that $\phi$ is really a pair of functions
$\phi_1$, $\phi_2$, and a constant $x$, taken modulo $2\pi$. In the notation of Paragraph \ref{boson}
$$
\phi_1=\sum_{k\neq0}a_k^Be^{ik\theta},\qquad \phi_2=\sum_{k\neq0}b_k^Be^{ik\theta}.
$$
For the disk,
\begin{equation}
  Z(\phi)=\exp(-g\sum_{k>0}|C_k|^2/k)\label{eq:4.3}
\end{equation}
and for the cylinder,
\begin{equation}
  Z(\phi_1,\phi_2,x)=Z(\phi_1)Z(\phi_2).\label{eq:4.4}
\end{equation}
Thus, as far as the variable $x$ is concerned, the measure is homogeneous, a behavior
that constrasts with that of the Ising model discussed in the previous section.

 For the Ising model on a triangular lattice 
the SOS-model constructed in Section \ref{dis1}
is almost the same as the SOS-model attached,
as in \cite{N} for example, to the $O(1)$-model on a hexagonal 
lattice,\footnote{In \cite{N} the partition function for the
$O(1)$-model is expressed as a sum over weighted closed curves in
the hexagonal lattice which is dual to the triangular
lattice. Every state of the Ising model leads also to 
a collection of closed curves, formed from the dual 
edges separating sites of different spin. The weight 
of the collection as a whole can be taken
as the mass of the set of Ising states that lead to it.
Our prescription leads, however, 
for the individual curves in the collection
to different Boltzmann weights
than the usual complex weights determined
locally as in \cite{N}. For the reasons explained in
the following section this does not affect the relation (\ref{eq:4.4a}).} and for this
model there are familiar arguments that suggest the behavior
(\ref{eq:4.1}) with  
$g=g_I=4/3$. We have not tested carefully the universality of the behavior
or of the constants.
Crude
experiments 
for the square and the triangular lattice suggest that the behavior
is universal but we are not certain that the constants do not vary slightly.

\begin{figure}
\begin{center}\leavevmode
\includegraphics[bb = 72 330 540 500,clip,width = 15cm]{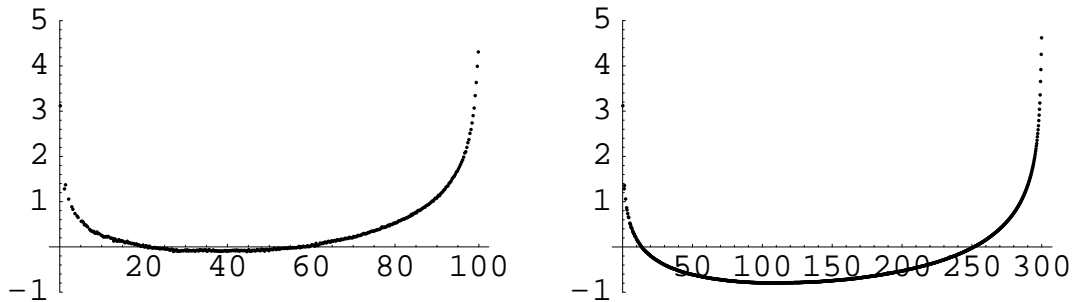}
\end{center}
\caption{The quantity $\langle(h(p)-h(0))^2\rangle-1.5\ln|p|$
measured on disks of radii $100$ and $300$
covered by a square lattice.\label{fig:6.1}}
\end{figure}

\begin{figure}
\begin{center}\leavevmode
\includegraphics[bb = 72 240 540 554,clip,width = 10cm]{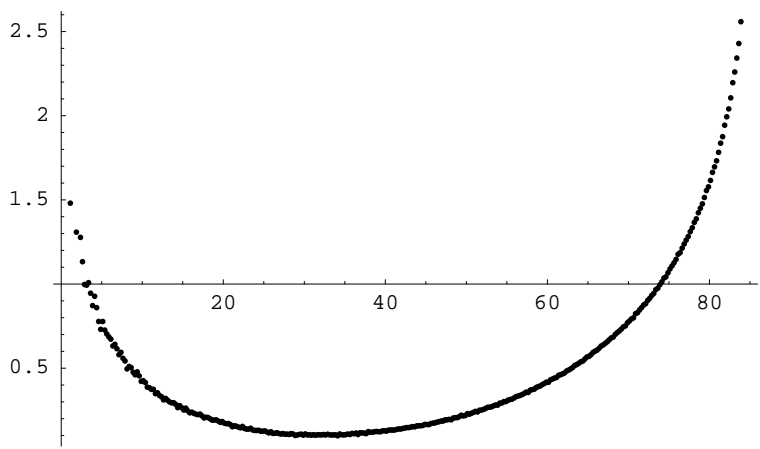}
\end{center}
\caption{The quantity $\langle(h(p)-h(0))^2\rangle-1.5\ln|p|$
measured on a disk of radius $90$ covered by a triangular lattice.\label{fig:6.C}}
\end{figure}

The function $\eta$ of the free-field theory
plays the same role as the function $h$ of our construction
so that to test (\ref{eq:4.1}) we
examine $\langle(h(p)-h(0))^2\rangle$.
For what they are worth,
the results for the plane appear in Figures \ref{fig:6.1} 
and \ref{fig:6.C} in which the value of 
\begin{equation}
    \langle(h(p)-h(0))^2\rangle-1.5\ln|p|\label{eq:4.4a}
\end{equation}
is plotted against $|p|$.  
For the square lattice in Figures \ref{fig:6.1}
the experiments are performed in disks
of radii $100$ and $300$, an edge of the lattice being taken in each case as unit.
The experiments are perhaps not
to be taken too seriously because the finite size leads to an
ambiguity. Not only are the states in a disk qualitatively different at the boundary
from those in the true bulk limit but also the jump lines that in a disk
terminate at the boundary could, in some sense, in the bulk turn
and pass once again through the disk, so that working in the disk increases
the statistical independence. 
The graphs, in which vertical distances
are drawn at a much larger scale, suggest that the function is approximately constant except
close to the origin and near the edge of the disk, where the effect
of the boundary manifests itself. The constant to which one might imagine
the difference (\ref{eq:4.4a}) tending has not yet stabilized 
in the diagrams. There is a difference of about $.4$ in the minimum of the
two curves. 
For comparison, a similar curve for the triangular lattice,
obtained once again in a small disk of radius $90$, an edge
of the lattice again being taken as unit, is shown in Figure \ref{fig:6.C}.
  
\begin{figure}
\begin{center}\leavevmode
\includegraphics[bb = 72 240 540 554,clip,width = 10cm]{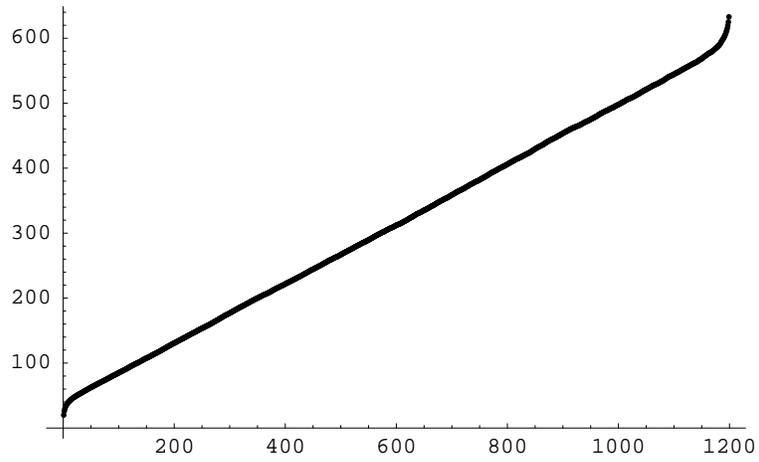}
\end{center}
\caption{The correlation function $\langle(h(p)-h(0))^2\rangle$
on a cylinder for the square lattice $120\times 2401$.\label{fig:6.D}}
\end{figure}

\begin{figure}
\begin{center}\leavevmode
\includegraphics[bb = 72 240 540 554,clip,width = 10cm]{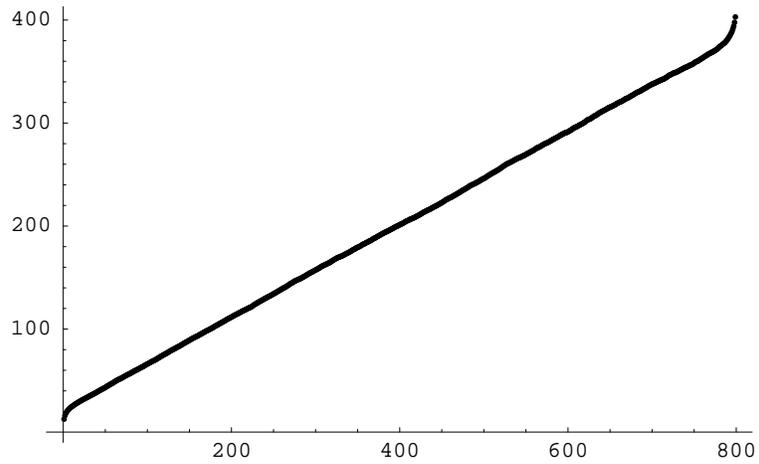}
\end{center}
\caption{The correlation function $\langle(h(p)-h(0))^2\rangle$
on a cylinder for the triangular lattice $160\times 1601$.\label{fig:6.E}}
\end{figure}

 For the cylinder and the same two lattices, square and
triangular, the graphs of
$\langle(h(p)-h(0))^2\rangle$ appear in Figures \ref{fig:6.D} and \ref{fig:6.E}.
We have not used the parameter $|p|$ in the
figures but rather the parameter $k$ because it is
then easier to explain which part of the curve we used
to calculate the slope (for the square lattice
$|p|=4\pi k/LV$, for the triangular $|p|=8\pi k/\sqrt{3}LV$).
None the less the data have been so normalized that 
if the behavior is, as in equation (\ref{eq:4.2}),
asymptotically $a+b|p|$ then the slope of the curves
in the figures on their middle, linear
parts and as functions of $k$
is also $b$. 
For the square lattice the cylinder is 
of circumference $120$ and length $2401$ in lattice units;
for the triangular of circumference $160$ and length $1601$.
The slope for the square lattice is about $.460$; for the triangular
it is about $.452$, which is not a number that we can deduce easily from $4/3$. 
These numbers are close; so universality of the slope is
strongly suggested. 

To obtain Figures \ref{fig:6.D} and \ref{fig:6.E} we
construct $h$ as in Sections \ref{dis1} and \ref{dis2} and
use the difference between the values of $h$ at points 
on generators of the cylinder symmetrically 
placed with respect to the central meridian
and at a distance of $k$ sites from it. Thus $k$ is necessarily less than
one-half the length of the cylinder. Since we use all generators
there is considerable statistical dependence. None the less this yields for a
cylinder of size $120\times2401$ a very regular graph and if we use
that part of it between $100$ and $1100$ we obtain a fit
$37.6275+.4597k$  from which the statistically generated values differ by no more
than two units at any point on this interval, so that the slope should be correct to
about two parts in a thousand. 
(The curve is in fact slightly concave and
the departure from linearity regular. With a quadratic fit and
a slightly shorter interval we would do much better with the fit but not with
the slope.)
The experiment repeated on the interval
$[200,1000]$ leads to a slope of $.4599$ but the same conclusions.
A similar experiment for a cylinder of size $120\times 1201$ yields
to a slightly better fit and similar conclusions with a slope of $.4593$.
An anisotropic lattice of size $78\times2401$ is roughly conformally equivalent
to a square lattice of size $120\times2401$. Using the points on
the interval $[200,1200]$ we obtain a fit of $39.8942+.4519k$ from
which the statistically generated values differ by no more than $1.5$ units.
The difference is again not random but not convex. We can again conclude that
the slope is correct to about two parts in a thousand. The difference
between the slopes in the symmetric and the 
anisotropic cases is $.0078$. In other words, it appears that we 
obtain the same constant.
A triangular lattice
of length $1601$ and circumference $160$ is conformally equivalent to
a square lattice of circumference $120$ and length a little shorter
than $1401$. Thus it is long enough. 
On using that part of Figure \ref{fig:6.E} 
in the interval $[100,700]$ we obtain a fit $20.6918 +.4515k$
that is as good as those for the square and anisotropic lattice
and suggests, for the same reasons, an error of two parts in a thousand. 

 We observe finally that the experiments described in Section \ref{dis1}, 
in which the analogue of
$Z(\phi)$ is studied, yield the behavior (\ref{eq:4.3}) and (\ref{eq:4.4}) 
with $g_B=2R_B^2=1.4710\neq g_I$. 

 For the convenience of the reader, we recall briefly the calculations that lead to
(\ref{eq:4.1}) and (\ref{eq:4.2}). 
The average $\langle(\eta(p)-\eta(0))^2\rangle$ is taken with
respect to the measure defined by the weights
\begin{equation}    \exp\left(-\frac{g}{8\pi}\sum_{p,\delta}
\left(\eta(p+\delta)-\eta(p)\right)^2\right)
   =\exp(-(Q\eta,\eta)),\qquad 
   \delta\in\{(\pm1,0),(0,\pm1)\}.\label{diridis}
\end{equation}
The operator $Q$ is obtained from the relation
$$
\sum_{p,\delta}\left(\eta(p+\delta)-\eta(p)\right)^2=
\sum_p\left(8\eta(p)-2\sum_{\delta}\eta(p+\delta)\right)\eta(p).
$$
We are calculating the second derivative of
$$
-\int\left(\exp(i\alpha\lambda(\eta)\right)\exp\left(-(Q\eta,\eta)\right)d\eta\big/
 \int\exp\left(-(Q\eta,\eta)\right)d\eta=
 -\exp\left(-\alpha^2(Q^{-1}\lambda,\lambda)/4\right)
$$
with respect to the parameter $\alpha$, where $\lambda$ is the linear form
$\eta\rightarrow \eta(p)-\eta(0)$ or the function $\delta_p-\delta_0$.
The second derivative
is
$$
  (Q^{-1}\lambda,\lambda)/2.
$$
 
 This expression is easier to treat when we pass to Fourier transforms. 
The two delta functions of $\lambda$ can be written as integrals of eigenfunctions
of $Q$. Since the operator $Q$ acts on $\eta$ so that $(Q\eta)(p)$
is $g/8\pi$ times
$$
8\eta(p)-2\eta(p+(1,0))-2\eta(p-(1,0))
                     -2\eta(p+(0,1))-2\eta(p-(0,1)),
$$
its eigenfunction $e^{2\pi i(p_1x+p_2y)}$
corresponds to the eigenvalue $g(\sin^2\pi x+\sin^2\pi y)/\pi$.
Therefore $\frac12(Q^{-1}\lambda,\lambda)$ becomes at $p=(p_1,p_2)$
$$
  \frac{\pi}{2g} \int_{-1/2}^{1/2}\int_{-1/2}^{1/2}\frac
  {|e^{2\pi i(p_1x+p_2y)}-1|^2}{\sin^2\pi x+\sin^2 \pi y}dxdy,
$$
or
\begin{equation}
\frac{2\pi}{g}\int_{-1/2}^{1/2}\int_{-1/2}^{1/2}\frac
{\sin^2(\pi(p_1x+p_2y))}{\sin^2\pi x+\sin^2 \pi y}dxdy.\label{eq:4.5}
\end{equation}
The integral outside a circle of small positive radius $\epsilon$ about
$0$ remains bounded as $|p|\rightarrow\infty$ and inside
this circle the denominator can be replaced by $\pi^2(x^2+y^2)$. The
result is
\begin{equation}
  \frac{2}{g\pi}\int_0^\epsilon
\frac{dr}{r}\int_0^{2\pi}\sin^2(ru\cos\theta)d\theta,\qquad u=|p|.\label{eq:4.6}
\end{equation}
The integral of (\ref{eq:4.6}) is the sum of
$$
\int_0^{1/u}\frac{dr}{r}\int_0^{2\pi}\sin^2(ru\cos\theta)d\theta=O(u^2\int_0^{1/u}rdr)=O(1)
$$
and
\begin{equation}
\int_{1/u}^\epsilon\frac{dr}{r}\int_0^{2\pi}\sin^2(ru\cos\theta)\ d\theta.\label{eq:4.7}
\end{equation}
Since $\sin^2\varphi=\frac12-\frac12\cos2\varphi$,
$$
  \int_0^{2\pi}\cos(z\cos\theta)d\theta=2\pi J_0(z),
$$
and $J_0(z)=O(|z|^{-1/2})$, (\ref{eq:4.6}) can be replaced by
$$
 \pi\int_{1/u}^\epsilon\frac{dr}{r}\approx\pi\ln u.
$$
Multiplying by $2/g\pi$ we obtain (\ref{eq:4.1}).

 For a cylinder we treat a lattice that is periodic in the vertical direction
(the $p_1$ direction)
with period $A$, which for 
simplicity we take to be even. If $p=(0,An)$, $n>0$, the analogue of (\ref{eq:4.5}) is
$$
\frac{2\pi}{g}\sum_{x=-A/2}^{A/2-1}\frac1A\int_{\-1/2}^{1/2}\frac{\sin^2(\pi Any)}
{\sin^2(\pi x/A)+\sin^2(\pi y)}\ dy.
$$
Once again we drop terms that remain bounded as $A$ approaches infinity. This yields
\begin{equation}
\frac2{g\pi}\left(\frac1A\int_{-1/2}^{1/2}\frac{\sin^2(\pi Any)}{y^2}dy+
\frac2A\sum_{x=1}^{A/2-1}\int_{-1/2}^{1/2}\frac{\sin^2(\pi Any)}
{x^2/A^2+y^2}dy\right).\label{eq:4.8}
\end{equation}

 We examine the second term of (\ref{eq:4.8}) using the identity
$$
  \sin^2(\pi Any)=\frac12-\frac12\cos(2\pi Any).
$$
The expression obtained from the term $1/2$ on the
right is independent of $n$ and on close examination is seen to behave 
like $\ln A$, but that is not pertinent here. Since
$$
\frac{1}{A}\sum_{x=1}^{A/2-1}\int_{\-1/2}^{1/2} \frac{\cos(2\pi Any)}{x^2/A^2+y^2}dy=
\sum_x\int_{\-A/2}^{A/2}\frac{\cos(2\pi ny)}{x^2+y^2}dy,
$$
which upon integration by parts becomes
$$
\frac{1}{2\pi n}\sum_x\frac{\sin(2\pi ny)}{x^2+y^2}\vrulesub{y=-A/2}{A/2}
  +\frac{1}{2\pi n}\sum_x\int_{\-A/2}^{A/2}\frac{2y\sin(2\pi ny)}{(x^2+y^2)^2}dy,
$$ 
the second term behaves --
independently of $A$ -- as $O(1/n)$. This leaves the first term of 
(\ref{eq:4.8})
which is $n$ times
$$
\frac{2}{g\pi}\left(\frac{1}{An}\int_{-1/2}^{1/2}\frac{\sin^2(\pi Any)}{y^2}dy\right).
$$
For large $An$ this expression is approximately $2\pi/g$. If, however, we measure
the distance between $p$ and $0$ not in terms of the circumference but in terms
of the radius of the cylinder, the constant $2\pi/g$ is replaced by $1/g$
as in (\ref{eq:4.2}). 

 Although we have inferred the relation (\ref{eq:4.4a})
from the corresponding relation for the SOS-model associated to
the $O(1)$-model by the construction of \cite{N}, our construction
of the measure on the set of functions $h$ is much more naive
and involves no complex weights. 
As a consequence the measure is no longer
gaussian. The relation (\ref{eq:4.1}), 
with $g=4/3$, applied to $h$ suggests that, if it were, the appropriate
gaussian would be
$$
  \frac{g}{4\pi}\{\left(\frac{df}{dx}\right)^2+\left(\frac{df}{dy}\right)^2\},
$$
thus that of (\ref{eq:4.0}). The usual formulas
for the expectation of the exponential $e^{i\lambda(h)}$
of the linear function $\lambda(h)=h(p)-h(0)$
then suggest, after renormalization, that the correlation function of the spins,
thus the expectation of $e^{ih(p)-ih(0)}$, is
$$
e^{-3/4\ln(p)}\sim1/p^{3/4}.
$$
The exponent is of course not correct. The explanation is presumably
similar to that of Section \ref{dis1}. It may be possible,
although we have made no attempt to do so,
to use the functions $h$ to construct in the limit a measure
on distributions in the plane and this measure may very well share some
basic properties with the usual gaussian measure, but it will not be gaussian.
 
 It should perhaps be observed that the random variable $\phi(p)-\phi(0)$
is not well defined on distributions, so that the expectation of
$e^{i\phi(p)-i\phi(0)}$ makes no sense. Strictly speaking, one should take a
smooth function $\lambda^\sigma=\lambda^\sigma_{p,0}$ approximating 
as $\sigma\rightarrow 0$ the difference
$\delta_p-\delta_0$ of two $\delta$-functions, calculate the expectation of 
$e^\sigma(p)$ of $e^{i\phi(\lambda^\sigma)}=e^{i\lambda^\sigma(\phi)}$, 
normalize by dividing by the value $e^{\sigma}(p_0)$ at a fixed $p_0$,
usually taken at a distance $1$ from the origin, and then pass to the limit
$\sigma\rightarrow 0$. This method was used in Paragraph \ref{arbitre}.
 
\section{Alternate constructions.}\label{alter}
In this section, we examine briefly other conventions and constructions 
that we could have chosen in our experiments.  

\subsection{SOS-model jumps.}
If, as indicated in the 
introduction, the aim is 
simply to develop the circle onto
the line, thereby turning the Ising model into an SOS-model, the particular
construction chosen is somewhat arbitrary. We could, apparently with equally
good reason, replace the jumps
of $\pm \pi$ by jumps from a set, 
$\{-(2k+1)\pi,-(2k-1)\pi,\dots,(2k+1)\pi\}$, $k\in \mathbb N$,
each choice being assigned
a probability on which the only conditions are that the sum of the
probabilities is one and that the probabilities of jumps by equal
amounts in opposite directions are equal. It is not, at first, clear what effect this has.

\begin{figure}
\begin{center}\leavevmode
\includegraphics[bb = 72 330 540 500,clip,width = 15cm]{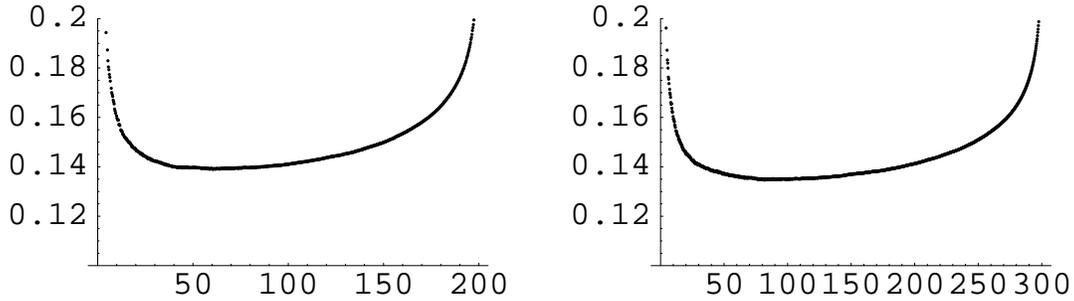}
\end{center}
\caption{The quantity $\langle N(p)\rangle/\ln|p|$ for the square
lattice on disks of radii $200$ and $300$.\label{fig:7.1}}
\end{figure}

 Thomas Spencer pointed out to
one of us that the behavior, for jumps of $\pm \pi$,
$$
\langle(h(p)-h(0))^2\rangle\sim 3/2\ln|p|
$$
is a consequence of a more geometric hypothesis.\footnote{The hypothesis
(\ref{eq:7.1}) refers only to the weights attached to curves without
regard to orientation and for them our weights
are the usual ones. When deducing (\ref{eq:4.4a}) from (\ref{eq:7.1}) the relative weights,
complex or not,
attached to the two possible orientations are irrelevant. All that
matters is that they be independent from curve to curve.}
To construct the function $h$ attached
to a particular state of the Ising model, we construct curves separating the
regions in which the spins take different values. Let, in the plane, $N=N(p)$ 
be the number of curves separating $p$ from the origin. The hypothesis
is that
\begin{equation}
           \langle N(p)\rangle\sim c_N\ln|p|.\label{eq:7.1}
\end{equation}
Since $h(p)$ is then obtained by assigning independent values to the jumps of $\pm\pi$,
it is clear that $c_N$ must be $2/g\pi^2$. For a cylinder 
the analogue of (\ref{eq:7.1}) is
\begin{equation}
    \langle N(p)\rangle\sim c|p|.\label{eq:7.2}
\end{equation}
Once again, out of curiosity, we tested
this hypothesis numerically for the square lattice. The results
are presented in Figure \ref{fig:7.1} in which $\langle N(p)\rangle/\ln|p|$
is plotted for the square lattice and two disks of radii $200$ and $300$. It appears
that except at the center and near the boundary the quotient is approximately constant
but that it is only very approximately equal to $3/2\pi^2\sim .15199$. 
There are
several possible causes -- in addition to a departure from gaussian behavior. 
As we saw in Paragraph \ref{simconf} the bulk state is approached only slowly
in a disk. Moreover
the finite-size effects that appear in the examination
of (\ref{eq:4.4a}) appear here too. The first consequence is that
there will be a tendency to overestimate the number $N(p)$ when $|p|$ is 
not small in comparison with the radius because the curves in a disk that reach
the boundary are not allowed to close. In principle, this effect should, for a given
$|p|$, be mitigated as the radius grows. On the other hand, rather than
increasing toward $.15$ as we pass from a radius of $200$ to one of $300$,
the minimum of the curve, 
decreases from about $.14$ to about $.13$. Since  the smallest
pertinent value of $|p|$ is about $75$ and $\ln(75)\sim 4.3$ 
and the 
difference in (\ref{eq:4.4a}) does not, as we saw in the previous chapter,
approach a limiting value rapidly, if it approaches one at all, a decrease in
the minimum of $.4/4.3\pi^2\sim .01$ is not completely unreasonable. No conclusions are
possible without further study. Our purpose
here is not, however, to examine (\ref{eq:7.1}) but rather to acquire a rough
understanding of what we might have discovered if we had chosen the jumps in
a different way. 

 The advantage of (\ref{eq:7.1}) and (\ref{eq:7.2}) is that they make  
clear that the behavior 
(\ref{eq:4.1}) and (\ref{eq:4.2}) does
not change when the definition of $h$ is modified. If there are jumps of
$(2k+1)\pi$ with probability $\varpi_k$, $k\in \mathbb Z$, then (\ref{eq:4.1}) persists
with a new constant
\begin{equation}
  c_N\sum_{k=-\infty}^\infty\varpi_k(2k+1)^2.\label{eq:7.2a}
\end{equation}
There is a similar change in (\ref{eq:4.2}). 

\begin{figure}
\begin{center}\leavevmode
\includegraphics[bb = 85 250 515 550,clip,width = 12cm]{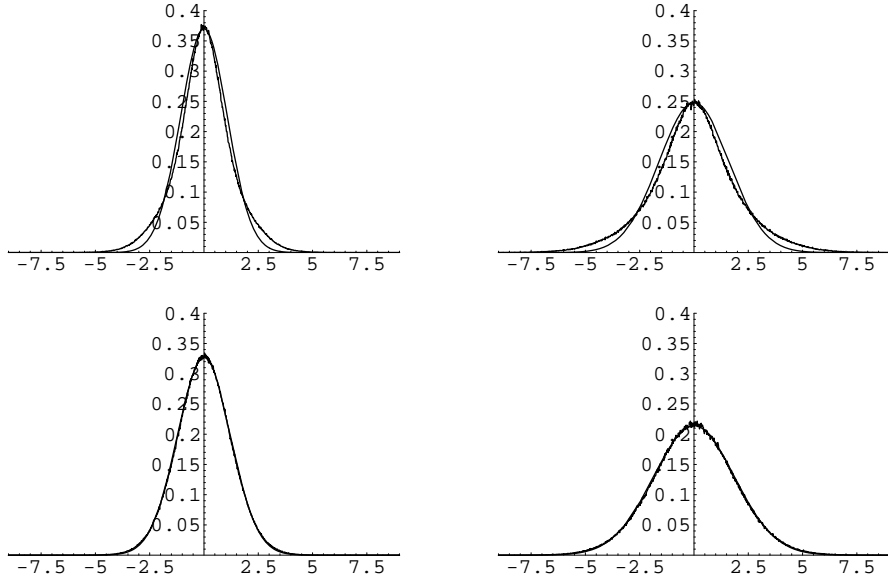}
\end{center}
\caption{The distribution as a function of $\Re A_1$ (first line)
and $\Re A_5$ (second line) with 
four jumps (first column) and six jumps (second column).\label{fig:7.3}}
\end{figure}

 The behavior of the functions $Z(\varphi)$ appears to be quite different.
We have performed a few rough experiments, replacing the jumps of $\pm\pi$ by jumps of
$-3\pi,-\pi,\pi,3\pi$, each with probability $1/4$ and by jumps of
$-5\pi,-3\pi,-\pi,\pi,3\pi,5\pi$, each with probability $1/6$. If 
the measures continue to exist, but with $g_B$ modified as
suggested by (\ref{eq:7.2a}) then the Fourier coefficients would
continue to be distributed as gaussians but with $g=g_B$ 
of (\ref{eq:4.3}) multiplied by $\frac15$ and by $\frac3{35}$ respectively,
so that the ideal value of $\sqrt{2R_B^2/\pi}\sim .68$ of the distribution at $0$
would be
multiplied by $\sqrt{1/5}$ or $\sqrt{3/35}$ yielding $\sim.31$ and $\sim.20$.
In the first row of Figure \ref{fig:7.3} (four and six jumps)
the distribution of the Fourier coefficients ${\mathfrak R}A_1$
for a cylinder of size $299\times599$ is compared in each of 
these cases with a gaussian with the same value at $0$.
There is some similarity but considerable difference. Moreover the value 
at $0$ is close to but different from the suggested value. 
For the higher coefficients the distribution looks more and
more like a gaussian. In the second row of Figure \ref{fig:7.3} 
the distributions
of ${\mathfrak R}A_5$, normalized so that the factor $\sqrt{k}$
with which we are familiar from Section \ref{dis1} are compared with gaussians.
Not only are they closer to gaussians, but the values at $0$ are closer
to those predicted by (\ref{eq:7.2a}).

\begin{figure}
\begin{center}\leavevmode
\includegraphics[bb = 85 250 515 550,clip,width = 12cm]{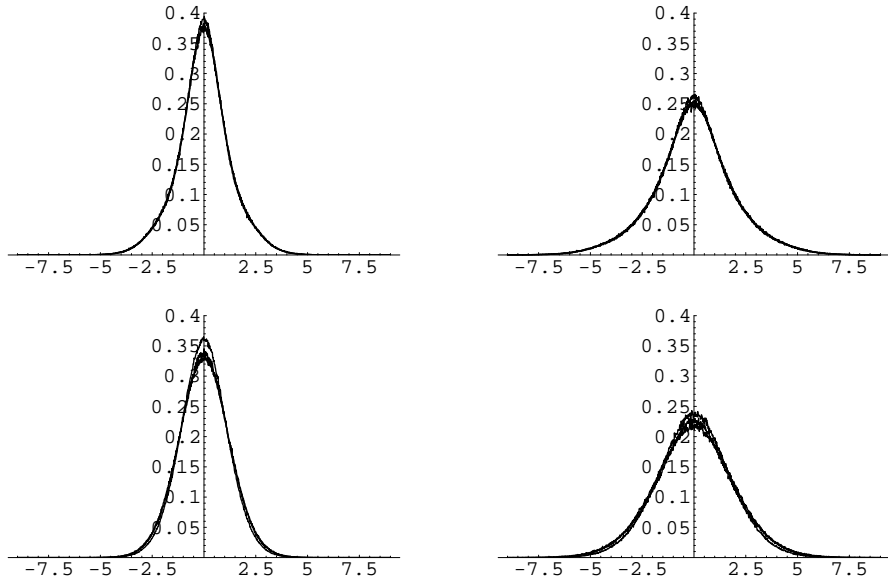}
\end{center}
\caption{The distribution as a function of $\Re A_1$ (first line)
and $\Re A_5$ (second line) with 
four jumps (first column) and six jumps (second column)
on the cylinders $99\times199$, $199\times399$, and $299\times599$.\label{fig:7.4}}
\end{figure}

 On the other hand, the first row of Figure \ref{fig:7.4}, in which 
the distributions of ${\mathfrak R}A_1$
for the three sizes $99\times199$, $199\times399$, and $299\times599$ 
are compared in each of the two cases, suggests that the limiting measures may
none the less exist. So does the second row of Figure \ref{fig:7.4} 
for ${\mathfrak R}A_5$. We have, however, as yet made no serious effort
to decide whether this is so, nor whether these measures could be conformally
invariant and universal.

\begin{figure}
\begin{center}\leavevmode
\includegraphics[bb = 72 330 540 500,clip,width = 15cm]{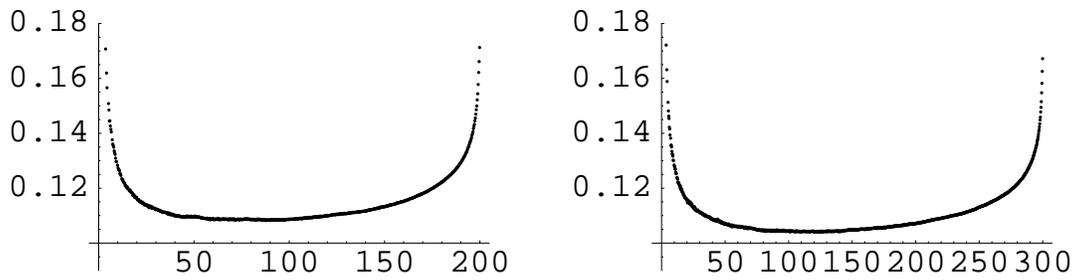}
\end{center}
\caption{The quantity $\langle(N(p)-\langle N(p) 
\rangle)^2\rangle/\ln|p|$ on disks of radii $200$ and $300$.\label{fig:7.5}}
\end{figure}

Another possible ``natural'' choice for the relative weights 
of the jumps $n\pi$, $n$ odd is given by the Dirichlet form. 
Its discretized form (\ref{diridis}) used in Section 6 suggests
that the weights $\varpi_{2k+1}$ and $\varpi_1$ of having a 
jumps $\pm (2k+1)\pi$ or $\pm \pi$ satisfy $\varpi_{2k+1}=\varpi_1^{2k+1}.$
If $\varpi_1$ is fixed by requiring that $\sum_{i\in\mathbb Z}\varpi_{2i+1}=1$,
then $\varpi_1=\sqrt{2}-1$. Then the constant $g_B$ would be multiplied
by the inverse of $\sum_{i\in\mathbb Z}\varpi_{2i+1}(2i+1)^2=3$, that is 
$\frac 13$.

 In addition to (\ref{eq:7.1}) we also examined, following a suggestion of Thomas
Spencer, the behavior of
$$
 \langle(N(p)-\langle N(p)\rangle)^2\rangle/\ln|p|,
$$
whose behavior is pertinent when attempting to establish
(\ref{eq:4.4a}) rigorously,
in disks of radii $200$ and $300$. Although the results are
not relevant to this paper, they are presented, for the
curious reader, in Figure \ref{fig:7.5}.
Once again, the curves are extremely flat, but there is a drop 
for the larger radius that has to be explained.

\subsection{The Fortuin-Kasteleyn construction.}
The Fortuin-Kasteleyn formulation of the Ising model can 
be used to map the partition function of the high-temperature phase of the model to a percolation-like sum over bond configurations.  To construct the F-K version of
an Ising model on a planar graph $\mathcal G$ with vertices $s\in S$ and 
bonds $b\in B$ we shall form the first barycentric
subdivision $\mathcal G'$ of $\mathcal G$. Thus associated to 
$\mathcal G$ are the vertices $s$, the bonds
$b$, each joining two sites, and the faces $f$, each face $f$ being bounded
by sites and vertices.
The sites $S'$ of $\mathcal G'$ are the sites in $S$
and points obtained by choosing arbitrarily from each bond $b$
and each face $f$ a point in its interior. Thus, set-theoretically,
$S'=S\cup B\cup F$. The bonds $B'$ are pairs consisting of a bond in
$B$ and one of its ends or a face in $F$ and a bond or vertex on its
boundary. In fact, the bonds in $B'$ joining a face to a vertex in its
boundary are for our purposes superfluous and are not included in our 
constructions.

 The partition function of the original model is taken in the
form
\begin{equation}
   Z=\sum_\sigma\prod_B\exp(J\delta_{\sigma(s),\sigma(t)}).\label{eq:7.3}
\end{equation}
Here $s$ and $t$ are the two sites joined by $b$. Thus,
for a square lattice, $\sinh(J)=1$,
$J=.881374$. For a given 
configuration $\sigma$, the clusters,
in the sense of this paper, are maximal connected subsets of $S$ on which $\sigma$ is
of constant sign. To obtain a Fortuin-Kasteleyn cluster we remove the bonds
of these connected clusters one by one with probability $1/\nu$,
$\nu=\exp J$ (for a square lattice $1/\nu=.414214$).
This replaces the sum (\ref{eq:7.3}) by a sum over decompositions of $\mathcal G$ into
subgraphs, each component being provided with a sign. A decomposition
is the subgraph obtained by keeping all vertices and removing some bonds.
\begin{equation}
\begin{aligned}
  Z&=\sum_\sigma\prod_B(1+(\exp(J\delta_{\sigma(s),\sigma(t)})-1))\\
  &=\sum\prod(\exp J-1)\\
  &=\sum\prod(\nu-1)\\
  &=\sum(1+(\nu-1))^r\{(\nu-1)^q/(1+(\nu-1))^r\}\\
  &=\sum\nu^r(1-\frac{1}{\nu})^q(\frac{1}{\nu})^{r-q}
\end{aligned}\label{eq:7.4}
\end{equation}
The sum in the second line runs over all decompositions into 
subgraphs, each component being signed, so that a constant spin
is assigned to each of its vertices, and so do the sums in the
remaining lines. 
From a signed decomposition we can of course
reconstruct, from the signs alone, the original state of the
Ising model. This state has $r$ bonds that join sites with the
same spin, so that its probability is $\nu^r$.
The number of bonds in the subgraph is $q$ and the factor 
$(1-\frac{1}{\nu})^q(\frac{1}{\nu})^{r-q}$ is the probability that we arrive
at it on removing bonds. If we
now ignore the spins, the final sum in (\ref{eq:7.4}) becomes
$$
  \sum(\nu-1)^q2^c,
$$
if $c$ is the number of connected graphs in the decomposition.

 To construct the function $h$ we associate to a decomposition 
a state $\sigma'$ on $S'$. The value of $\sigma'$ is $1$ at the vertices
of $\mathcal G$, at the bonds of $\mathcal G$ that belong to the subgraph, but is $-1$ at all
other vertices of $\mathcal G'$. Now $h$ can be constructed as before, except
that the jumps are to be $\pm\pi/2$ and not $\pm\pi$. It turns out to
be instructive, at least for the crossing
probabilities, to replace the probability $1/\nu$ by a variable
probability $1-\mu$ between $0$ and $1$. Thus $\mu_{FK}=.585786$.

We have considered only graphs formed by square lattices on either a 
cylinder (for distributions and correlations) or a rectangle 
(for crossing probabilities).  Our aim was not to establish 
conformal invariance and universality for the F-K construction, but 
rather to acquire a provisional understanding of
the way the various objects introduced in this paper 
behave under an alternative description of the model.

There are two ways to define crossings in the F-K construction.  If 
cluster signs are taken into account, a crossing is a cluster 
of sites with positive spins that joins one side
of the rectangle to the opposite one. 
The crossing probabilities considered earlier are recovered if $\mu=1$ but 
the crossing probabilities are zero if $\mu=0$.
 
If clusters are unsigned, crossings are defined as in bond percolation.  
This is more in the spirit of the F-K formalism and 
we shall use this definition.  Note that both conventions are 
linked:  if $\pi_+$ and $\pi_-$ are the crossing 
probabilities over a positively or negatively signed cluster, 
and $\pi_{+-}$ the probability that there are spanning clusters of both 
positive and negative sign, then the 
probability $\pi$ that an unsigned cluster crosses is given 
by the following obvious relation,
$$
\pi = \pi_+ + \pi_- - \pi_{+-}.
$$

\begin{figure}
\begin{center}\leavevmode
\includegraphics[bb = 72 240 540 554,clip,width = 10cm]{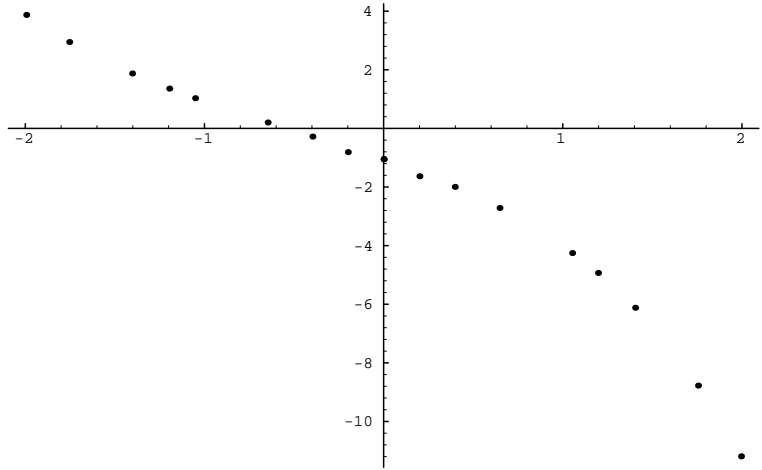}
\end{center}
\caption{$\log \pi_h(\mu_{FK})/(1-\pi_h(\mu_{FK}))$ as
a function of $\log r$.\label{fig:ma1}}
\end{figure}

\begin{figure}
\begin{center}\leavevmode
\includegraphics[bb = 72 240 540 554,clip,width = 10cm]{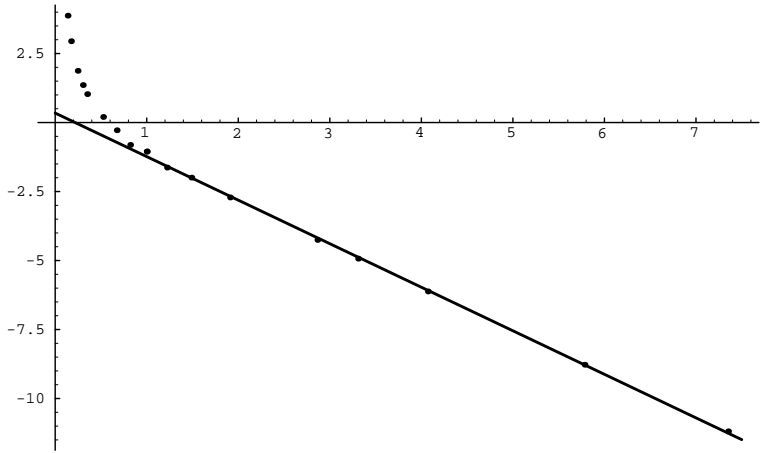}
\end{center}
\caption{Fits of the asymptotic behavior of $\log \pi_h(r,\mu_{FK})$ as 
a function of $r$. \label{fig:ma4}}
\end{figure}

Our crossing probabilities now depend on two variables: the 
aspect ratio of the rectangle $r$ and the probability $\mu$ of not 
removing a link.  We studied each of these variables separately, varying one and
keeping the other fixed.  We first took $1-\mu=1-\mu_{FK}\equiv 1/\nu$ 
and studied the dependence 
on the aspect ratio.  Results for $\pi_h(r,\mu_{FK})$, the
probability of a horizontal crossings in $\mathcal G'$ on
either $+$ or $-$ clusters, are shown in Figure \ref{fig:ma1}.
The numbers of sites in $\mathcal G$ inside the rectangles were
around 40000 and the samples 250000.
The absence of symmetry implies that duality fails,
$$
\pi_h(r,\mu_{FK})+\pi_h(1/r,\mu_{FK})\neq 1.
$$
The asymptotic behaviour of $\log \pi_h(r,\mu_{FK})$, shown on Figure
\ref{fig:ma4}, is found to 
be $$\log \pi_h(r,\mu_{FK})\quad
\underset{r\rightarrow\infty}{\longrightarrow}\quad
- 0.502 \pi r+\text{constant},$$
a number reasonably close to $\pi/2$, despite the rather low statistics.

\begin{figure}
\begin{center}\leavevmode
\includegraphics[bb = 72 240 540 554,clip,width = 10cm]{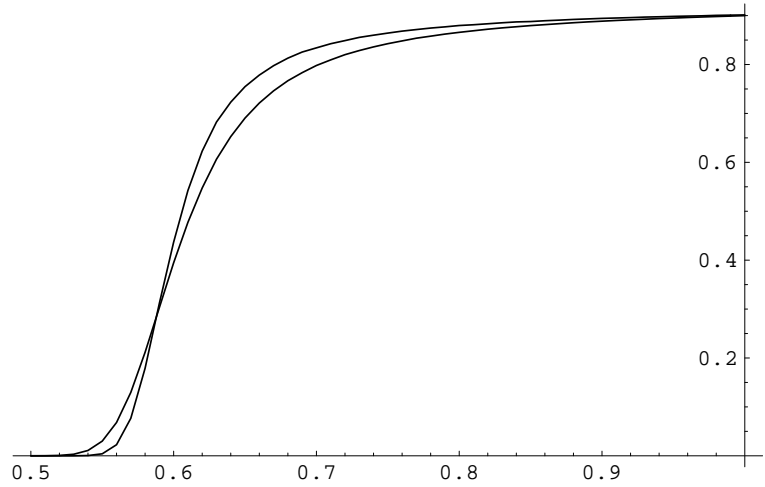}
\end{center}
\caption{$\pi_{h}(1,\mu)$ as a function of $\mu$ for $100\times 100$ and $200\times 200$ square lattices $\mathcal G$. (The curve of the
larger lattice is the top one for large $\mu$.) \label{fig:ma2}}
\end{figure}

In the second experiment we measured the dependence 
on $\mu$ of $\pi_h(1,\mu)$.  
The results presented in Figure \ref{fig:ma2} raise the question whether
$\mu=\mu_{FK}$ might be a critical value of the function $\pi_h(r,\mu_{FK})$
or at least of $\pi_h(1,\mu_{FK})$, that is $\pi_h(1,\mu)$ would be
zero for $0<\mu<\mu_{FK}$ and nonzero for $\mu>\mu_{FK}$.
{\invisible{The results presented in figure \ref{fig:ma2} 
suggest that $\mu=\mu_{FK}$ might be a critical value for $\pi_h(r,\mu_{FK})$, 
meaning that 
$\pi_{h}(1,\mu)=0$ for $0<\mu<\mu_{FK}$ and that it becomes nonzero 
for $\mu>\mu_{FK}$.}}  It is not obvious from the numbers obtained 
what the limit of the function $\pi_h(1,\mu_{FK}), \mu>\mu_{FK},$
is when the mesh goes to zero.

\begin{figure}
\begin{center}\leavevmode
\includegraphics[bb = 72 330 540 500,clip,width = 15cm]{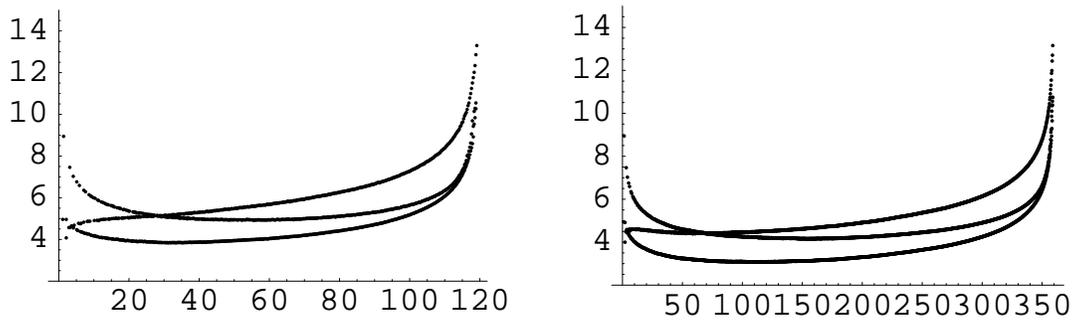}
\end{center}
\caption{The correlation $\langle(h(p)-h(0))\rangle_{\mu_{FK}}^2-\frac23\ln |p|$
on disks of radii $60$ and $180$.\label{fig:7.6}} 
\end{figure}

We examined the correlation functions
$$
\langle (h(p)-h(q))^2\rangle_\mu,
$$
both on a cylinder and on the plane.
According to \cite{N} one should expect (provided that an
analogue of (\ref{eq:7.1}) is valid) that for correlations in the plane
$$
\langle(h(p)-h(0))^2\rangle_{\mu_{FK}}\sim \frac{2}{3}\ln|p|.
$$
This is confirmed by the graphs of
Figure \ref{fig:7.6}. In the first the radius 
of the disk considered is relatively 
small, about 60 bond units; for the second it is 180 bond units.
The presence of three 
distinct curves, corresponding to the cases that
$p$ is a new site on an old site, an
old bond, or an old face, while $0$ is
taken to be a new site on an old,
is curious. It appears that they remain
distinct in the limit of an infinite radius,
but their separation remains bounded.
Once again the scale in the vertical direction is
very large; the curves of these diagrams are,
in fact, extremely flat except near the ends.

On the cylinder, the correlation functions behave as
$$
\langle (h(p)-h(q))^2\rangle_\mu\sim a(\mu) + b(\mu)|p-q|,
$$
at least if $p$ and $q$ lie on a common generator.
As observed in Section \ref{hh2}, the quantity $a(\mu)$ is a 
constant that depends on the mesh and on the nature of the pair
$\{p,q\}$, on whether $p$ or $q$ is a site, bond or 
face of the graph $\mathcal G$.   
If the conventions of the equation (\ref{eq:4.2}) are used,
the value of $b(\mu_{FK})$, estimated on a cylinder
of size $99\times 699$, is close to $.26$.

\begin{figure}
\begin{center}\leavevmode
\includegraphics[bb = 72 240 540 554,clip,width = 10cm]{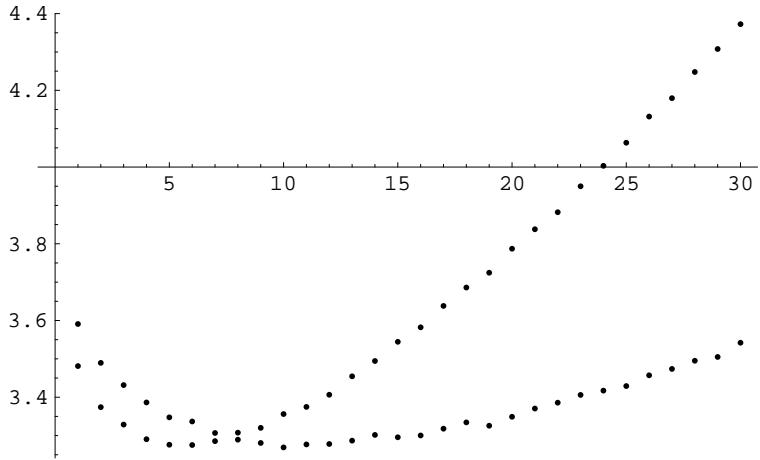}
\end{center}
\caption{The numbers $\hat\omega_k$ for $1\le k\le 30$ for
the cylinders $149\times339$ and $299\times679$.\label{fig:7.7}}
\end{figure}

\begin{figure}
\begin{center}\leavevmode
\includegraphics[bb = 72 240 540 554,clip,width = 10cm]{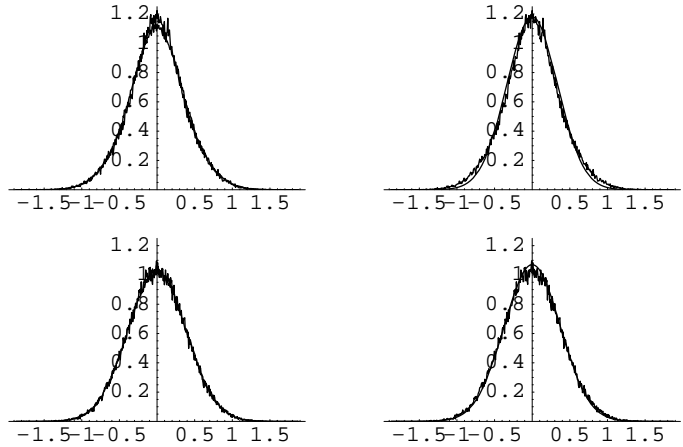}
\end{center}
\caption{The distribution constructed according to the FK-definition
as a function of $\Re A_1$ or $\Re A_5$.\label{fig:7.8}}
\end{figure}

 We studied the distribution of the function $h$ for two sizes of cylinder, 
$149\times339$ and $299\times679$, 
but only on the boundaries, not on inner circles.
These cylinders are a little short, so that about $8/1000$
of the samples are such that the sum of the jumps on a
circumference are not $0$, but passing to longer cylinders
of size $149\times449$ and $299\times899$, although it reduces this fraction
to $2/1000$ does not change the conclusions.  
The measures on the boundary appear to be gaussian once again, but with a new
value of $g_B$ that is a little greater than $3$. (Given the behavior of
$\omega_k$ of Figure \ref{fig:7.7} it is not so clear what $g_B$ is to be.
Further study might suggest defining it by the asymptotic behavior
of $\omega_k$.) 
We plot the values
of $\hat\omega_k$ for $1\le k\le 30$ on Figure \ref{fig:7.7}. 
The results, coarse as they are, are similar to those
described in Section \ref{dis1}, although there are curious features that
advise against hasty conclusions. The collection of values
for the two cylinders cross at $k=7$. Graphs of the distributions of
${\mathfrak R}A_1$ and ${\mathfrak R}A_5$ appear in Figure \ref{fig:7.8}. 
On the left the results for the cylinders
of different sizes are compared with each other;
on the right the results for the largest of the
two cylinders are compared with gaussians. Figures \ref{fig:7.7} and \ref{fig:7.8}
together
suggest
that the behavior of the function $h$ constructed according to the FK-definition
might have similarities with that of the function constructed by the methods of this paper,
but we have not examined the matter carefully. 
In particular, we tested neither conformal
invariance nor universality.

\vspace{0.5 cm}
 
\subsection{Infinite temperature.}

\begin{figure}
\begin{center}\leavevmode
\includegraphics[bb = 85 260 522 540,clip,width = 15cm]{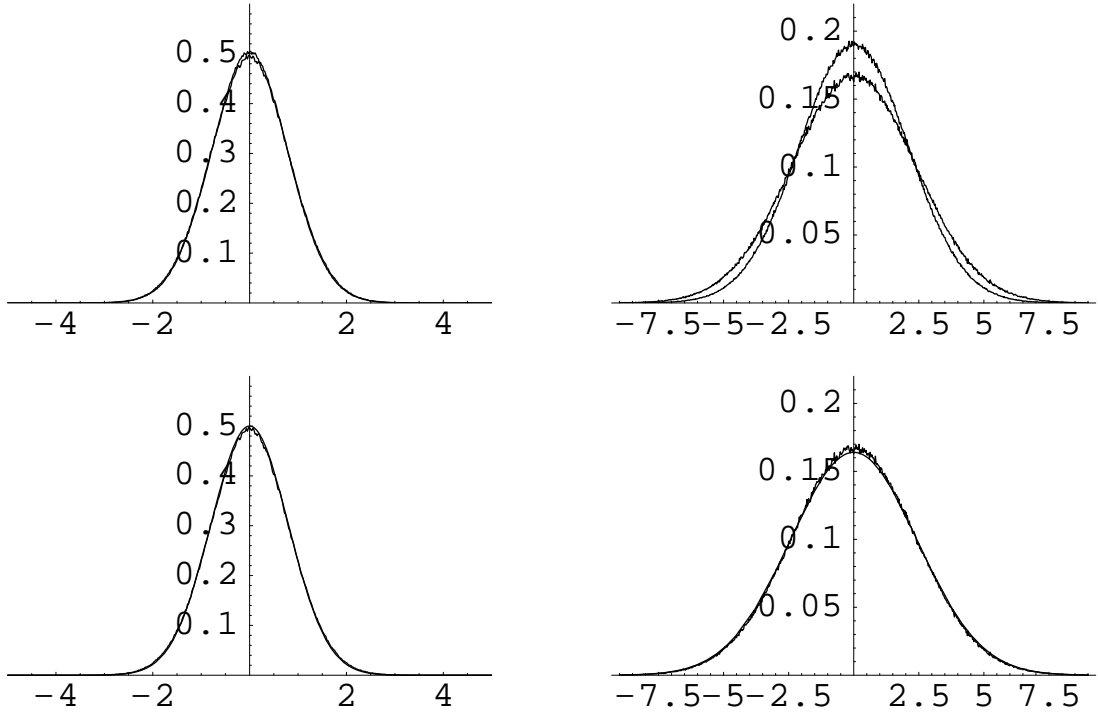}
\end{center}
\caption{The distributions $\Re A_1$ and $\Re A_{10}$ at infinite
temperature.\label{infty.ABCD}}
\end{figure}

\begin{figure}
\begin{center}\leavevmode
\includegraphics[bb = 72 330 540 500,clip,width = 15cm]{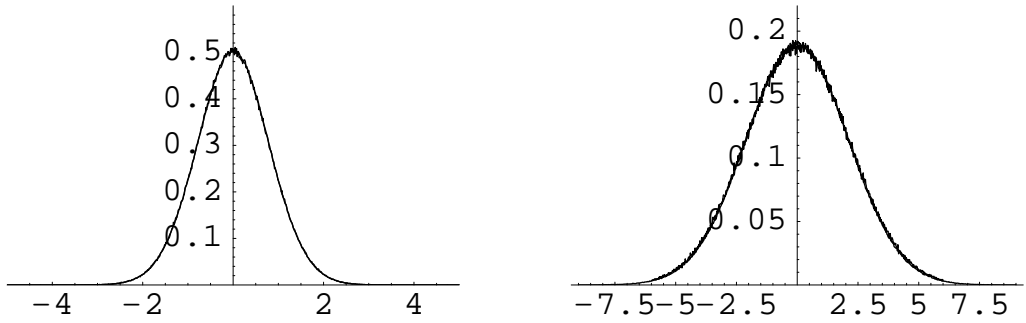}
\end{center}
\caption{The distributions $\Re A_1$ and $\Re A_{10}$ at infinite temperature compared 
for square and triangular lattices.\label{infty.GH}}
\end{figure}

\begin{figure}
\begin{center}\leavevmode
\includegraphics[bb = 72 330 540 500,clip,width = 15cm]{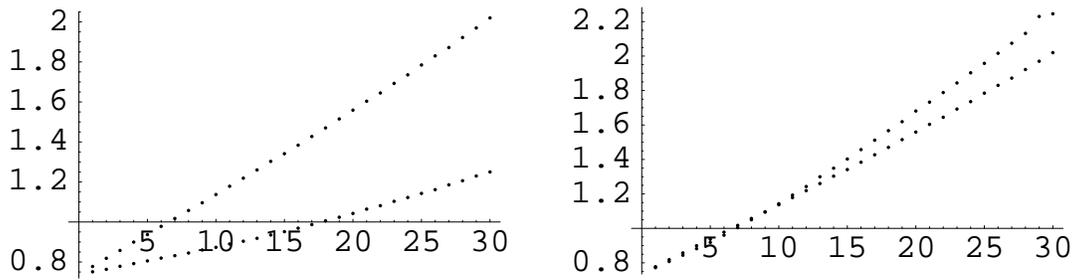}
\end{center}
\caption{The numbers $\hat \omega_k$ at infinite temperature.\label{infty.EF}}
\end{figure}

 For the Ising model at infinite temperature, thus for
site percolation in which each site is open with probability $.5$,
the crossing probabilities cease to be of interest. They are all
$0$ or, in exceptional and trivial cases, $1$. On the other hand,
the partition functions $Z^\infty(\psi_1,\psi_2,x,q)$ seem to behave much like
those at the critical temperature.
In Figure \ref{infty.ABCD} we present results for the square lattice
on cylinders of size $99\times399$
and size $299\times1199$. On the top, the results for 
${\mathfrak R}A_1$ and ${\mathfrak R}A_{10}$
for these cylinders are compared with each other.
On the bottom the results for the largest of the two
cylinders are compared with a gaussian.
As in Figure \ref{fig:2.1}, there has been no renormalization
of these distributions, so that if the distributions were similar
to those of Section \ref{dis1} the ratio of the heights of the two
curves would be $1/\sqrt10\approx.32$. It is about $.34$, but the cylinders
are still fairly small. 
Although this has no perceptible consequences, these cylinders are
short enough that about $15/10000$ of the sample states yield jumps
whose sum along a circumference is not $0$, so that the states
at the two ends are certainly not independent.
In Figure \ref{infty.GH} the results
for the smallest cylinder are compared with those
for a cylinder on a triangular lattice of size $116\times401$. 
This is a very stubby cylinder, but, curiously enough, once again only 
about $15/100000$ of the states are such that the sum of the jumps
along a circumference is not $0$. Figure \ref{infty.EF}
is analogous to 
Figure \ref{fig:2.2}: the two sequences of points on the left are for the square lattice,
the upper for the smaller of the two cylinders, the lower for the larger;
the two sequences of points 
on the right are for the smaller of the cylinders with a square
lattice (lower set) and for the cylinder with 
the triangular lattice (upper set). If Figure \ref{infty.EF} is to be believed
the constant $2R_B$ changes
and becomes approximately one-half its previous value, but, 
as with the other examples of this section, our aim was more qualitative than
quantitative.

 We can also introduce, as in Section \ref{huit}, the partition functions
$Z^\infty_{++}$ and $Z^\infty_{+-}$
or the measures $Z^\infty(\psi,x)$. A little reflection shows
that the two numbers do not
depend on $q$ and are now both equal to $\frac12$.
The analogue 
$$
Z^\infty_{++}\delta_0+Z^\infty_{+-}\delta_\pi=m_q^\infty(\{b_k=0\},x)
$$
of an equation deduced from equations (\ref{eq:5.1}) and
(\ref{eq:5.2}) is not valid, rather
the simulations suggest that, if
$Z^\infty(0,x,q)$ exists, then it peaks at
$\pi/2$ and $3\pi/2$. On the other hand, $Z^\infty(0,0,x,q)$
has quite a different behavior and a relation 
between $Z^\infty_{q=0}(\{a_k\},\{b_k\},x)$
and $Z^\infty_{q=0}(\{a_k\},x)$ like that of equation
(\ref{eq:old}) is difficult to ascertain because the functions 
$h$ do not very often have level lines that encircle the cylinder,
even when the cylinder is very long.

\begin{figure}
\begin{center}\leavevmode
\includegraphics[bb = 72 250 540 540,clip,width = 14cm]{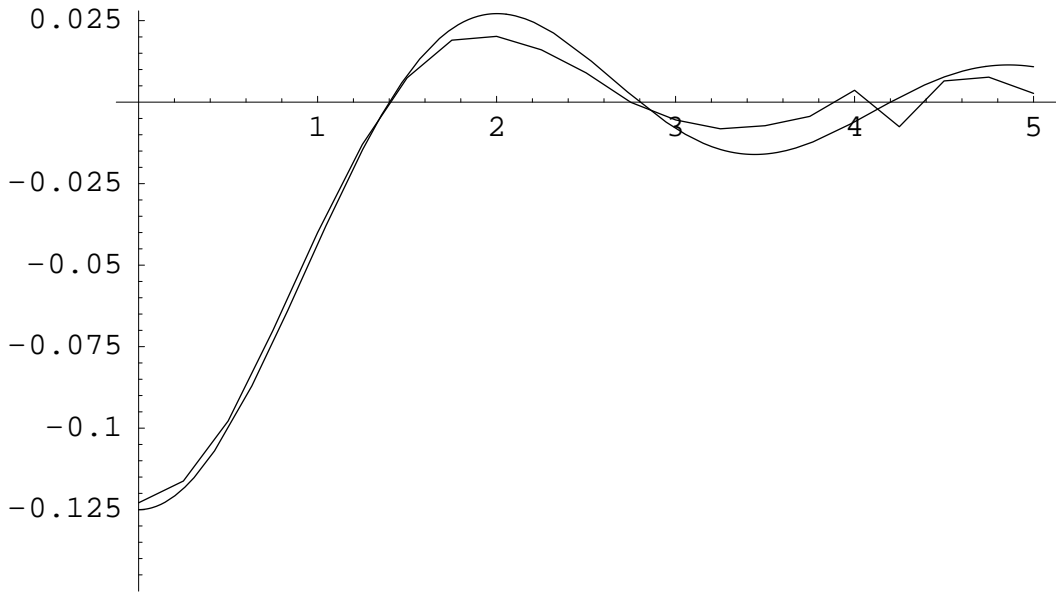}
\end{center}
\caption{An unsatisfactory but curious approximation.\label{infsin}}
\end{figure}

 More pertinent to the study of the measures at criticality is that
the behavior of 
$$
\int Z^\infty(\psi_1,\psi_2,x)d\psi_2/\int Z^\infty(\psi_1,\psi_2,x)d\psi_2dx
=\sum\mu_k^\infty(\psi)\exp(ikx)
$$
is similar to that of $Z_0(\psi,x)$. This may be of some advantage
for numerical studies since at infinite temperature no
thermalizations are necessary. Consider for example the analogue $f^\infty$ of
the function defined by equation (\ref{eq:5.b}). 
There is, once again, a simple, rough, but inexact -- as is clear from Figure \ref{infsin}
-- approximation to this function,
$$
   f^\infty(x)\sim a\frac{\sin(b\pi x)}{b\pi x},
$$
but, as before, we were unable to improve upon it in a useful fashion.

\section{Afterword.}\label{concl}

Most of the phenomena we have examined
in the paper are manifestations of the influence of the boundary, so 
that to some extent the 
thrust of the paper runs counter to the notion that statistical mechanics,
especially as it refers to critical phenomena,
is the study of bulk properties. Since
critical behavior appears when the appropriate equilibrium between
the strength of the interactions and the number of paths by which the
interaction is transmitted over long distances is achieved it is not, from 
a mathematical standpoint, such a bad idea to study criticality
by examining the consequences
of systematically blocking large numbers of these paths.
On the other hand, renormalization is usually conceived in bulk terms.
So it is a relief that the distributions investigated here, whose ultimate purpose is
the introduction of a concrete notion of fixed point,
do not become trivial when the boundary moves off to infinity.

For the Ising model, however, in contrast to percolation
or to the free boson, there are formal difficulties
in the introduction of a closed renormalization transformation
that we are still unable to overcome, even
with measures that continue to have a meaning in the bulk.
 
One connection that we would like to make,
and that is suggested by some of the
experimental results of the paper, is with
the notion of conformally invariant field theory in the strongly
geometric form envisaged by Graeme Segal (\cite{GS1,GS2}).
It may be that the basic objects of that theory are
constructible from the measures examined in this paper.
Recall that in that theory one of the first objects to
construct is a Hilbert space
$$
\mathcal H=\sum_\alpha \mathcal H_\alpha\otimes\mathcal H_{\bar \alpha}
$$
associated to a circle with parametrization. In addition,
suppose that we are given a Riemann surface $\Sigma$ with boundary $C$
consisting of disjoint parametrized circles $C_1,\dots,C_m$ and $C'_1,\dots,C'_n$,
the parametrizations being given by real 
analytic functions. Then (equation (1.4) of \cite{GS1})
the theory is provided with an operator
$$
  U_\Sigma:\quad\mathcal H^m\rightarrow\mathcal H^n.
$$

We might suppose that $\mathcal H$ is the $L^2$-space of a measure $\mu$
on the space
of distributions on the parametrized circle.
One such measure whose existence is suggested by the experiments of this
paper is the measure $\mu$ on distributions on a circle in the plane or on
the central circle of an infinite cylinder described at the end of
Paragraph \ref{simconf}, thus the measure defined in the
bulk. It is possible that $L^2(\mu)$
is, if not $\mathcal H$, then the vacuum sector 
$\mathcal H_{\alpha_0}\otimes\mathcal H_{\bar\alpha_0}$ or
some other subspace of $\mathcal H$.

Consider the annulus $\Sigma_q$ of inner radius $q$ and outer
radius $1$ and the operator $U_q$ associated to this 
surface. Take $C_2$ to be the outer circumference with the
natural parametrization, $C_1$ the inner,
and $C$ to be their union. We consider the annulus as imbedded in the plane
or, if we treat it as a cylinder of finite length, as being imbedded
in a cylinder extending to infinity in both directions.
The construction of bulk measures suggested in Section \ref{dis2} yields
experimentally a measure $m_q=m_{\Sigma_q,C}$
on the product of the spaces of distributions on $C_1$ and $C_2$.
If, as we might suppose, $m_q$ is absolutely continuous with respect
to $\mu\times\mu$ then it is given by a kernel
$$
   dm_q(\psi_2,\psi_1)=K_q(\psi_2,\psi_1)d\mu(\psi_2)d\mu(\psi_1).
$$
It is not impossible that the operator $U_q$,
or rather its restriction to the sector
represented by $L^2(\mu)$,
is given by
$$
  U_qF(\psi_2)=\int K_q(\psi_2,\psi_1)F(\psi_1)d\mu(\psi_1).
$$

An essential feature of these operators would be the relation $U_q=U_{q_1}U_{q_2}$ 
when $q=q_1q_2$ which would follow from a relation
\begin{equation}
  \int K_{q_2}(\psi_2,\psi)K_{q_1}(\psi,\psi_1)d\mu(\psi).\label{eq:conv}
\end{equation}
Let $C$ be the circle that separates the annulus of parameter
$q$ into annuli of parameters $q_1$ and $q_2$.
We apply the notions of conditional probability and the markovian
property, as well as the obvious symmetry of $K_q$, to the bulk measures. Thus
\begin{align*}
K_q(\psi_2,\psi_1)d\mu(\psi_2)d\mu(\psi_1)&=dm_q(\psi_2,\psi_1)\\
  &=\int dm_q(\psi_2,\psi_1|\psi)d\mu(\psi)\\
  &=\int dm_{q_1}(\psi_1|\psi_2|\psi)dm_{q_2}(\psi_2|\psi)d\mu(\psi)\\
  &=\int dm_{q_1}(\psi_1|\psi)dm_{q_2}(\psi_2|\psi)d\mu(\psi)\\
  &=\{\int K_{q_1}(\psi,\psi_1)K_{q_2}(\psi_2,\psi)d\mu(\psi)\}d\mu(\psi_1)d\mu(\psi_2),
\end{align*}
from which the equation (\ref{eq:conv}) would follow.

These are tentative suggestions, and we only make them
to confess that we have not yet had an opportunity
to test them experimentally. That may not be an easy matter.
Nor do we know whether they are confirmed by
the conventional wisdom.
To construct some analogue of Segal's operators on the whole $\mathcal H$
it may be necessary to utilise the phase
of Section \ref{huit}, but here again more reflection is necessary. 

Another set of experiments waiting to be performed, although
here the outcome is more certain
and the experiments therefore less tempting, is an examination of the
behavior of the measures in a neighborhood of the critical point
as we vary $J$ (or the temperature) and introduce a small
magnetic field. The limits 
as the mesh goes to $0$ are expected to exist no longer,
but the behavior of the measures, of their moments
for example, should yield the usual critical exponents
$\nu$ and $\Delta$
and should correspond to the usual intuition. We are nevertheless
curious to see how the geometry of the fixed point is reflected
in the coordinates introduced in this paper and
to see, in particular, which linear combinations become 
irrelevant.

We have also not pursued the study of other models,
the Potts model, the $n$-vector model and so on. The examples 
of Section \ref{alter} indicate a surprising sensitivity to
the definitions that it would be useful to examine further. For the Ising model we made,
more by good luck than good management, 
a particularly happy choice which it is not utterly clear
how to generalize to other models. 

\section*{Appendix.}

The present work contains simulations of both qualitative
and quantitative nature. We aimed in most of Sections
\ref{dis1} to \ref{hh2} to provide numerical
results reproducible to the precision of statistical errors.
It is therefore important that we be precise
about our conventions. Though many details are given in the text we
complete them here with technical additions. In Paragraphs
\ref{condprob} and \ref{sec53}
and in Section \ref{alter} the work is mostly qualitative
and the reader who wants to examine these matters further will need
to devise his own experiments.

\subsection*{Distribution $m_D$.}

\begin{figure}
\begin{center}\leavevmode
\includegraphics[bb = 72 210 540 580,clip,width = 10cm]{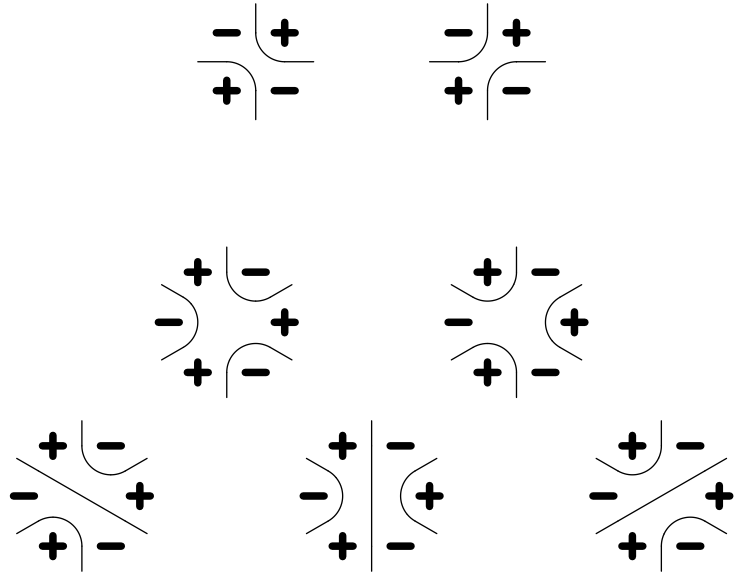}
\end{center}
\caption{Configurations of jump lines on the square 
and the hexagonal lattices.\label{fig:a.2}}
\end{figure}

 According to the principles of the introduction, each possible function
$h$ lying above a given $\sigma$ is to be assigned the same measure. This
principle has to be incorporated into the programs locally. For example, 
there are two possibilities for the configuration of jump lines 
(or level curves) passing through the
center of the configuration appearing
in the first row of Figure \ref{fig:a.2}.
They are chosen with equal probability. Since the curves are
constructed one at a time by adjoining edges, when we first adjoin an edge
passing through the center we then turn to the left or right with equal probability
$1/2$. The next time we pass through the center there is no choice; there is
only one unused successor remaining. For a triangular lattice, there are no
ambiguous configurations. For the hexagonal lattice, 
all combinations of $+$ and $-$ around vertices of the dual lattice
lead to at most two possible choices of jump lines (and they are
then treated as in the square lattice) except for the configurations
in the two last rows of Figure \ref{fig:a.2} for which there are five
possible local configurations of jump lines.
Each will then have the probability $\frac15$. 
As a consequence, when a curve first 
passes through the
center of this configuration it continues on a
straight line with probability $\frac15$ (which then leads
necessarily to one of the configurations in the
bottom row) or makes a sharp reverse turn to the
left or the right with equal probabilities
$\frac25$. If the first curve through
the center is straight, the following
curves are determined. Otherwise the next curve, which may very well
be a continuation of the first, returning after perhaps 
extensive wandering, has two options, each chosen with probability
$\frac12$. 

\begin{figure}
\begin{center}\leavevmode
\includegraphics[bb = 72 240 540 554,clip,width = 10cm]{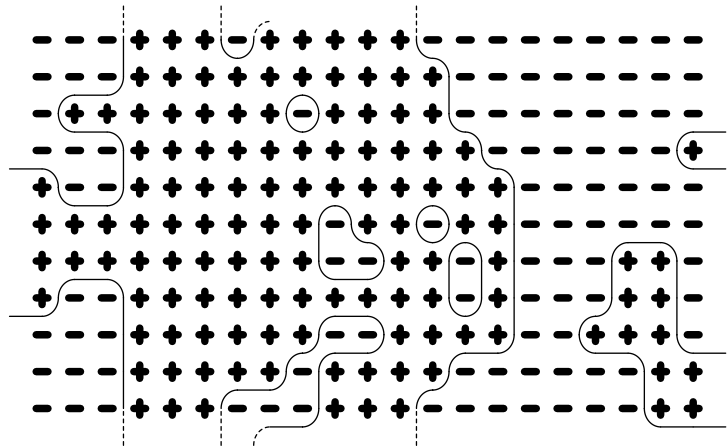}
\end{center}
\caption{A configuration on a $11\times 21$ cylinder with the jump lines
of $h$.\label{fig:a.1}}
\end{figure}

 For the square lattice, two examples of the random determination
of $h$ occur in 
Figure \ref{fig:a.1} where a configuration was drawn together with the 
jump lines of $h$. If the site at the bottom-left corner has coordinate
$(1,1)$, then four clusters meet at $(9\frac12,2\frac12)$ and at 
$(13\frac12,5\frac12)$. In the first occurence, the two minus-clusters 
are joined and, in the second, they are separated. 
By definition the jump lines occur on edges dual to lattice bonds. Their
vertices were rounded in this figure to show clearly the difference between
joining and separating. The jump lines that 
wrap around the cylinder are indicated by dashed lines. 

The restrictions of $h$ (on both the cylinders and the squares) were taken
along several curves $C$. For the square lattices the curves were taken along
lines of sites so that the intersection with dual bonds is unambiguous. 
The triangular lattices were oriented
such that longitudinal lines with sites had one site per mesh unit. The
longitudinal lines with sites of the hexagonal lattices had the 
pattern site-site-vacant repeated over every three-mesh cycles. For the
triangular and hexagonal lattices the conformal images of the curves $C_i$ 
on the cylinder never contained dual bonds parallel to them. They were
however moved slightly to the closest position where
their intersections with dual bonds were equally spaced. For the curves
$C=C_0$ at the boundary they were chosen as the curve closest from 
the boundary satisfying the previous requirement. 

We also measured the distributions on a disk of radius $r=300.2$.
The center of the disk was a site. All the sites inside the disk of radius
$r$, and only these, were thermalized. Some of the boundary sites had three
neighbors, others only two. We then determined an effective radius 
$r_{\text{eff}}$ as the radius of the largest circle that intersects only dual
bonds associated to sites in the disk. It turned out to be $r_{\text{eff}}=
299.50$. The restriction of $h$ at the boundary was obtained along the
circle of radius $r_{\text{eff}}-\epsilon$ with $\epsilon=0.001$.
The jumps in $H$ are of the form $\pm \pi \delta(\theta-\theta_0)$
where $\theta_0$ is the position of the intersection on the circle
of the dual bond with the curve $C$. The exact positions of all the
intersections with dual bonds were determined and used to compute
Fourier coefficients. The radius of inner circles were determined as
fractions of $(r_{\text{eff}}-\epsilon)$. For example $C_0$ and $C_1$
are at 8 mesh units from one another on the $397\times 793$ cylinder
and the radii of the corresponding curves on the disk should be
$(r_{\text{eff}}-\epsilon)$ and $0.8811(r_{\text{eff}}-\epsilon)$
since $e^{-2\pi \cdot 8/397}\sim 0.8811$.

Initial thermalization was provided by a few thousand Swendsen-Wang
sweeps starting from a random configuration for the smaller cylinders, 
by 5000 sweeps for $397\times793$ and by 10000 or more for $793\times1585$.
After the initial thermalizations, measurements were taken every third 
Swendsen-Wang sweep for all the cylinders, except for the $793\times 
1585$ for which we used a 5-sweep cycle. A quick time-series analysis 
indicated that these cycle lengths insured proper statistical independence
of consecutive measurements. The pseudo-random number generator was the one
proposed by Tezuka and L'\'Ecuyer in \cite{TL}.

Programs for the square lattice on the cylinders and on the disks were
written independently by at least two of the authors and errors were
chased down until measurements agreed within the statistical errors
reported in the text.

It might not be clear, on reading the main text, why certain data 
are given for some of the geometries studied and not for others. 
It is because  
the results for the various runs were kept in two different formats. For the 
first format, the observed values of each random 
variable $\Re A_k$ or $\Im A_k$ were grouped into 401 bins of equal size.  
Thus for each random variable, 401 nonnegative 
integers were stored. The width of the bins 
used during the first months 
was a little too narrow and some of the values  
fell 
outside the range covered. Later, in the final months,  
the bin width was adjusted to avoid this difficulty. When we
used the first format, we also kept, most of the time, the sum, the sum of the 
squares and the sum of mixed products of the random variables. These allowed
us to calculate accurately the two first moments of the distribution and the 
correlation coefficients. The second format was
more thorough. For each configuration of spins $\sigma$, we constructed 
one possible $h$ and recorded its restriction to the various  
curves $C$, not simply to the extremities 
of the cylinder or its median. When we realized that the conformal invariance 
might hold in the sense of Section \ref{dis2}, we kept the restriction
of $h$ to more curves. These data bases, with only the boundary as $C$ or with
several curves $C$ for each
configuration, are sizeable even when compressed (a few Gigabytes in
all). We generated one for the cylinder $397\times793$ with eleven curves 
$C$. For $59\times401$ and $157\times1067$ we only kept the 
restriction of $h$ to one extremity. With the second format it was
possible to test various assertions that we could
not have anticipated without the experience gained
from the experiments, but the first required far less memory, so that
more sizes were examined.

\subsection*{Crossings.}

To determine the aspect ratio of the rectangles where the crossings
are measured we have used the width and the height of the smallest rectangle
that contains the sites of the lattice considered. For example, for
the rectangles of $LV\times LH$ sites of the square lattice,
the ratio is $LH/LV$. We used here the orientations of the
lattices used for the measurement of $m_D$ (see above).

For $\pi_h, \pi_v$ and $\pi_{hv}$ on rectangles, crossings started on $+$
spins from one boundary and ended on $+$ spins on the other. For $\pi_h^A$
and $\pi_v^A$, the crossings were required to reach the central
meridian if it contained sites or, if it did not, to reach the line of sites just 
before. The dimensions of the rectangles for the square lattice were the
same as those used for percolation crossings in \cite{LPPS}. The results
for the triangular lattices are given in Table VII. The dimensions $\{LV, LH\}$
that were used are 
$$\begin{matrix}
   \{586, 69\}, & \{596, 73\}, & \{566, 73\}, & \{582, 79\}, & \{540, 77\}, & \{554, 83\}, \\ 
   \{566, 89\}, & \{616, 103\}, & \{508, 89\}, & \{474, 87\}, & \{504, 97\}, & \{526, 107\}, \\ 
   \{456, 97\}, & \{476, 107\}, & \{438, 103\}, & \{440, 109\}, & \{424, 111\}, & \{410, 113\}, \\ 
   \{512, 147\}, & \{420, 127\}, & \{400, 127\}, & \{400, 133\}, & \{354, 125\}, & \{348, 129\}, \\ 
   \{398, 155\}, & \{336, 137\}, & \{314, 135\}, & \{386, 175\}, & \{318, 151\}, & \{302, 151\}, \\
   \{310, 163\}, & \{276, 153\}, & \{366, 213\}, & \{270, 165\}, & \{310, 199\}, & \{356, 239\}, \\
   \{258, 183\}, & \{248, 185\}, & \{282, 221\}, & \{324, 267\}, & \{232, 201\}, & \{232, 211\}, \\
   \{210, 201\}, & \{200, 201\}, & \{224, 237\}, & \{196, 219\}, & \{196, 229\}, & \{190, 233\}, \\
   \{184, 237\}, & \{184, 249\}, & \{176, 251\}, & \{288, 431\}, & \{164, 259\}, & \{176, 291\}, \\
   \{152, 265\}, & \{156, 287\}, & \{148, 285\}, & \{210, 425\}, & \{152, 323\}, & \{278, 625\}, \\
   \{152, 359\}, & \{148, 367\}, & \{132, 345\}, & \{140, 381\}, & \{126, 361\}, & \{182, 551\}, \\
   \{132, 421\}, & \{110, 367\}, & \{116, 409\}, & \{112, 413\}, & \{126, 491\}, & \{106, 433\}, \\
   \{110, 471\}, & \{154, 691\}, & \{108, 515\}, & \{112, 561\}, & \{96, 505\}, & \{116, 641\},  \\
   \{92, 535\}, & \{108, 661\}, & \{90, 573\}. &&& 
\end{matrix}$$
The ratio $r$ is given by $r=2 LH/\sqrt{3} LV$
as $LH$ and $LV$ count the number of lines and columns of sites.

For $\pi_h, \pi_v$ and $\pi_{hv}$ on the disk, crossings started from and ended
on sites in the annulus between $r=300.2$ and $r-\sqrt2$. The crossings
for $\pi_h^A$ and $\pi_v^A$ had to reach the central diameter.

On cylinders the crossings between the curves $C_i$ started from and ended on the curves. On the disk the five curves were chosen at
radii $\hat r, 0.8811\hat r, 0.7763\hat r, 0.6026\hat r, 0.3632\hat r$
with $\hat r=300.2 - \sqrt2$. The crossings from $C_i$ to $C_j$ ($r_i>r_j$)
started outside the outer curve $C_i$ and ended inside the inner $C_j$.

The programs for all lattices and geometries were written by two of
us and checked until they agreed within the statistical errors for a sample
larger than $10^6$ even though most crossings were measured with samples
of $\sim 200$K. (See Section \ref{trav} for the samples used for the various
lattices and geometries.)

\subsection*{The phase $x$.}

The phase $x$ measured by the experiments is described in Section \ref{huit}.
For Figure \ref{fig:5.3}, results from cylinders of the following sizes were plotted 
$$\begin{matrix}
\{59,  27\}, &
\{59,  37\}, &
\{59,  47\},&
\{59,  61\}, &
\{59,  73\}, &
\{59,  93\}, \\
\{59, 119\}, &
\{59, 147\},&
\{59, 179\}, &
\{59, 211\},&
\{59, 249\}, &
\{59, 283\},
\end{matrix}$$
all with at least
$400$K configurations each, and 
$$\begin{matrix}
\{117,  53\},&
\{117,  73\},&
\{117,  95\},&
\{117, 123\},&
\{117, 145\},&
\{117, 187\},\\
\{117, 239\},&
\{117, 293\},&
\{117, 357\},&
\{117, 421\},&
\{117, 499\},&
\{117, 565\},
\end{matrix} $$
with at least $600$K configurations.

The distribution of the random variable $x$ is also used
to obtain the ratios $b/a$ through the constrained integrals
(\ref{intconst}) and (\ref{intconst2}). The errors on the
ratios $b/a$ appearing in Table VIII are difficult to evaluate
as the numbers $a$ and $b$ are the local maxima of a
smoothed distribution. For the integral (\ref{intconst}), the
most difficult to measure, the samples varied between $31$K
and $50$K. After experimentation with various smoothing parameters
we think that the two first digits of the ratios $b/a$ for
the case {\it constrained/constrained} are exact. The accuracy
for the other cases is far better, the samples being at least
$85$K for the {\it constrained/fixed} and $300$K for the
{\it fixed/fixed}.

\subsection*{The correlation $\langle (h(p)-h(0))^2\rangle$.}

These correlations can be measured in a straightforward way using the
above definition of $h$ and the details in the text.

\include{caption}

\end{document}